\documentclass[preprint]{siamart190516}

\usepackage{pdfpages}
\graphicspath{{figures/img-lo-res/}}  

\usepackage{amsmath}
\usepackage{bigints}
\usepackage{amssymb}
\usepackage{mathtools}
\usepackage{algorithm}
\usepackage{algpseudocode}
\usepackage{dsfont}

\usepackage{tikz}
\usetikzlibrary{calc}
\usepackage{pgfplots}
\usetikzlibrary{arrows,decorations.markings}
\usepackage{subcaption}
\usepackage[export]{adjustbox} 
\usepackage{rotating}
\usepackage{array}  
\usepackage{wrapfig}  

\usepackage{soul} 
\usepackage{float}
\usepackage{appendix}
\usepackage{enumitem}
\usepackage{refcount}
\usepackage{marginnote}
\usepackage{scrextend} 
\usepackage{hyperref} 
\usepackage{cleveref}[2012/02/15]  
\crefformat{footnote}{#2\footnotemark[#1]#3}

\DeclareMathOperator*{\argmin}{arg\,min}

\DeclareMathOperator{\Hess}{Hess}
\DeclareMathOperator{\cof}{cof}

\newcommand{\one}{\mathds{1}}
\newcommand{\eye}{\mathds{I}}
\newcommand{\zero}{\mathds{O}}

\renewcommand*{\thefootnote}{\fnsymbol{footnote}}  

\newsiamthm{assumption}{Assumption}

\def\failurecaption{-16pt}         

\usepackage{amsfonts}
\usepackage{graphicx}
\usepackage{epstopdf}
\ifpdf
  \DeclareGraphicsExtensions{.eps,.pdf,.png,.jpg}
\else
  \DeclareGraphicsExtensions{.eps}
\fi

\newsiamremark{remark}{Remark}
\newsiamremark{hypothesis}{Hypothesis}
\crefname{hypothesis}{Hypothesis}{Hypotheses}
\newsiamthm{claim}{Claim}

\headers{Simulating Cascading Transmission Line Failure}{Roth, Barajas-Solano, Stinis, Weare, and Anitescu}

\title{A Kinetic Monte Carlo Approach for Simulating Cascading Transmission Line Failure\thanks{Submitted to the editors DATE.
\funding{This material was based upon work supported by the U.S. Department of Energy, Office of Science, Office of Advanced Scientific Computing Research (ASCR) under Contract DE-AC02-06CH11347. We acknowledge partial NSF funding through awards FP061151-01-PR and CNS-1545046 to MA.}}}

\author{Jacob Roth\thanks{Mathematics and Computer Science Division, Argonne National Laboratory, Lemont, IL 60439
  (\email{jmroth@mcs.anl.gov, anitescu@mcs.anl.gov}).}
\and David A. Barajas-Solano\thanks{Pacific Northwest National Laboratory, Richland, WA 99354
	(\email{David.Barajas-Solano@pnnl.gov, panos.stinis@pnnl.gov}).}
\and Panos Stinis\footnotemark[3]
\and Jonathan Weare\thanks{Courant Institute of Mathematical Sciences, New York University, New York, NY 10012
	(\email{weare@cims.nyu.edu})}
\and Mihai Anitescu\footnotemark[2]}

\usepackage{amsopn}

\ifpdf
\hypersetup{
  pdftitle={Simulating Cascading Transmission Line Failure},
  pdfauthor={Jacob Roth, David Barajas-Solano, Panos Stinis, Jonathan Weare, and Mihai Anitescu}
}
\fi

\newcommand{\pre}[2]{#1_{st}^{\tau}(#2)}

\begin{document}

\maketitle

\begin{abstract}
In this work, cascading transmission line failures are studied through a dynamical model of the power system operating under fixed conditions.
The power grid is modeled as a stochastic dynamical system where first-principles electromechanical dynamics are excited by small Gaussian disturbances in demand and generation around a specified operating point.
In this context, a single line failure is interpreted in a large deviation context as a first escape event across a surface in phase space defined by line security constraints.
The resulting system of stochastic differential equations admits a transverse decomposition of the drift, which leads to considerable simplification in evaluating the quasipotential (rate function) and, consequently, computation of exit rates.
Tractable expressions for the rate of transmission line failure in a restricted network are derived from large deviation theory arguments and validated against numerical simulations.
Extensions to realistic settings are considered, and individual line failure models are aggregated into a Markov model of cascading failure inspired by chemical kinetics.
Cascades are generated by traversing a graph composed of weighted edges representing transitions to degraded network topologies. Numerical results indicate that the Markov model can produce cascades with qualitative power-law properties similar to those observed in empirical cascades.
\end{abstract}






\renewcommand{\thefootnote}{\arabic{footnote}}

\section{Introduction}
\label{sec:introduction}
Because of recent blackout events, cascading failure in the power grid---whereby a local network disturbance may cause system-wide disruptions that result in loss of power to significant parts of the network \cite{dobson_complex_2007,simonsen_transient_2008,hines_topological_2010,witthaut_nonlocal_2015,tamrakar_propagation_2018}---has become an increasingly important field of study.
The uncertain nature of power demand fluctuations in electrical transmission networks, together with the increasing penetration of uncertain power sources such as renewable energy, has made the threat of cascading failure more pronounced.
The ability to quantify the probability of infrastructure component failure due to generation and demand variation is critical to the safe operation of the power grid.
Furthermore, it is important to understand how these failure probabilities are related to the controllable properties of the network.

Previous work in power system reliability has developed methods for studying both individual component and cascading grid failures.
In \cite{wadman_large_2016,nesti_temperature_2019}, Monte Carlo simulation and large deviation arguments were employed to estimate the probability of individual component failure.
Among the various models of cascading failure that have been proposed, a central theme is that of a power law governing the relationship between cascade severity and frequency \cite{dobson_complex_2007,rohden_cascading_2016}.
In previous work, Dobson \cite{dobson_complex_2007} identified economic and engineering factors that counterbalance each other to drive the system's operating state toward a critical point.
The ``slow'' dynamics help explain the prevalence of power laws describing cascade severity and have inspired recent work studying the cascading failure mechanism and its relation to power grid operating conditions.
In particular, recent approaches have analyzed cascading failures with rare event simulation methodologies (such as splitting and importance sampling) or specialized models that address certain properties of the grid (such as loading levels, as in \cite{sloothaak_robustness_2018}).

In this work, we continue recent trends that analyze cascading failures with rare event methodologies, but we rely on the theory of large deviations to computationally streamline the calculation of failure probabilities.
In particular, we model the dynamics of the power grid from a first-principles, physically inspired approach and reduce the problem of dynamical Monte Carlo simulation at low temperatures to the solution of a nonlinear program (NLP).
Drawing inspiration from \cite{demarco_1987_security}, we introduce stochastic fluctuations in demand and generation to a standard dynamical model of the classical power transmission network~\cite{bergen_1981_structure,bergen_1984_application,demarco_1985_small,zheng_2010_bistable,zheng_2016_new,yang-motter_2017_continuous} and interpret transmission line failures as first exit events for the resulting stochastic dynamics.
Despite several assumptions of these classical system models, which include disregarding transmission losses, they have favorable properties such as port-Hamiltonian structure \cite{vanderchaft_port_2014} that render them amenable to analytical treatment.

We employ the theory of small noise transitions for irreversible diffusion processes~\cite{bouchet_2016_generalisation,wf_random_2012,kushner_1984_approximation} to derive asymptotically valid expressions that describe the rate of failure of transmission lines in a restricted network.
Our analytic expression (i) characterizes the probability density of the time to failure for system dynamics initialized at a stable equilibrium, or ``operating point,'' (ii) is computable, and (iii) provides a clear and interpretable relation between line failure probability and the properties of the power grid.
Based on the analytic approach to describing line failure, we adapt our model to more realistic scenarios (including finite-temperature regimes and more complex networks), and we develop a composite model of cascading failure that aggregates the models of individual component failure.

This manuscript is structured as follows.
In \cref{sec:rate} we study the individual line failure problem.
We present the stochastic dynamical model of the power grid, define line failure in a restricted model of the power grid, derive expressions for the asymptotic failure rate in this context, and present numerical results that justify our analysis.
In \cref{sec:adapt} we study potential pathologies of the line failure problem that complicate the straightforward application of results derived in \cref{sec:rate} to real systems.
We find that the numerical evidence supports the claim that the simplified, tractable problem studied in \cref{sec:rate} may be a reasonable approximation of the line failure rates of practical interest.
From our findings, in \cref{sec:markov}
we employ our model of individual line failure to develop a chemical-kinetics-inspired Markov simulator of cascading line failures.
In this section, we present validation results in the context of observed cascade data and discuss numerical and theoretical properties of the discrete state space Markov model.
In \cref{sec:conclusions} we present our conclusions and briefly outline areas for future research.

\subsection*{Contributions}
The key contributions of our work are (i) the development of a tractable expression for the failure rate of an individual transmission line due to small perturbations in power demand and generation and (ii) the assembly of individual failures in a kinetic Monte Carlo (KMC) setting to describe macroscopic properties similar to real systems.


\section{Unconditional failure rate}
\label{sec:rate}
In this section we proceed to define the unconditional line failure problem (\cref{def:unconditional-failure}), derive analytic expressions for the rate of failure (\cref{eq:0-order-qst-final,eq:1-order-qst-final}), and present validation results (\cref{fig:rate-best-worst,fig:rate-modification-fit}) justifying their correctness.
\subsection{Transmission model}
\label{subsec:rate-transmission}
We model the power transmission network as the undirected graph $(\mathcal{B}, \mathcal{E})$, where $\mathcal{B}$ denotes the set of $N$ network buses (graph nodes) and $\mathcal{E}$ denotes the set of transmission lines (graph edges).
The set of buses $\mathcal{B}$ is split into three disjoint subsets
corresponding to three types of bus: (i) electricity-producing \emph{generator} buses, $\mathcal{G}$, (ii) electricity-consuming \emph{load} buses, $\mathcal{L}$, and (iii) network-balancing \emph{slack} or reference buses, $\mathcal{S}$ (here we follow standard terminology from the power systems literature \cite{bergen_1999_power}\footnote{In load flow analyses, the \textit{slack} bus serves as a source/sink node capable of satisfying any network power imbalances. In power system dynamics, the state space evolution is typically measured relative to a systemwide average phase angle (center of inertia) or a particular component; in our case we give the slack bus the dual role of both balancing the network and serving as the system's angular reference.}).
For simplicity in this work we assume that $|S| \coloneqq 1$ corresponds to a generator bus.

To model the short-term dynamics of the transmission network around a synchronous operating point, we employ a structure-preserving, energy-based model of port-Hamiltonian form \cite{vanderchaft_port_2014,stegink_2019_energy} based on the so-called swing equations (see \cite{bergen_1999_power} and the supplemental material for more detail).
Similar models have been employed for stability analysis~\cite{bergen_1984_application,demarco_1985_small,demarco_1987_security} and cascading failure analysis~\cite{zheng_2010_bistable,yang-motter_2017_continuous} of power grid systems.
In our case, we assume that the short-term dynamics remain applicable over indefinitely long time horizons, since we are interested in studying operation around the synchronous point \cite{stegink_2019_energy}.
To present the dynamics model, we employ the following notation.
For a vector $\mathbb{R}^{N}$, $[ \cdot ]_X$ denotes the subvector corresponding to the subset of indices $X \subseteq [1, N]$.
Additionally, $\eye_X$ denotes the identity matrix of dimension $|X|$, $\zero_{X \times Y}$ denotes the rectangular zero matrix of $|X|$ rows and $|Y|$ columns, $\one_X$ denotes the column vector of ones of dimension $|X|$, and $x^*$ denotes complex conjugation of $x$.

The network state at time $t$ is described by the bus voltage magnitude, $[V_t]_i$, the bus phase angle with respect to the angular reference, $[\theta_t]_i$, and the bus angular frequency $[\omega_t]_i$, for each bus $i \in [1, N]$.
The electromechanical dynamics of the power transmission network are modeled by the set of (singularly perturbed) ordinary differential equations  \cite{bergen_1984_application,demarco_1985_small}
\begin{equation}
  \label{eq:port-hamiltonian}
  \dot{x}_t = -K \nabla \mathcal{H}^y(x_t), \quad x_0 = \bar{x}, \quad K \coloneqq S - J
\end{equation}
where $x_t~\coloneqq~([\omega_t]_{\mathcal{S} \cup \mathcal{G}}^{\top}, \, [\theta_t]_{\mathcal{G} \cup \mathcal{L}}^{\top}, \, [V_t]_{\mathcal{L}}^{\top}) \in \mathbb{R}^d$, $d = 2 N - 1$, is the state vector of the network dynamics, composed of the vectors of generator angular frequencies, $[\omega_t]_{\mathcal{S} \cup \mathcal{G}}$, generator and load bus relative phase angles, $[\theta_t]_{\mathcal{G} \cup \mathcal{L}}$, and load bus voltage magnitudes, $[V_t]_{\mathcal{L}}$; $J$ is a skew-symmetric ``connection'' matrix, and $S$ is a diagonal ``damping'' matrix with non-negative entries such that \cref{eq:port-hamiltonian} is of port-Hamiltonian form \cite{vanderchaft_port_2014}.
Historically, singular perturbation models have been introduced to address limitations in static load models, particularly at low voltages, and here we follow \cite{bergen_1984_application}.
By $\mathcal{H}^y(\cdot)$ we denote the system's energy function parameterized by the network parameters $y~\coloneqq~\{[\delta]_{\mathcal{S}}, [V]_{\mathcal{S} \cup \mathcal{G}}, \, B^y, \, P^y_0, \, Q^y_0, M^g, D^g, D^d, D^{\epsilon} \}$ and given by \cite{narasimhamurthi_1984_generalized,pai_energy_1989}
\begin{equation}
  \label{eq:energy-func}
  \begin{aligned}
  \mathcal{H}^y(x) \coloneqq &\frac{1}{2} \left \langle [\omega_t]_{\mathcal{S} \cup \mathcal{G}}, \, M^g \, [\omega_t]_{\mathcal{S} \cup \mathcal{G}} \right \rangle + \frac{1}{2} \left \langle [v_t]_{\mathcal{L}}, \, B^y \, [v_t]_{\mathcal{L}} \right \rangle \\ &+ \left \langle [P^y_0]_{\mathcal{G} \cup \mathcal{L}}, \, [\theta_t]_{\mathcal{G} \cup \mathcal{L}} \right \rangle + \left \langle [Q^y_0]_{\mathcal{L}}, \, \log [V_t]_{\mathcal{L}} \right \rangle,
  \end{aligned}
\end{equation}
where $M^g$ is the diagonal matrix of generation inertia constants, $B^y$ is the network's (often sparse) susceptance matrix (weighted Laplacian), $[P^y_0]_{\mathcal{G} \cup \mathcal{L}}$ is the vector of net active power at the $\mathcal{G} \cup \mathcal{L}$ buses at operating conditions, $[Q^y_0]_{\mathcal{L}}$ is the vector of net reactive power at the load buses at operating conditions, $[v_t]_{\mathcal{L}}$ is the vector of voltage phasors defined elementwise at the load buses as $[v_t]_i \coloneqq [V_t]_i \exp j [\theta_t]_i$, $[\delta]_{\mathcal{S}}$ is the angular reference, and $\langle \cdot, \, \cdot \rangle$ denotes the (Hermitian) inner product.
Since we do not consider variations of the network parameters $y$ in this work, for the remainder of the article we drop the dependence on $y$.
This model, like other structure-preserving models, disregards transmission losses so that the line connectivity information is fully encoded in the susceptance matrix.
The matrices $J$ and $S$ are given by \cite{zheng_2016_new} as
\begin{align}
  \label{eq:J-S-matrices}
  J &=
  \begin{bmatrix}
    0 & -(M^g)^{-1} T_1^{\top} & 0 \\
    T_1 (M^g)^{-1} & 0 & 0 \\
    0 & 0 & 0
  \end{bmatrix},\\
  S &=
  \begin{bmatrix}
    (M^g)^{-1} D^g (M^g)^{-1} & 0 & 0 \\
    0 & T_2^{\top} (D^d)^{-1} T_2^{\top} & 0 \\
    0 & 0 & (D^{\epsilon})^{-1} \eye_{{\mathcal{L}}}
  \end{bmatrix},
\end{align}
where $D^g$ is the diagonal matrix of generator damping constants, $D^d$ is the diagonal matrix of real load damping constants, and $D^{\epsilon}$ is a singular perturbation parameter governing the system's proximity to instantaneously satisfying  circuit law constraints (i.e., proximity to the power flow manifold \cite{bolognani_fast_2015}).
Here, the matrices $T_1$ and $T_2$ are given by \cite{zheng_2016_new} as
\begin{equation}
  \label{eq:t-matrices-apx}
  T_1 =
  \begin{bmatrix}
    -\one_{\mathcal{G}} & \eye_{\mathcal{G}}\\
    -\one_{\mathcal{L}} & \zero_{\mathcal{L} \times \mathcal{G}}
  \end{bmatrix}, \quad
  T_2 =
  \begin{bmatrix}
    \zero_{\mathcal{G} \times \mathcal{L}} \\ \eye_{\mathcal{L}}
  \end{bmatrix}.
\end{equation}
We remark that $K \coloneqq S - J$ is invertible and positive definite (though not symmetric) \cite{zheng_2010_bistable}, and hence the system qualifies as a ``quasigradient'' system \cite{demarco_new_1988} and is only a linear coordinate-transformation away from a gradient system.
A more detailed derivation of our port-Hamiltonian model from the governing equations of power transmission network dynamics is summarized in the supplemental material.

It remains to specify the initial conditions $\bar{x}$ in~\cref{eq:port-hamiltonian}.
For given network parameters $y$, the equilibrium condition $\nabla \mathcal{H}(x) = 0$, or equivalently, $\argmin_x \mathcal{H}(x)$, defines a set of metastable critical points for the deterministic dynamics \cref{eq:port-hamiltonian} differing from one another by a factor of $2 n \pi$, $n \in \mathbb{Z}$, along each angular direction (see \cref{fig:energy-contour-wide-3bus}).
We denote by $\bar{x}$ the stable critical point closest to the origin in the angular directions, that is,
\begin{equation}
  \label{eq:xbar}
  \bar{x} \coloneqq \argmin_{x \in \mathcal{X}} |\theta(x)|, \quad \mathcal{X} = \{ x : \nabla \mathcal{H}(x) = 0,\, \nabla^2 \mathcal{H}(x) \succeq 0 \}.
\end{equation}

To account for fluctuations in power demand and generation, we introduce noise into \cref{eq:port-hamiltonian} through scaled Gaussian perturbations at appropriate indices.
Load perturbations are assumed to be the aggregate result of independent consumer device behavior, and generator perturbations are assumed to be small fluctuations in active generation capacity.
Since the perturbations directly affect net real ($P_0$) and net reactive ($Q_0$) power, which enter the energy gradient $\nabla \mathcal{H}$ linearly, they can be gathered into a simple additive term (see supplemental material for more detail).
The noise model proposed here differs slightly from previous work~\cite{demarco_1987_security,matthews_2018_simulating} through the inclusion of generator perturbations and damping (introduced for numerical purposes), but its specific form allows us to perform the analysis presented in \cref{subsec:rate-qst}.
Our stochastic port-Hamiltonian model for the grid dynamics then becomes \cite{demarco_1987_security}
\begin{equation}
  \label{eq:stochastic-port-hamiltonian}
  \mathrm{d} x^{\tau}_t = (J - S) \nabla \mathcal{H} \left (x^{\tau}_t \right ) \mathrm{d} t + \sqrt{2 \tau} S^{1 / 2} \mathrm{d} W_t, \quad x^{\tau}_0 = \bar{x},
\end{equation}
where $W_t \in \mathbb{R}^d$ is a vector of $d$ independent Wiener processes and the parameter $\tau$ defines the strength of the scaled load and generation fluctuations.

We mention three important properties of the stochastic differential equation (SDE)~\cref{eq:stochastic-port-hamiltonian}, which we summarize in the following remark.
\begin{remark}[SDE properties]\label{rmk:sde-properties}%
	\begin{enumerate}
    \item The SDE is \emph{irreversible} because of the presence of the antisymmetric ``connection'' matrix $J$ \cite{touchette_2018_introduction}, and it represents a form of nonequilibrium Markovian dynamics.
		Typically, irreversibility makes a system difficult to analyze; however, this is mitigated in our case by the following property.
    \item System \cref{eq:stochastic-port-hamiltonian} admits a transverse decomposition of the drift~\cite{wf_random_2012,bouchet_2016_generalisation,bouchet_2016_perturbative}; see \cref{apx:rate-transverse}. 
    \item The stochastic port-Hamiltonian model \cref{eq:stochastic-port-hamiltonian} has the Gibbs measure
		\begin{equation}
		\label{eq:gibbs}
		\mu^{\tau} = (Z^{\tau})^{-1} \exp \left [ -\mathcal{H}(x) / \tau \right ], \quad Z^{\tau} \coloneqq \int \exp \left [ -\mathcal{H}(x) / \tau \right ] \, \mathrm{d} x,
		\end{equation}
		(for $\tau > 0$) as its stationary distribution~\cite{hwang_accelerating_1993}.
	\end{enumerate}
\end{remark}
We choose a stochastic model for the grid dynamics with such properties (specifically the final two) because they will greatly simplify analyzing transmission line failures as first exit events~\cite{wf_random_2012,bouchet_2016_generalisation}.

\subsection{Unconditional failure problem}
\label{subsec:rate-failure-problem}
\begin{figure}[H]
	\begin{subfigure}[b]{0.48\textwidth}
		\centering
		\includegraphics[width=\linewidth]{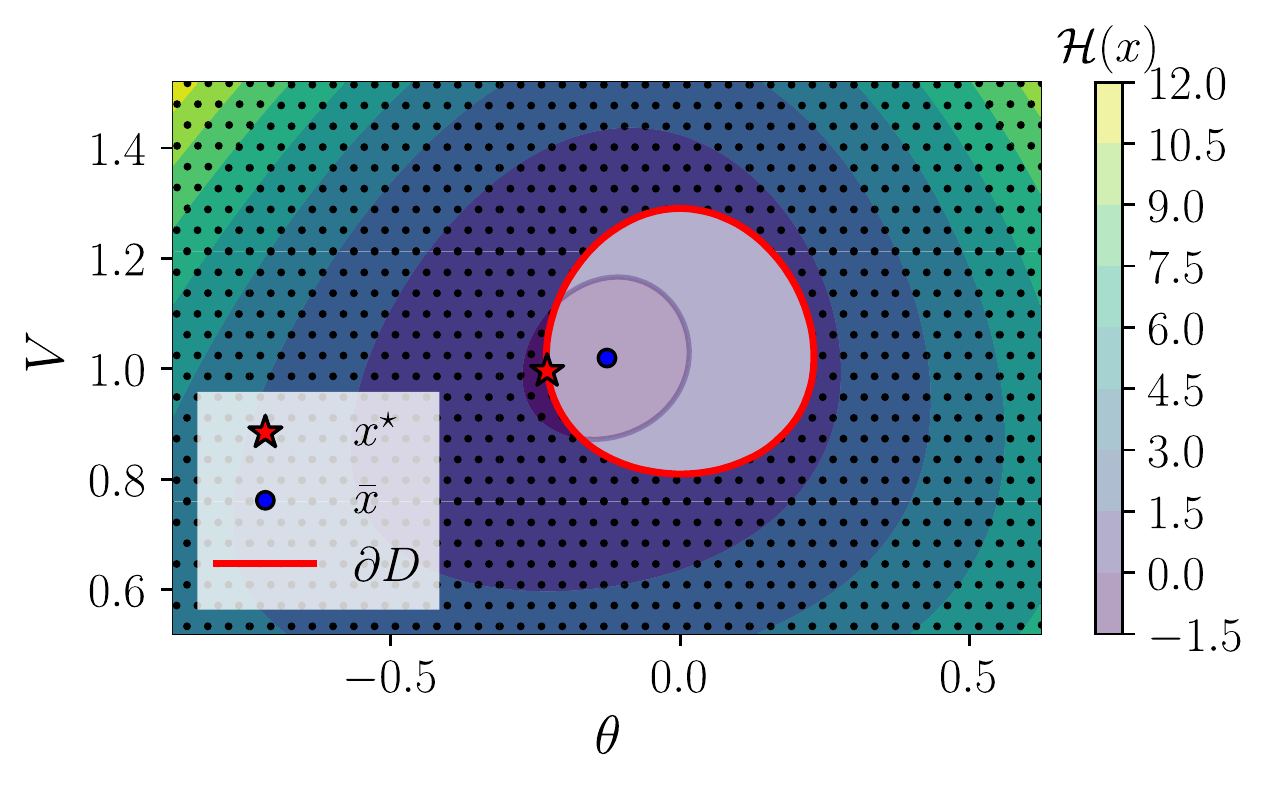}
		\caption{}
		\label{fig:energy-contour-local-3bus}
	\end{subfigure}
	\begin{subfigure}[b]{0.48\textwidth}
		\centering
		\includegraphics[width=\linewidth]{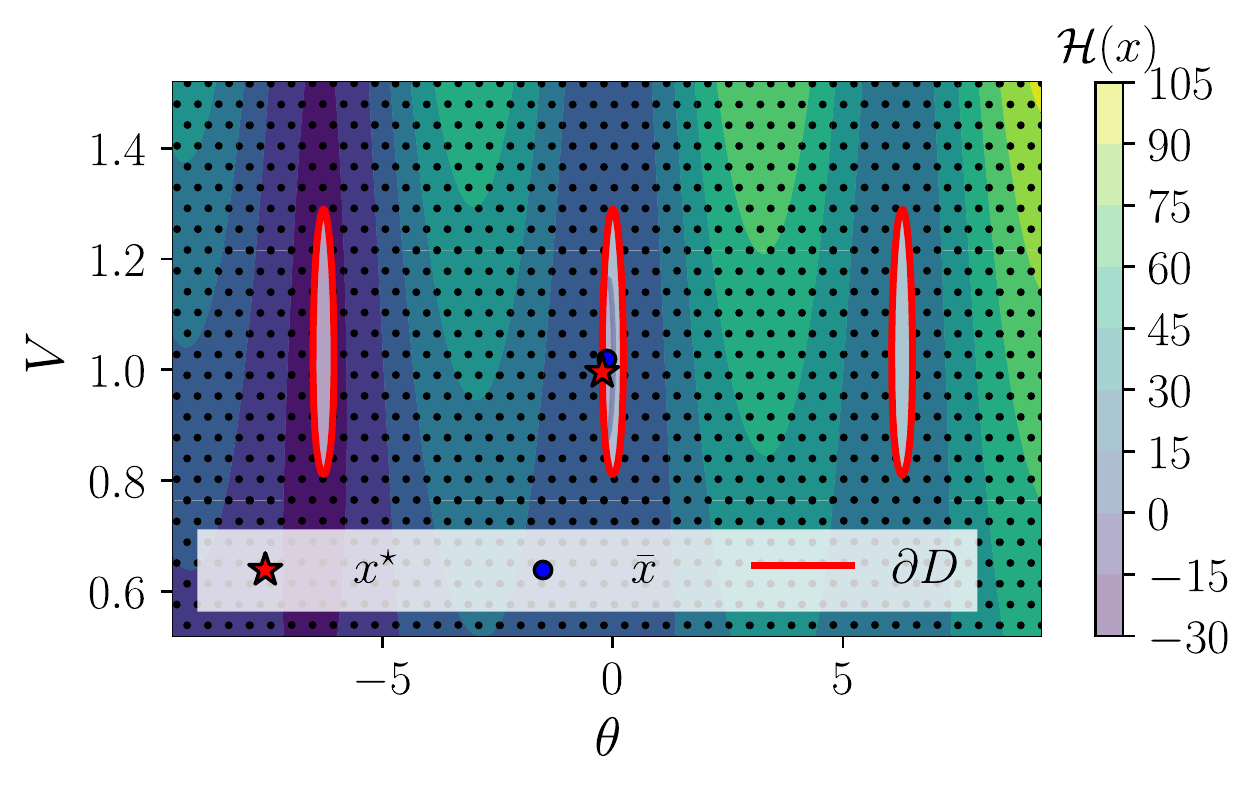}
		\caption{}
		\label{fig:energy-contour-wide-3bus}
	\end{subfigure}
	\vspace{-0.6cm}
	\caption{\small{Energy surface contour and failure region (dotted area) for a slack-load transmission line in a 3-bus power transmission system (see supplemental material for more detail).}
		(\subref{fig:energy-contour-local-3bus}) \small{Local basin of attraction of $\bar{x}$ and safety region.}
		(\subref{fig:energy-contour-wide-3bus}) \small{Various metastable configurations with their respective basin of attraction and safety regions.}}
	\label{fig:energy-contour-3bus}
\end{figure}

We now motivate our underlying problem, formally introduced in \cref{def:unconditional-failure}.
For the power transmission network described by the model~\cref{eq:stochastic-port-hamiltonian}, we are interested in quantifying the time it takes for the line energy of the $l$th line, denoted $\Theta_l$, to exceed a given safety threshold $\Theta^{\max}_l$.
Disregarding transmission losses, the line energy $\Theta_l$ is given in terms of the network's state $x$ by
\begin{align}
  \label{eq:current}
  i_{l} &= (v_i - v_j) \, B_{ij} , \\
  \label{eq:line-failure-criteria}
  \Theta_l(x) &= |i_{l}|^2 = i_{l} i_{l}^* = B_{ij}^2 \left [ V_i^2 - 2V_i V_j \cos(\theta_i - \theta_j) + V_j^2 \right ],
\end{align}
where $(i, j)$ denote the start and end buses of the $l$th line, respectively; $B_{ij}$ denotes the susceptance of the $l$th line; and $i_l$ denotes the current phasor of the $l$th line.
The safety criterion $\Theta_l(x) < \Theta^{\max}_l$ defines a safety region in state space
\begin{equation}
  \label{eq:safety-region}
  D_l \coloneqq \left \{ x \colon \Theta_l(x) < \Theta^{\max}_l \right \},
\end{equation}
which contains the equilibrium point $\bar{x}$.
When the $l$th line energy exceeds the corresponding threshold, the line is said to fail irreparably, resulting in the disconnection of the line from the network.
Therefore, we can interpret the $l$th line failure as the first exit of the stochastic dynamics \cref{eq:stochastic-port-hamiltonian} from the basin of attraction of $\bar{x}$ through the boundary $\partial D_l$ of the safety region (see \cref{fig:energy-contour-local-3bus}).

In a power system, the automatic line relay trigger event effectively removes that line from the network, which we model by setting its susceptance to zero.
This results in a discontinuous change to the system matrix $B$ and the corresponding energy function.
In this way, the systemwide energy function is continuous until a failure occurs, at which point there is a discontinuous change in the energy function (though continuous formulations have been proposed \cite{zheng_2010_bistable}).
When following the network's dynamics, the trajectory may wander into a region of state space where $\Theta_k \geq \Theta_k^{\max}$ for $k \ne l$, thereby triggering the relay corresponding to line $k$ before triggering the relay corresponding to line $l$ of interest.
In \cref{sec:rate}, we assume that line $l$ of interest is the only fallible line in the network.
This consideration is summarized in the following definition.
\begin{definition}[Unconditional and conditional failure events]\label{def:unconditional-failure}
	An \emph{unconditional} failure event for line $l$ is defined as an event at time $t$ such that $\Theta_l(x_t) \geq \Theta_l^{\max}$, regardless of any previous failure events $\Theta_k(x_s) \geq \Theta_k^{\max}$ for $k \ne l$ and $s < t$.
	This terminology arises in contrast to \emph{conditional} failure events in which the event of interest is the line $l$ failure (transgression through $\partial D_l$) conditioned on this transgression being the \emph{first} transgression through \emph{any} boundary for the full network.
\end{definition}
The conditional problem is the appropriate framework for computing the failure rate of practical interest, but drawing a mapping between the two types of failure problems is difficult.
We return to the conditional-unconditional failure distinction in \cref{sec:adapt} and study the unconditional failure problem until then.

\subsection{Quasistationary exit rate}
\label{subsec:rate-qst}

For the stochastic process $x^{\tau}_t$ defined by~\cref{eq:stochastic-port-hamiltonian}, we are interested in the probability distribution of the first exit time, or ``first passage time,''
\begin{equation}
  \label{eq:fpt}
  T^{\tau}_{\partial D_l} \coloneqq \inf \left \{ t > 0 : x^{\tau}_t \in \partial D_l \right \} ,
\end{equation}
where the boundary $\partial D_l$ of the safety region $D_l$ satisfies the following conditions:
\begin{enumerate}[label=\emph{(A.\roman*)}]
\item \label{assn:non-characteristic} $\left \langle (J - S) \nabla \mathcal{H}(x), n(x) \right \rangle < 0$ for all $x \in \partial D_l$, where $n(x)$ is the outward vector normal to $\partial D_l$ at $x$.
  Equivalently, the surface $\partial D_l$ is \emph{noncharacteristic}.
\item \label{assn:name-tbd} $\left \langle \nabla \mathcal{H}(x), n(x) \right \rangle > 0$ for all $x \in \partial D_l$.
\end{enumerate}
These assumptions ensure that the deterministic dynamics started at points along $\partial D_l$ converge toward the stable point $\bar{x}$.
Numerical analysis of various IEEE power transmission system test cases indicates that these assumptions are in general valid for line failure criteria of the form~\cref{eq:line-failure-criteria}.

For the stochastic port-Hamiltonian system and line failure criteria satisfying assumptions~\ref{assn:non-characteristic} and~\ref{assn:name-tbd}, we have that, in the limit $\tau \to 0$, the mean first passage time (MFPT) $\mathbb{E} T^{\tau}_{\partial D_l}$ satisfies the limit~\cite{kushner_1984_approximation}
\begin{equation}
  \label{eq:mfpt-ld}
  \lim_{\tau \to 0} \tau \log \mathbb{E} T^{\tau}_{\partial D_l} = \min_{x \in \partial D_l} V(\bar{x}, x),
\end{equation}
where $V(\bar{x}, x)$ is the so-called Freidlin-Wentzell quasipotential~\cite{wf_random_2012}.
Since our port-Hamiltonian system admits a transverse decomposition (\cref{rmk:sde-properties}), calculation of the quasipotential is simplified considerably \cite{kushner_1984_approximation}.
In this case, by definition, the quasipotential satisfies the properties $V(\bar{x}, \bar{x}) = 0$ and $V(\bar{x}, x) \geq 0$ for all $x \in D_l$ and solves the Hamilton-Jacobi equations \cite{kushner_1984_approximation} for \cref{eq:stochastic-port-hamiltonian}
\begin{equation}
  \label{eq:hamilton-jacobi-V-full}
  \left \langle \nabla V(x), S V(x) \right \rangle + \left \langle (J - S) \nabla \mathcal{H}(x), \nabla V(x) \right \rangle = 0, \quad x \in D_l.
\end{equation}
One can verify by substitution that
\begin{equation}
	\label{eq:quasipotential}
  V(\bar{x}, x) \coloneqq \mathcal{H}(x) - \mathcal{H}(\bar{x})
\end{equation}
satisfies \cref{eq:hamilton-jacobi-V-full} and the properties of the quasipotential listed above.
Here we pause to make the following remark.
\begin{remark}[Quasipotential evaluation]
Since \cref{eq:quasipotential} satisfies the definition of a quasipotential for dynamics \cref{eq:stochastic-port-hamiltonian}, quasipotential evaluation is equivalent to point evaluation of the energy function and \emph{does not} require path optimization to compute the Freidlin-Wentzell action functional \cite{wf_random_2012} (which forms the basis for most quasipotential computations).
Fundamentally, this is due to the existence of a transverse decomposition for the drift vector.
\end{remark}

The limit \cref{eq:mfpt-ld} indicates that the MFPT is of order $\exp [ V(\bar{x}, x^{\star}) / \tau]$, where we define $x^{\star}$ (the unconditional ``exit point'') as the solution to the following nonlinear program:
\begin{equation}
  \label{eq:xstar}
  x^{\star} = \argmin_{x \in \partial D_l} \mathcal{H}(x),
\end{equation}
which corresponds to the most likely exit point of the stochastic dynamics along the surface $\partial D_l$ in the limit $\tau \to 0$.
For the remainder of this section, we assume that $x^{\star}$ is unique, although we return to this issue in \cref{sec:adapt}.

Over the period of time	$1 \ll t \ll \exp [ V(\bar{x}, x^{\star}) / \tau]$, the diffusion process is characterized by a ``quasistationary'' (metastable) equilibrium in which the process forgets its initial condition and the probability mass is concentrated around the equilibrium point $\bar{x}$ according to the quasistationary density \cite{bouchet_2016_generalisation}.
Probability mass is then lost through the boundary $\partial D_l$ at an approximately constant rate $\lambda^{\tau}$ known as the ``quasistationary exit rate.''
It is known that for Markov processes starting from quasistationary distributions, the distribution of first passage times is exponential with mean given by the reciprocal of the exit rate~\cite{collet_2013_qst}.
Thus, in order to characterize the distribution of first passage times for the quasistationary range in the limit $\tau \to 0$, it suffices to approximate $\lambda^{\tau}$.

\subsection{Asymptotic expression for the exit rate}
\label{subsec:rate-asymptotic}
We derive zeroth- and first-order approximations for the unconditional rate of transmission line failure.

\subsubsection{Zeroth-order approximation}
We derive an approximation to the quasistationary exit rate asymptotically valid in the small-noise limit.
For this purpose we follow the procedure presented in~\cite{bouchet_2016_generalisation} for irreversible diffusion processes.
Here, we note that the theory of \cite{bouchet_2016_generalisation} was derived for the case of a full-rank diffusion matrix, which in our case would correspond to a full-rank matrix $S$.
Unfortunately, $S$ is diagonal with non-negative, rather than strictly positive, entries, and thus the assumption of full rank does not hold.\footnotemark\footnotetext{Zero elements correspond to the generator angle equations due to $T_2^{\top} (D^d)^{-1} T_2$}
Therefore, to employ the theory of \cite{bouchet_2016_generalisation}, we must introduce the following additional assumption on the exit surface:
\begin{enumerate}[label=\emph{(A.iii)}]
\item \label{assn:noise} $\left \langle n(x), S n(x) \right \rangle > 0$ for all $x \in \partial D_l$;
\end{enumerate}
that is, we require there to be a diffusive probability flux in the direction orthogonal to the exit surface.
This assumption limits the types of transmission lines that can be analyzed with the proposed approach.
In particular, \textit{generator-generator} (including \textit{slack-generator}) lines do not satisfy this assumption for the proposed dynamics and line failure models.
Fortunately, \textit{generator-generator} lines are relatively insignificant in practice, since most \textit{generator-generator} lines contain an intermediary substation.
In the IEEE cases, these lines are either uncommon or can be addressed by introducing a fictitious bus between the two nodes.
Only one such line exists in the IEEE 30-bus system (see \cref{fig:30bus-ieee}), and across sixteen IEEE systems ranging from nine buses to 9,241 buses (with a mean of 2,302 buses), the proportion of such lines ranges between 0\% and 30\% (with each case falling under 7\% except for the 118-bus system at 30\%).

Having addressed this technical challenge, we can employ the results of~\cite{bouchet_2016_generalisation}.
Specifically, injecting the transverse decomposition of the drift of \cref{eq:stochastic-port-hamiltonian}, derived in~\cref{apx:rate-transverse}, into (4.25) of \cite{bouchet_2016_generalisation}, we obtain the following zeroth-order expression for the quasistationary exit rate in integral form:
\begin{equation}
  \label{eq:qst-exit-rate-prefinal}
  \lambda_0^{\tau} = \int_{\partial D_l} \sqrt{ \frac{\det \operatorname{Hess} \mathcal{H}(\bar{x})}{(2 \pi \tau)^d}} \exp \left [ - \frac{\mathcal{H}(x) - \mathcal{H}(\bar{x})}{\tau}\right ]
  \left \langle (J + S) \nabla \mathcal{H}(x), n(x) \right \rangle \, \mathrm{d} A(x).
\end{equation}
We can see that in the limit $\tau \to 0$, the value of this integral is dominated by the contribution of the exponential term $\exp \left \{ -[\mathcal{H}(x) - \mathcal{H}(\bar{x})] / \tau \right \}$ in the vicinity of the exit point.
Therefore, we can employ the Laplace method for multiple integrals~\cite{wong_2001_asymptotic} to obtain a formula for the leading asymptotic behavior of the exit rate.
This calculation is summarized in \cref{apx:rate-qst-exit-rate-derivation}.
The resulting expression for the quasistationary exit rate is
\begin{equation}
  \label{eq:qst-exit-rate-final}
  \lambda_0^{\tau} \underset{\tau \to 0}{\sim} \nabla^{\top} \mathcal{H}(x^{\star}) S \nabla \mathcal{H}(x^{\star}) \sqrt{ \frac{ \det \Hess \mathcal{H}(\bar{x}) }{2 \pi \tau |B^{\star}|}} \exp \left [ - \frac{\mathcal{H}(x^{\star}) - \mathcal{H}(\bar{x})}{\tau} \right ],
\end{equation}
where $B^{\star}$ is a factor accounting for the curvature of $\partial D_l$ in the vicinity of $x^{\star}$ (see \cref{eq:xstar}), given by
\begin{equation}
  \label{eq:B-star}
  B^{\star} \equiv \nabla_x \mathcal{H}(x^{\star})^{\top} L^{-1} \nabla_x \mathcal{H}(x^{\star}) \det L, \quad L = \Hess \mathcal{H}(x^{\star}) - k \Hess \Theta_l (x^{\star}),
\end{equation}
and $k$ is the Lagrange multiplier on the $\Theta_l$ constraint in the constrained optimization problem of the exit point \cref{eq:xstar}.\footnote{In numerical computation, we compute \cref{eq:qst-exit-rate-final} and related expressions using the \emph{magnitude} of $\operatorname{det} \operatorname{Hess} \mathcal{H}(\bar{x})$.}
For convenience, we refer to the subexponential term as the \textit{prefactor} and the exponential term as the \textit{energy factor} and write
\begin{equation}
	\label{eq:qst-exit-rate-pf-ef}
	\lambda_0^{\tau} \underset{\tau \to 0}{\sim} \frac{C^{\star} C_0(\bar{x})}{\tau^{1/2}} \exp \left[ -\frac{\Delta \mathcal{H}}{\tau} \right] ,
\end{equation}
where we define
\begin{equation}
\label{eq:qst-exit-rate-pf}
C^{\star}  \coloneqq \frac{ \nabla^{\top} \mathcal{H}(x^{\star}) S \nabla \mathcal{H}(x^{\star}) }{ \sqrt{2 \pi |B^{\star}|} } , \quad
C_0(\bar{x}) \coloneqq \sqrt{ \det \Hess \mathcal{H}(\bar{x}) } ,
\end{equation}
and for convenience we also define (with implicit dependence on a line of interest)
\begin{equation}
\label{eq:qst-exit-rate-ef}
\Delta \mathcal{H} \coloneqq \mathcal{H}(x^{\star}) - \mathcal{H}(\bar{x}).
\end{equation}
It follows that the MFPT is given by $\mathbb{E} T^{\tau}_{\partial D_l} \underset{\tau \to 0}{\sim} 1 / \lambda_0^{\tau}$.

Our zeroth-order expression for $\lambda_0^{\tau}$ has the following properties.
First, it is straightforward to  evaluate, since it requires solving only two optimization problems (an unconstrained optimization problem for $\bar{x}$ and a constrained optimization problem for the exit point $x^{\star}$, \cref{eq:xstar}) and computing the curvature factor.
Second, it shows how the exit rate (and thus the probability distribution of line failure times) depends on the properties of the network.
Namely, the rate depends on the properties of the network directly through the energy function $\mathcal{H}(x)$ and the matrix $S$, and indirectly through the values of $\bar{x}$, $x^{\star}$, and $B^{\star}$.

\subsubsection{First-order approximation}
Next we extend the applicability of the zeroth order failure rate to finite temperature regimes.
Temperature enters the expression for line failure rate \cref{eq:qst-exit-rate-pf-ef} in two competing ways.
As temperature decreases, the subexponential prefactor ($C^{\star} C_0(\bar{x}) \tau^{-1/2}$) increases and the exponential energy factor ($\exp[ - \Delta \mathcal{H} \tau^{-1} ]$) decreases.
In the small noise limit, the effect of the prefactor is dwarfed by the contribution of the energy factor.
In the large noise limit, the $\tau^{-1/2}$ dependence in the prefactor leads to nonphysical behavior.
Specifically, the failure rate decreases as temperature exceeds a ``critical'' temperature $\tau$ of the same order of magnitude as $\Delta \mathcal{H}$ where the energy factor contribution is approximately unity, and the system is expected to fail more easily in this regime.

We note that the subexponential prefactor stems from the normalization constant of the (quasi)stationary distribution $p^{\tau}_{st}$ (see \cref{eq:gibbs}).
A higher-order analysis of $p^{\tau}_{st}$ results in a modification to the failure rate, and we offer a brief physical justification of its appropriateness.

First, we examine the higher orders of the WKB approximation \cite{bender2013advanced} for the stationary distribution $p_{st}^{\tau}$ given by
\begin{equation}
\label{eq:n-order-stationary-wkb}
p_{st}^{\tau}(x) = \frac{C_{st}^{\tau}(x)}{\tau^{1/2}} \, \exp \left[ \frac{\sum_{j=0}^{\infty} \mathcal{H}_j(x) \tau^j}{\tau}  \right],
\end{equation}
where $\mathcal{H}_0(x)=-\mathcal{H}(x)$ and $\mathcal{H}_j$ for $j=1,\ldots$ are unknown functions of $x.$
We show in \cref{apx:rate:higher} that for the power grid model we consider, the higher-order terms in the exponential are constant in $x$ (depending only on $\bar{x}$) and can be absorbed into the subexponential prefactor.
The argument is built order by order in $\tau$ using the Fokker-Planck equation for the stationary density $p_{st}^{\tau}(x).$
For each order, we use the result that the previous order approximation is constant in $x$ and the two properties of the power grid model: (i) the matrix $S$ is constant and (ii) $F(x) \coloneqq {\operatorname{div}} J \nabla \mathcal{H}(x) + \langle A(x), S\nabla \mathcal{H}(x) \rangle=0,$ where $A_i(x)=\sum_{j=1}^d \partial_{ij} S_{ij}(x)$ (see \cref{apx:rate:higher} and \cref{apx:rate-transverse} for more details).

Second, we examine the corrections that come from the expansion of the prefactor $C_{st}^{\tau}(x).$
We have chosen an expansion in integer powers of $\tau$ given by
\begin{equation}
\label{eq:n-order-prefactor-prefinal}
C_{st}^{\tau}(x) = \sum_{j=0}^n C_j(x) \tau^j.
\end{equation}
We show in \cref{apx:rate:higher} that
\begin{equation}
\label{eq:n-order-prefactor-final}
C_{st}^{\tau}(x) = \sum_{j=0}^n C_j(\bar{x}) \tau^j,
\end{equation}
where the $C_j(\bar{x})$ in \cref{eq:n-order-prefactor-final} are constants.
As in the case of the higher order terms in the exponent, the argument is again built order by order in $\tau$ and utilizes the same properties of the power grid model (see \cref{apx:rate:higher} for more details).

Using a zeroth-order approximation of the prefactor (as in \cref{eq:qst-exit-rate-final}), we obtain for the failure rate
\begin{equation}
\label{eq:0-order-qst-final}
\lambda_0^{\tau} \underset{\tau \to 0}{\sim} C^{\star} C_0(\bar{x}) \tau^{-1/2} \exp \left[ -\frac{\Delta \mathcal{H}}{\tau} \right].
\end{equation}
Similarly, using a first-order approximation of the prefactor, we obtain for the failure rate
\begin{equation}
\label{eq:1-order-qst-final}
\lambda_1^{\tau} \underset{\tau \to 0}{\sim} \left( \frac{C^{\star} C_0(\bar{x})}{\tau^{1/2}} + \frac{C^{\star\star} C_1(\bar{x}) \tau}{\tau^{1/2}} \right) \exp \left[ -\frac{\Delta \mathcal{H}}{\tau} \right]
\end{equation}
for new constants $C^{\star\star}$ and $C_1(\bar{x})$.

Determining analytic approximations of $C_1(\bar{x})$ and $C^{\star\star}$ requires a more detailed analysis of the quasistationary distribution.
Instead, we present a heuristic supported by numerical evidence (\cref{fig:rate-modification-fit}).
In particular, at $\tau \approx \Delta\mathcal{H}$, the exponential factor is of order 1 and does not significantly influence the failure rate, which is dominated by $C^{\star} C_0(\bar{x}) \tau^{-1 / 2}$.
To interpolate between the low-temperature asymptotics dominated by the energy factor and the high-temperature asymptotics dominated by $\alpha\tau^{1 / 2}$ (for $\alpha > 0$), we take $\alpha = C^{\star} C_0(\bar{x}) \tau^{-1 / 2}$ to match the coefficient of the zeroth-order failure rate \cref{eq:0-order-qst-final} at $\tau = \Delta \mathcal{H}$.
This amounts to taking
\begin{equation}
\label{eq:first-order-prefactor}
C_1(\bar{x}) = C_0(\bar{x}), \quad \text{and} \quad C^{\star\star} = \frac{C^{\star}}{\Delta \mathcal{H}},
\end{equation}
which has the added benefit of not requiring estimation of further parameters.
From a physical perspective, we expect the failure rate to increase with temperature.
The form we propose aligns with this intuition and interpolates between two regimes at the hypothesized critical point.


\subsection{Numerical validation of the asymptotic exit rate}
\label{subsec:rate-validation}
The goal of our numerical experimentation is to study the asymptotic properties of the failure rate theory expressed in \cref{eq:0-order-qst-final,eq:1-order-qst-final}.
In particular, our experiments explore the theory's ability to measure system stability directly via initial operating conditions.
The theory employed in \cref{subsec:rate-asymptotic} provides guidance only for the  small-noise limit, and we focus first on exploring asymptotic behavior before turning to the question of quantifying \textit{when} such a regime might begin (if at all).
In the remainder of this section, we present our simulation methodology and discuss our numerical findings.

\subsubsection{Methodology}
As introduced in \cref{sec:introduction}, the unconditional failure problem that we are concerned with measures the system's first exit through $\partial D_l$ for a particular line $l$ of interest, ignoring any previous transgressions through boundaries associated with lines other than the line of interest.
The related \emph{conditional} problem is the appropriate framework for computing the failure rate of practical interest, but here we focus on the unconditional failure problem and return to the unconditional-conditional failure issue in \cref{sec:adapt}.

\begin{figure}[H]
	\begin{subfigure}[T]{0.33\textwidth}
		\vspace{-5pt}
		\centering
		\includegraphics[width=0.9\linewidth]{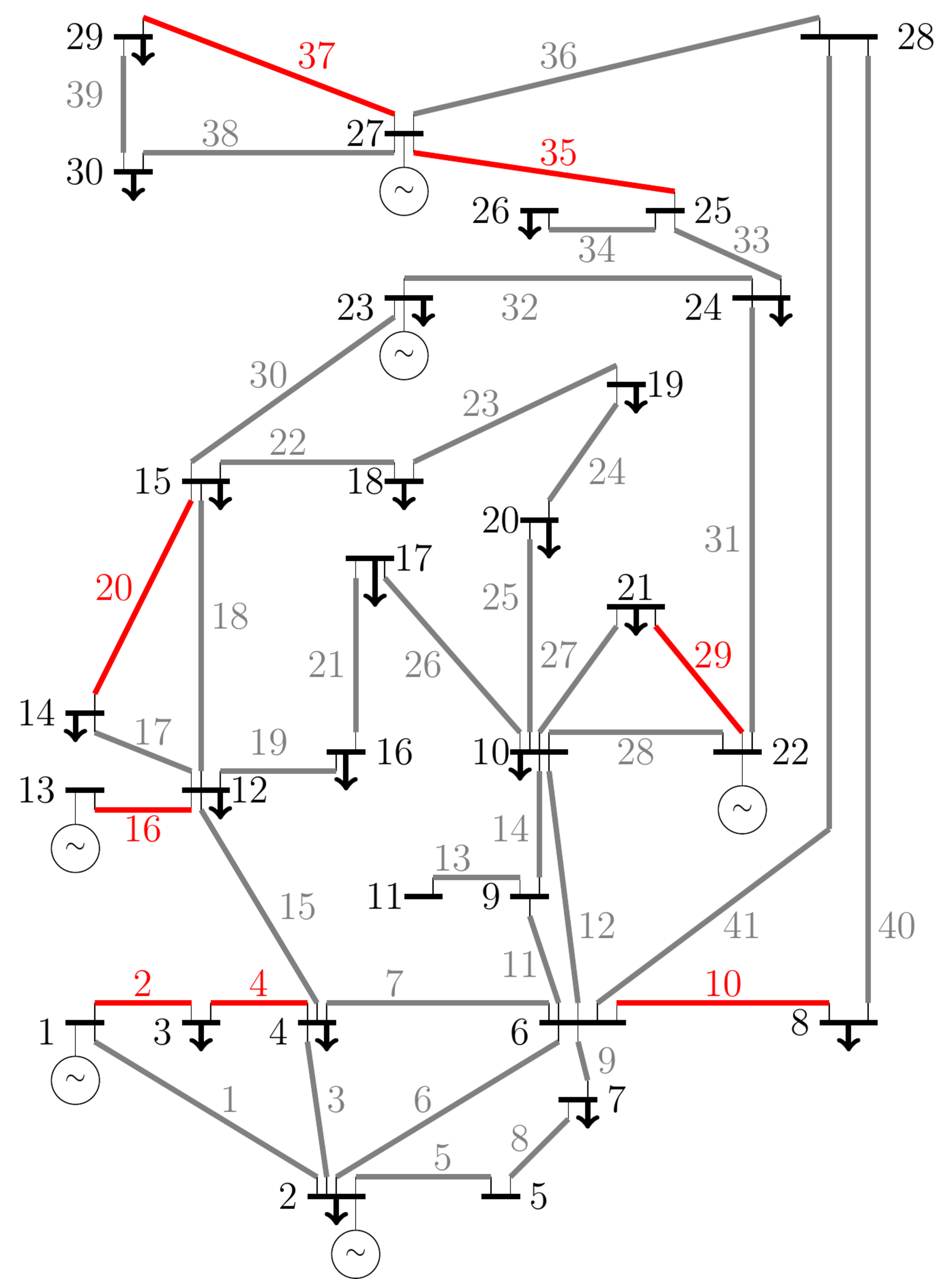}
	\end{subfigure}%
	\begin{subfigure}[T]{0.33\textwidth}
		\centering
		\vspace{-10pt}
		\includegraphics[width=0.9\linewidth]{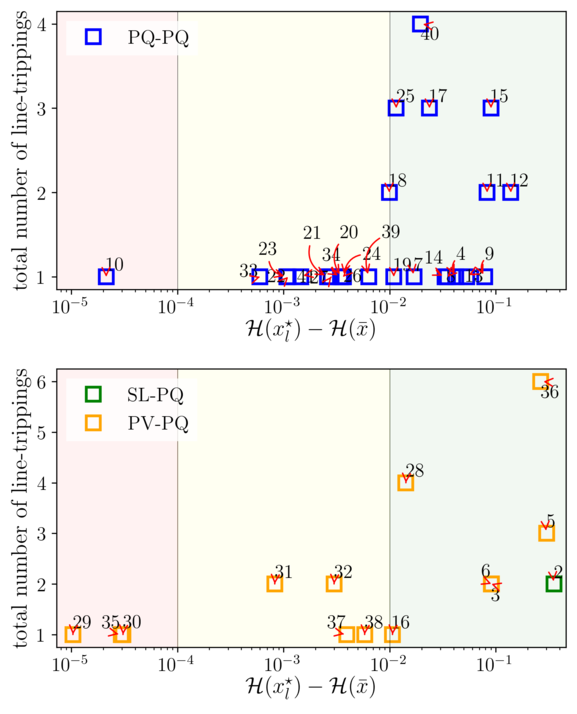}
	\end{subfigure}%
	\begin{subfigure}[T]{0.33\textwidth}
		\centering
		\begin{minipage}[T]{\linewidth}
			\centering
			\resizebox{\linewidth}{!}{
\bgroup
\scriptsize
\def\arraystretch{1.25}%
\begin{tabular}{||>{\centering\arraybackslash}p{2.5cm}|>{\centering\arraybackslash}p{3.5cm}||}
	\hline
	\textbf{parameter} & \textbf{value} \\ [0.5ex] 
	\hline\hline
	$b_l,\, l \in \mathcal{E}$ & from \texttt{Matpower}\footnote{{From lossless assumption $G=0$, \texttt{Matpower} resistive elements (\texttt{r}) are set to zero. For simplicity, so too are \texttt{shunt} and \texttt{tap} elements.}} \\
	\hline
	$D^{\epsilon}$ & 0.01\footnote{{Parameter set experimentally to ensure that voltages can adjust on sub-minute timescales.}} \\
	\hline
	$M^g_{i},\, i \in \mathcal{G} \cup \mathcal{S}$ & 0.0531\footnote{\label{fn:demarco-parms}{Parameter value from \cite{zheng_2016_new}.}} \\
	\hline
	$D^g_{i},\, i \in \mathcal{G} \cup \mathcal{S}$ & $0.05$\footref{fn:demarco-parms} \\ 
	\hline
	$D^d_i,\, i \in \mathcal{L}$ & $0.005$\footref{fn:demarco-parms} \\ 
	\hline		
	$\gamma$ & 1.0 \\
	\hline
	$\tau$ & \textit{varies} \\
	\hline
	${\Delta}t$ & \textit{varies}, but $\approx 1e^{-6}$ \\
	\hline
	$\Theta_l^{\max},\, l \in \mathcal{E}$ & 1.2 $\times$ \texttt{Matpower rateA} \\
	\hline
\end{tabular}
\egroup

}\\\vspace{-1mm}
		\end{minipage}%
	\end{subfigure}%
	\hfill\\%
	\begin{subfigure}{0.32\textwidth}
		\caption{}
		\label{fig:30bus-ieee}
	\end{subfigure}
	\hfill%
	\begin{subfigure}{0.32\textwidth}
		\caption{}
		\label{fig:30bus-line-classification}
	\end{subfigure}
	\hfill%
	\begin{subfigure}{0.32\textwidth}
		\caption{}
		\label{fig:30bus-dynamics-parameters}
	\end{subfigure}
	\hfill%
\vspace{-0.4cm}
\caption{(\subref{fig:30bus-ieee}) \small{IEEE 30-bus network diagram \cite{30bus}; black lines/labels correspond to buses, red lines/labels correspond to lines studied in \cref{subsec:rate-validation}, and grey lines/labels correspond to lines not studied.}
(\subref{fig:30bus-line-classification}) \small{Energy difference between equilibrium and line $l$ failure point ($\mathcal{H}(x^{\star}_l) - \mathcal{H}(\bar{x})$) for each acceptable load-load line (\textit{top}) and generator-load line (\textit{bottom}) plotted against the total number of failed lines at $x^{\star}_l$. Particularly at dense connection points (such as buses 6 and 10), failure of a certain line may require simultaneous failures at other lines.}
(\subref{fig:30bus-dynamics-parameters}) \small{Parameter table for dynamics simulations. The singular perturbation parameter $D^{\epsilon}$ governs the deviation from the power-flow manifold. Smaller values indicate closer proximity to the manifold but are associated with larger load voltage magnitude fluctuations (both deterministic and stochastic).}}
\label{fig:30bus-system-diagnostics}
\end{figure}

To simulate an unconditional problem, we consider one boundary at a time and treat the line $l$ of interest as an isolated component in a modified network where $\Theta_k^{\max} \equiv \infty$ for lines $k \ne l$.
Given specified operating conditions consisting of fixed demand, generation, and network topology (which can be summarized in parameters $y$\footnote{For fixed demand ($P$ and $Q$) data, only $[V]_{\mathcal{G} \cup \mathcal{S}}$ and $[P^g]_\mathcal{G}$ need to be specified; these can be obtained by solving an optimal power flow (OPF) \cite{bergen_1999_power,frank_primer_2012}.}), the system is initialized at equilibrium point $\bar{x}$ \cref{eq:xbar}, the minimizer of the resulting energy $\mathcal{H}^y(x)$.
At equilibrium, we assume that the grid state is synchronized, meaning that the momentum variables $[\omega]_{\mathcal{G} \cup \mathcal{S}}$ are identically zero.
Operating parameters are then augmented with dynamics parameters (including generator damping and masses) to fully specify $y$.
Next, we assume that emergency line limits ($\Theta^{\max}$) are scaled as a constant factor of the base limits\footnote{Base limits are given by field \texttt{RateA} in a \texttt{Matpower} case file.} (see \cref{fig:30bus-dynamics-parameters}).
We select an individual line $l$ for study and integrate the dynamics \cref{eq:stochastic-port-hamiltonian} with a modified Euler-Maruyama scheme (the Leimkuhler-Matthews method from \cite{leimkuhler_2015_md, matthews_2018_simulating}) until line $l$ failure is observed; the procedure is summarized in \cref{alg:line-l-uncond-failure}.

We use the IEEE 30-bus system (see \cref{fig:30bus-ieee}) as a specific problem instance.
It is a standard test system that contains six generators, 24 loads, and 41 transmission lines and was chosen in part because it also has specified line limits.
This avoids the nontrivial issue of setting limits appropriately \cite{oren_constructing_2017}, although a reasonable proxy might be to ensure a standard level of system security, such as $N$-1 stability \cite{alsac_optimal_1974,vonmeier_electric_2006}.
Of the 41 total system lines in the IEEE 30-bus system, 40 satisfy the required assumptions for our exit analysis (assumptions \ref{assn:non-characteristic}--\ref{assn:noise}; see also Section~2.2.1 in \cite{bouchet_2016_generalisation}).
For the set of 40 lines, we compute failure points, scan for pathologies (\cref{def:nested,def:non-isolated,def:inaccessible}), and stratify lines based on type (\textit{load-load}, \textit{generator-load}, or \textit{slack-load}) and risk class (\textit{high}, \textit{medium}, or \textit{low}) using each line's computed energy difference $\Delta \mathcal{H}$ as a proxy for risk.
We then choose eight representative lines to study: one (nonpathological) line from each risk class for both \textit{load-load} and \textit{generator-load} types, one \textit{slack-load} line (the only such line in the network that is a pathological line), and one additional (nonpathological) high-risk line.
For each simulation, we initialize the dynamics at the same equilibrium point $\bar{x}$ determined from the \texttt{Matpower} case file\footnote{In the 30-bus system, we use demand and topology data from the case file and solve a lossless OPF problem to determine dispatch settings. Under the prespecified line limits, a few lines are heavily stressed at the resulting $\bar{x}$ (lines 10, 29, and 35 have initial line energy loadings that are close to 100\% of their capacity, as specified by the \texttt{RateA} limits in the case files). For comparison, we also solve the OPF without transmission losses and observe similar congestion, indicating that the lossless assumption is not the cause.}
and integrate until line $l$ failure.
We repeat failure experiments across temperatures ranging from $\tau = 1.0$ to $\tau \ll 1.0$ (as limited by computational cost).

Briefly, we draw attention to a few details regarding the numerical implementation.
First, we are limited in gathering low-temperature data.
Using 40 parallel processors, Monte Carlo simulations for an individual line were performed with an approximate computation budget of 24 hours per line per temperature value.
Given the exponential scaling of the failure rate \cref{eq:1-order-qst-final} at low temperatures, the limiting experiment is the lowest temperature for each line.
Second, since the drift term in the port-Hamiltonian model \cref{eq:port-hamiltonian} is independent of the state variables $x$, the Euler-Maruyama--based integration scheme is identical to the Milstein method and has order-1 strong convergence \cite{milstein_approximate_1975,higham_algorithmic_2001}.
We also note that because the system \cref{eq:stochastic-port-hamiltonian} originates from a singularly perturbed differential algebra equation model, the dynamics are stiff and require a ``small'' timescale discretization in order to handle the singular perturbation using simple explicit integrators.
Determining an appropriate ${\Delta}t$ is case dependent, and in what follows we take steps on the order of $10^{-6}$ seconds.

\subsubsection{Experimental results}
The purpose of this section is threefold.
First, we present and discuss empirical convergence and applicability results from our unconditional failure simulations.
Second, we return to the question of adjusting the failure rate at high temperatures.
Third, we discuss conclusions and limitations of our experimental framework and note that ancillary simulation results are included in the supplemental material.

\subsubsection*{Failure simulations and convergence diagnostics}
Fundamentally, we are concerned with understanding the approximation quality of \cref{eq:0-order-qst-final,eq:1-order-qst-final} to characterize unconditional failures rates for our model of grid dynamics \cref{eq:stochastic-port-hamiltonian}.
Relying on the asymptotic theory of \cref{subsec:rate-asymptotic}, we use absolute logarithmic error (henceforth $\delta$, defined as $|\log(\lambda_{sim}^{\tau} / \lambda^{\tau})|$) as a metric for convergence.
Absolute logarithmic error symmetrically penalizes deviations from 1 and is measured in logarithmic units.
We compute $\delta$ for each line across the range of temperatures studied (findings are summarized in \cref{tab:convergence-summary}) and present detailed diagnostics for two particular simulation cases representing the best and worst lines (as measured by $\delta$ averaged across all temperature experiments).
\cref{fig:rate-best-worst} displays the detailed convergence results for analytic zeroth order and the heuristically determined first-order failure rate approximations.

In \cref{fig:rate-best-worst}, we observe good agreement between the analytic rates and the simulated rate as temperature decreases (panels (a)--(b) of \cref{fig:rate-best-worst}), as predicted by theory.
In both cases, the ratio between simulated and approximate failure rates shows encouraging progress toward unity as temperature decreases.
The relationship between approximation quality and temperature is moderated by numerical parameters such as integration step size, but we find that numerical error can be controlled by setting integration parameters appropriately.
Accordingly, we expect agreement to improve at smaller step sizes and lower temperatures.
Further, we remark that this behavior is typical across all lines studied, as evidenced by the average (first-order) error of less than one logarithmic unit (see \cref{tab:convergence-summary}).

Panels (c)--(d) of \cref{fig:rate-best-worst} present additional diagnostics.
In (c), the distribution of exit times approximately follows an exponential distribution parametrized by the reciprocal of the mean of the empirical data, the simulated mean failure rate.
This is encouraging because it indicates that expressions \cref{eq:0-order-qst-final,eq:1-order-qst-final} describe the distribution of failure times in addition to the mean failure time.
Further, at low temperatures, the zeroth- and first-order failure rate expressions appear to provide a conservative approximation of the simulated failure rate.
In (d), we display the value of voltage, angle, and frequency state variables at simulated failure points.
Since observed failures tend to cluster around the lowest energy failure point $x^{\star}$, in our opinion this provides evidence of the reasonableness of approximating the flux through the failure boundary by the first-order Laplace approximation at $x^{\star}$.

Next we turn to the question of determining when the approximation belongs to a ``small-noise'' or ``large-noise'' asymptotic regime.
In \cref{fig:rate-best-worst}, panels (a)--(b)  show that $\tau \approx \Delta \mathcal{H}$ is an approximate critical point for each line.
We remark that since $\Delta \mathcal{H}$ is a line-dependent quantity, regime classification by this measure is also line dependent.
For example, as depicted in panel (b), the exponential factor begins to contribute for $\tau < \Delta \mathcal{H}$ in both simulations, and the failure ratio begins to improve soon thereafter.
Conversely, the prefactor contribution begins to dominate for $\tau > \Delta \mathcal{H}$.
In the zeroth-order approximation, this regime results in divergence of theory and simulation since the prefactor is dominated by $\tau^{- 1 / 2}$ but failure rate increases with increasing temperature.
The first-order approximation captures this intuition, however, and we observe that failure rate scaling of $\lambda^{\tau} \propto \tau^{1 / 2}$ generally well approximates the simulated behavior (except for line 37, which exhibits scaling  $\lambda^{\tau} \propto \tau^{r}$ for $r > 1/2$).
\newcommand{\failurewidth}{0.245}

\newcommand{\BestFailrateComparison}{figures/img-lo-res/rate-30bus-uncond-line-10-failrate_comparison}
\newcommand{\BestFailratioComparison}{figures/img-lo-res/rate-30bus-uncond-line-10-failratio_comparison}
\newcommand{\BestFailtimeHistogram}{figures/img-lo-res/rate-30bus-uncond-line-10-failtime_histogram}
\newcommand{\BestStatespaceHistogram}{figures/img-lo-res/rate-30bus-uncond-line-10-statespace_histogram_bus-8}
\newcommand{\BestNsims}{5,000}

\newcommand{\WorstFailrateComparison}{figures/img-lo-res/rate-30bus-uncond-line-16-failrate_comparison}
\newcommand{\WorstFailratioComparison}{figures/img-lo-res/rate-30bus-uncond-line-16-failratio_comparison}
\newcommand{\WorstFailtimeHistogram}{figures/img-lo-res/rate-30bus-uncond-line-16-failtime_histogram}
\newcommand{\WorstStatespaceHistogram}{figures/img-lo-res/rate-30bus-uncond-line-16-statespace_histogram_bus-13}
\newcommand{\WorstNsims}{500}

\begin{figure}[H]
\begin{minipage}{\linewidth}
\begin{subfigure}[t]{\failurewidth\linewidth}%
	\centering
	\resizebox{\linewidth}{!}{\includegraphics[trim={0 0 3cm 1.5cm}, clip=true]{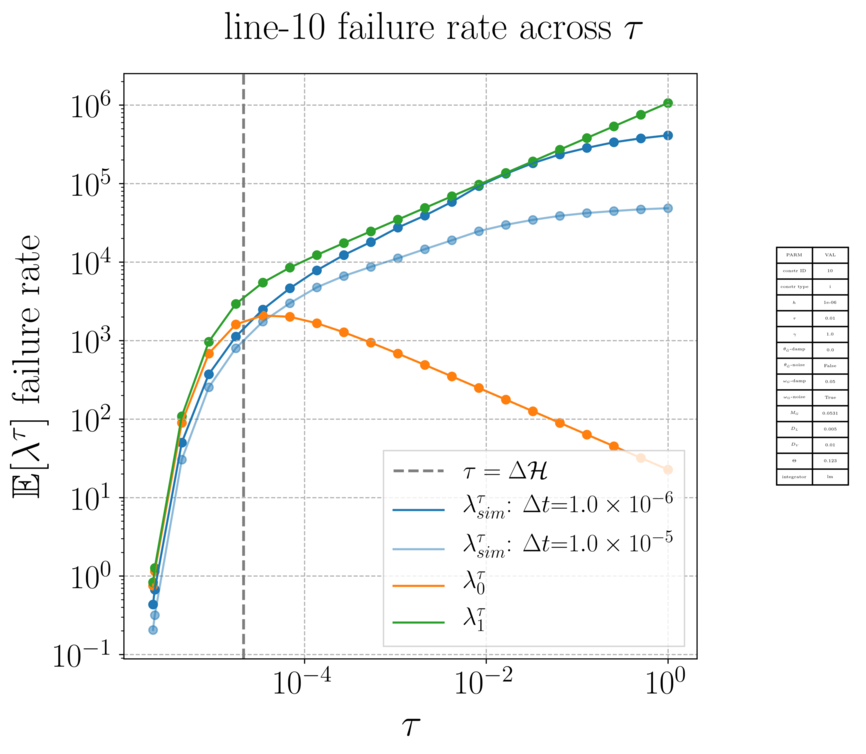}}%
\end{subfigure}
\begin{subfigure}[t]{\failurewidth\linewidth}%
	\centering
	\resizebox{\linewidth}{!}{\includegraphics[trim={0 0 3cm 1.5cm}, clip=true]{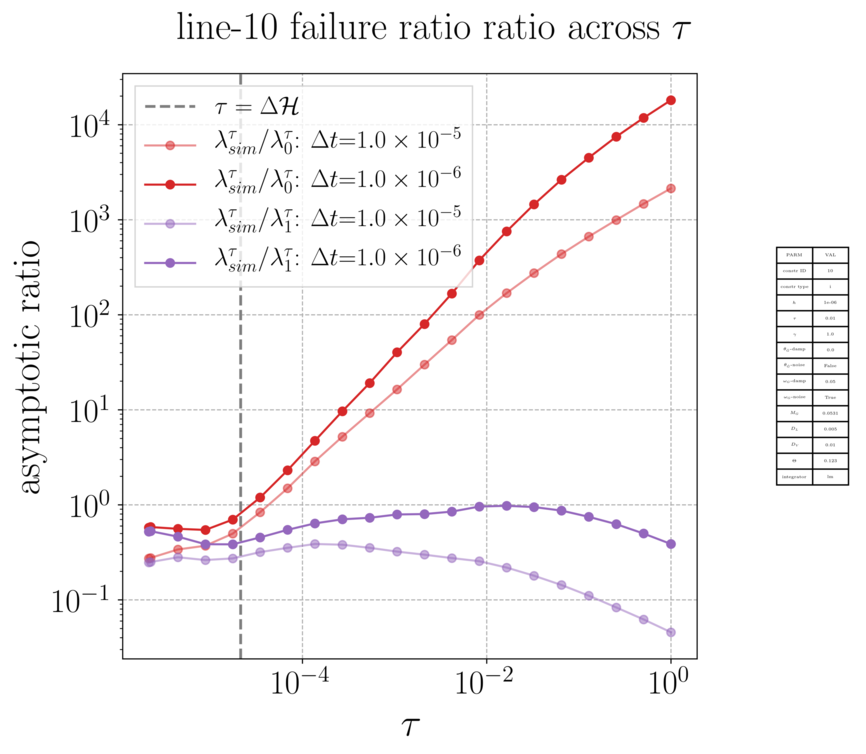}}%
\end{subfigure}
\begin{subfigure}[t]{\failurewidth\linewidth}%
	\centering
	\resizebox{\linewidth}{!}{\includegraphics[trim={0 0 3cm 1.5cm}, clip=true]{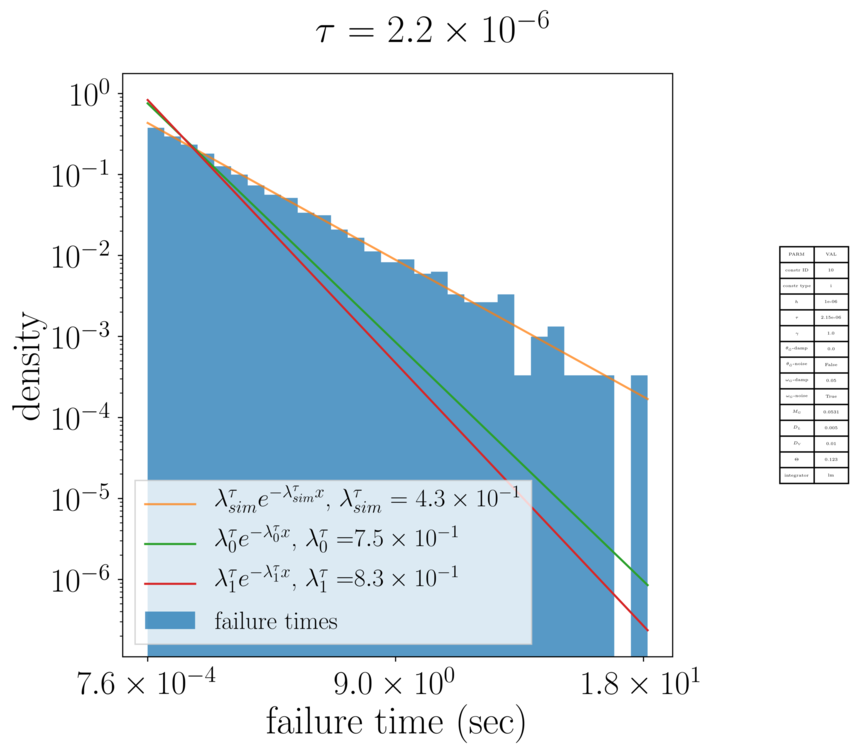}}%
\end{subfigure}
\begin{subfigure}[t]{\failurewidth\linewidth}%
	\centering
	\resizebox{\linewidth}{!}{\includegraphics[trim={0 0 3cm 1.5cm}, clip=true]{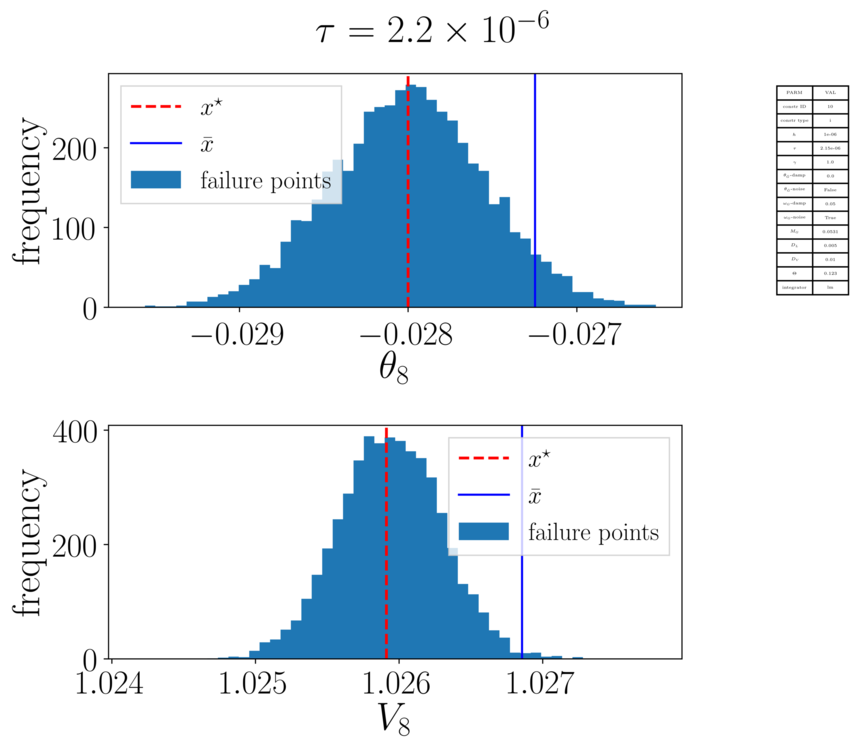}}
\end{subfigure}
\begin{subfigure}[t]{\failurewidth\linewidth}
	\vspace{\failurecaption}%
	\renewcommand\thesubfigure{I.\alph{subfigure}}
	\caption{}
	\label{fig:rate-uncond-best-failrate_comparison}%
\end{subfigure}%
\hfill%
\begin{subfigure}[t]{\failurewidth\linewidth}
	\vspace{\failurecaption}%
	\renewcommand\thesubfigure{I.\alph{subfigure}}
	\caption{}
	\label{fig:rate-uncond-best-failratio_comparison}%
\end{subfigure}%
\hfill%
\begin{subfigure}[t]{\failurewidth\linewidth}
	\vspace{\failurecaption}%
	\renewcommand\thesubfigure{I.\alph{subfigure}}
	\caption{}
	\label{fig:rate-uncond-best-failtime_histogram}%
\end{subfigure}%
\hfill%
\begin{subfigure}[t]{\failurewidth\linewidth}
	\vspace{\failurecaption}%
	\renewcommand\thesubfigure{I.\alph{subfigure}}
	\caption{}
	\label{fig:rate-uncond-best-statespace_histogram}%
\end{subfigure}%
\hfill%
\label{fig:rate-best}
\end{minipage}
\begin{minipage}{\linewidth}
\begin{subfigure}[t]{\failurewidth\linewidth}%
	\centering
	\resizebox{\linewidth}{!}{\includegraphics[trim={0 0 3cm 1.5cm}, clip=true]{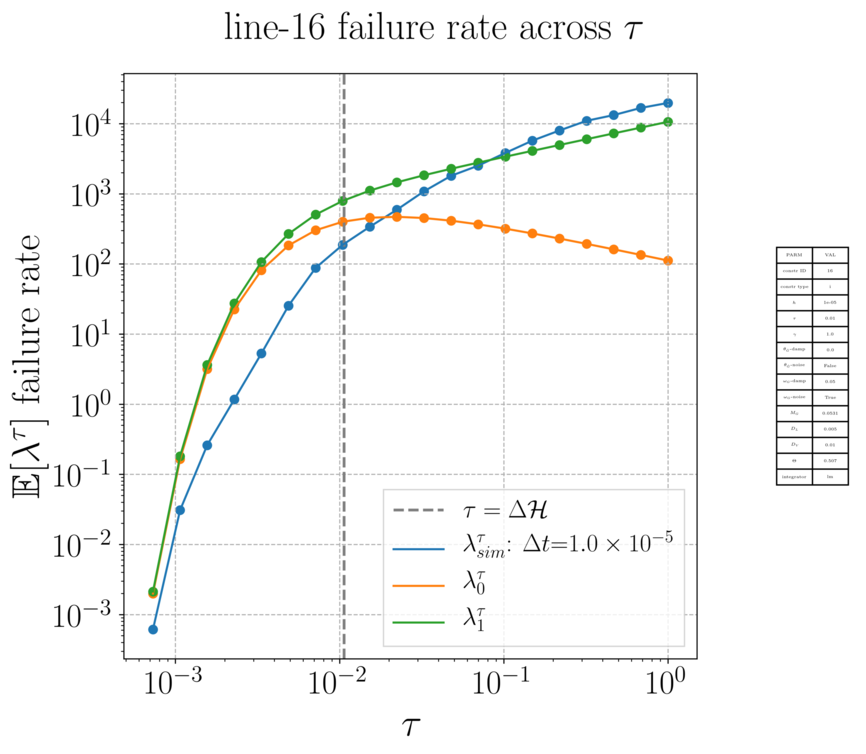}}%
\end{subfigure}
\begin{subfigure}[t]{\failurewidth\linewidth}%
	\centering
	\resizebox{\linewidth}{!}{\includegraphics[trim={0 0 3cm 1.5cm}, clip=true]{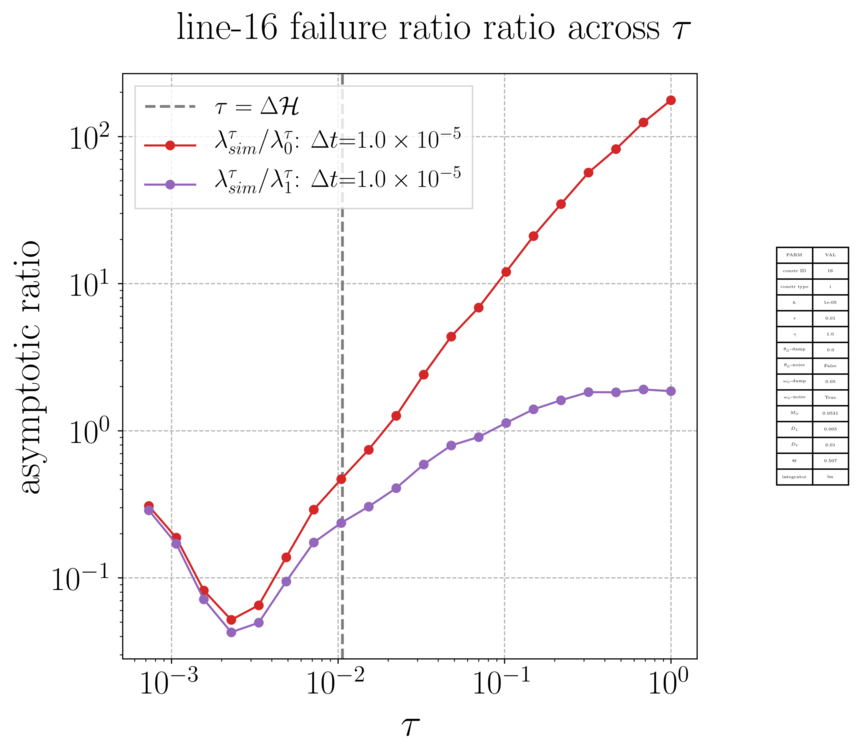}}%
\end{subfigure}
\begin{subfigure}[t]{\failurewidth\linewidth}%
	\centering
	\resizebox{\linewidth}{!}{\includegraphics[trim={0 0 3cm 1.5cm}, clip=true]{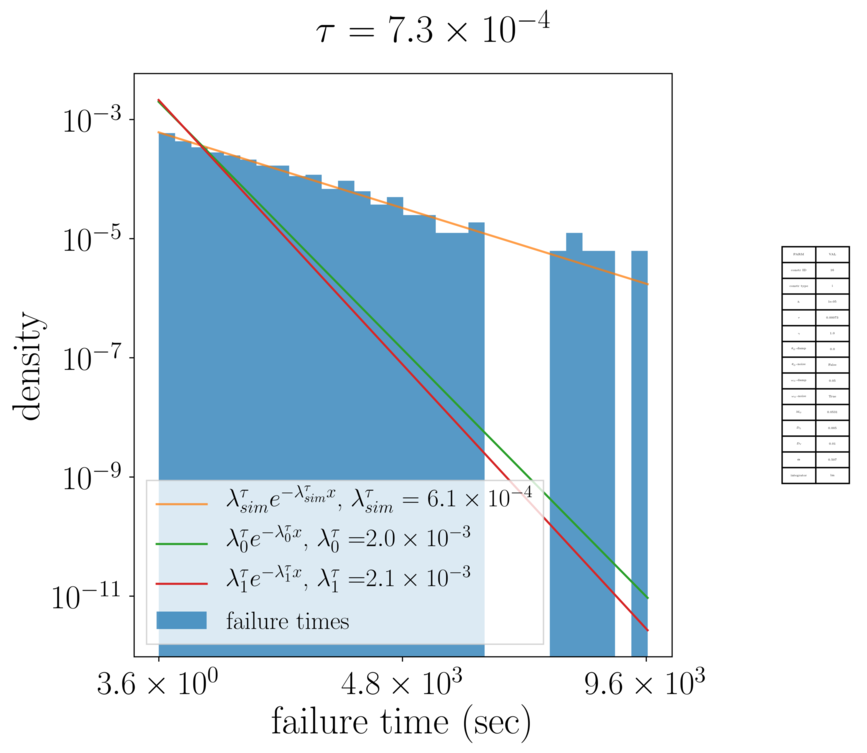}}%
\end{subfigure}
\begin{subfigure}[t]{\failurewidth\linewidth}%
	\centering
	\resizebox{\linewidth}{!}{\includegraphics[trim={0 0 3cm 1.5cm}, clip=true]{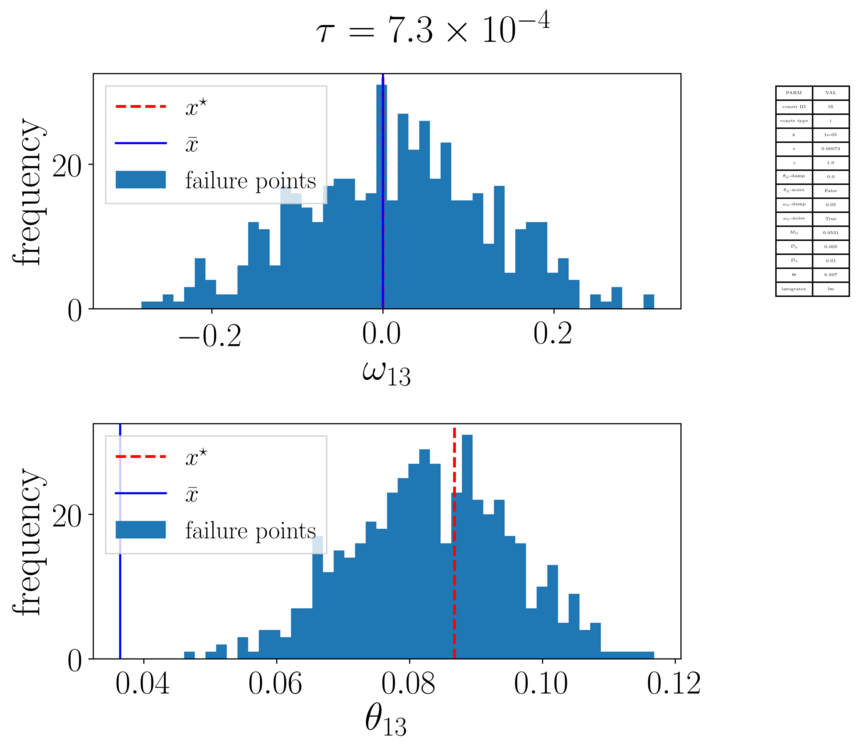}}
\end{subfigure}
\addtocounter{subfigure}{-4}
\begin{subfigure}[t]{\failurewidth\linewidth}
	\vspace{\failurecaption}
	\renewcommand\thesubfigure{II.\alph{subfigure}}
	\caption{}
	\label{fig:rate-uncond-worst-failrate_comparison}%
\end{subfigure}%
\hfill%
\begin{subfigure}[t]{\failurewidth\linewidth}
	\vspace{\failurecaption}
	\renewcommand\thesubfigure{II.\alph{subfigure}}
	\caption{}
	\label{fig:rate-uncond-worst-failratio_comparison}%
\end{subfigure}%
\hfill%
\begin{subfigure}[t]{\failurewidth\linewidth}
	\vspace{\failurecaption}
	\renewcommand\thesubfigure{II.\alph{subfigure}}
	\caption{}
	\label{fig:rate-uncond-worst-failtime_histogram}%
\end{subfigure}%
\hfill%
\begin{subfigure}[t]{\failurewidth\linewidth}
	\vspace{\failurecaption}
	\renewcommand\thesubfigure{II.\alph{subfigure}}
	\caption{}
	\label{fig:rate-uncond-worst-statespace_histogram}%
\end{subfigure}%
\label{fig:rate-worst}
\end{minipage}
\vspace{\failurecaption}
\caption{\small{Unconditional line failure diagnostics for the best ({5,000} simulations of line 10, top, (I)) and worst ({500} simulations of line 16, bottom, (II)) experiments, based on $\delta$ (see \cref{tab:convergence-summary}).}
\small{(a) Failure rate comparison across temperature.}
\small{(b) Asymptotic ratio across temperature.}
\small{(c) (Scaled) histogram of unconditional failure times at lowest temperature tested.}
\small{(d) Histogram of voltage, angle, or frequency state variable at observed failures relative to $\bar{x}_i$ and $x_i^{\star}$ where $i$ corresponds to the bus index of (one of) the line's connecting buses; shown at lowest temperature tested.}
}
\label{fig:rate-best-worst}
\end{figure}

\subsubsection*{First-order approximation justification}
In this section, we study the first-order approximation of the failure rate from \cref{eq:1-order-qst-final} (which we call the \emph{first-order heuristic method}) and estimate the prefactor constants $C_0, C^{\star}, C_1$, and $C^{\star\star}$ based on least-squares fits from simulation data (which we call the \emph{first-order empirical method}).
To estimate the prefactor constants, we first transform the simulated failure rate data to the ``prefactor'' space by removing the contribution of the analytic exponential factor.
Since the analysis that results in the zeroth-order rate \cref{eq:qst-exit-rate-final} is valid in the small-noise limit, we fit $\hat{C}_0$ from numerical simulation data only for temperatures such that $\tau \ll \Delta \mathcal{H}$.
We define this regime as $5\tau < \Delta \mathcal{H}$ and estimate
\begin{equation}
\label{eq:C0-estimate}
\hat{C}_0 = \argmin_c E_{C_0}, \quad  E_{C_0} \coloneqq \left\| \lambda^{\tau}_{sim} \times \exp \left[ \frac{\Delta \mathcal{H}}{\tau} \right] - C^{\star} \,c\, \tau^{-1/2} \right\|_2^2.
\end{equation}
Across the seven lines studied in our experiments, we find that our estimate of $\hat{C}_0$ is within an order of magnitude of $C_0(\bar{x})$ (see \cref{fig:rate-modification-fit}).

Similarly, we estimate $\hat{C}_1$ from simulation data in the large-noise limit in the regime where $\tau \gg \Delta \mathcal{H}$.
We define this regime as $\tau > 5 \Delta \mathcal{H}$ and estimate
\begin{equation}
\label{eq:C1-estimate}
\hat{C}_1 = \argmin_c E_{C_1}, \quad E_{C_1} \coloneqq \left\| \lambda^{\tau}_{sim} \times \exp \left[ \frac{\Delta \mathcal{H}}{\tau} \right] - (C^{\star} \hat{C}_0 +  C^{\star} \Delta \mathcal{H}^{-1} \,c\, \tau) \tau^{-1/2} \right\|_2^2
\end{equation}
after estimating $\hat{C}_0$ from \cref{eq:C0-estimate}.
Across all lines studied in our experiments (excluding the slack-load line for which we did not have high-temperature observations), we find agreement to within an order of magnitude indicating that $\hat{C}_1(\bar{x}) \approx C_0(\bar{x})$ and that our $C^{\star\star}$ approximation is reasonable.
Across the range of lines and temperatures studied, we note that the benefit from the empirical method is marginal, less than 0.2 logarithmic units (see \cref{tab:convergence-summary}) when compared with the heuristic method.
We also studied the inclusion of a fractional power $\tau^{1/2}$ in the prefactor estimate of \cref{eq:n-order-prefactor-final}, but empirical estimation did not meaningfully improve the fit over the first order method, so we did not pursue higher fractional orders (3/2, 5/2, $\ldots$).
Because of this and the convenience of not requiring any additional parameter estimation, we use \cref{eq:1-order-qst-final} with prefactor estimates from \cref{eq:first-order-prefactor} in the remainder of this document.
\bgroup
\def\arraystretch{1.75}%
\begin{figure}[H]
\begin{minipage}[t]{0.65\linewidth}
\vspace{0pt}
\footnotesize
\resizebox{\linewidth}{!}{%
\begin{tabular}{||ccc|>{\centering\arraybackslash}p{1.15cm}>{\centering\arraybackslash}p{1.15cm}|>{\centering\arraybackslash}p{1.15cm}>{\centering\arraybackslash}p{1.15cm}|>{\centering\arraybackslash}p{1.15cm}>{\centering\arraybackslash}p{1.15cm}||}
	\hline
	\multicolumn{3}{||c|}{\textbf{line}}                              & \multicolumn{2}{c|}{\textbf{$0^{th}$ order}} & \multicolumn{2}{c|}{\textbf{$1^{st}$ order empirical}} & \multicolumn{2}{c||}{\textbf{$1^{st}$ order heuristic}} \\ 
	\hline
	\multicolumn{1}{||c|}{number} & \multicolumn{1}{c|}{type} & risk & $\bar{\delta}$          & \textit{(sd)}         & $\bar{\delta}$          & \textit{(sd)}         & $\bar{\delta}$          & \textit{(sd)}         \\ \hline
	\hline
	\multicolumn{1}{||c|}{2\footnote{No ``high-temperature'' observations for line 2; $C_1$ taken as \cref{eq:first-order-prefactor}. Note also that line 2 is pathological and is pathological (\cref{def:nested}).}}         & \multicolumn{1}{c|}{sl-pq}     & low     & 0.82         & \textit{(0.49)}         & 0.29         & \textit{(0.19)}         & 0.64         & \textit{(0.45)}         \\ \hline
	\multicolumn{1}{||c|}{4}         & \multicolumn{1}{c|}{pq-pq}     & low     & 1.20         & \textit{(1.00)}         & 0.31         & \textit{(0.20)}         & 0.49         & \textit{(0.32)}         \\ \hline
	\multicolumn{1}{||c|}{10}         & \multicolumn{1}{c|}{pq-pq}     & high     & 4.18         & \textit{(3.37)}         & 0.33         & \textit{(0.20)}         & 0.47         & \textit{(0.31)}         \\ \hline
	\multicolumn{1}{||c|}{16}         & \multicolumn{1}{c|}{pv-pq}     & low     & 2.37         & \textit{(1.44)}         & 0.91         & \textit{(0.82)}         & 1.19         & \textit{(0.94)}         \\ \hline
	\multicolumn{1}{||c|}{20}         & \multicolumn{1}{c|}{pq-pq}     & med     & 2.59         & \textit{(2.25)}         & 0.51         & \textit{(0.40)}         & 0.70         & \textit{(0.31)}         \\ \hline
	\multicolumn{1}{||c|}{29}         & \multicolumn{1}{c|}{pv-pq}     & high     & 4.32         & \textit{(3.43)}         & 0.46         & \textit{(0.31)}         & 0.64         & \textit{(0.30)}         \\ \hline
	\multicolumn{1}{||c|}{35}         & \multicolumn{1}{c|}{pv-pq}     & high     & 4.58         & \textit{(3.64)}         & 0.50         & \textit{(0.35)}         & 0.55         & \textit{(0.35)}         \\ \hline
	\multicolumn{1}{||c|}{37}         & \multicolumn{1}{c|}{pv-pq}     & med     & 2.76         & \textit{(2.52)}         & 0.91         & \textit{(0.58)}         & 0.86         & \textit{(0.56)}         \\ \hline
	\hline
	\multicolumn{1}{||c|}{aggregate}         & \multicolumn{1}{c|}{$-$}     & $-$     & 2.88         & \textit{(2.87)}         & 0.53         & \textit{(0.49)}         & 0.69         & \textit{(0.53)}         \\ \hline
\end{tabular}
}
\end{minipage}
\hfill%
\begin{minipage}[t]{0.32\linewidth}
	\vspace{-1.5pt}
	\centering
	\resizebox{0.835\linewidth}{!}{\includegraphics[trim={0.25cm 0 0 0.75cm}, clip=true]{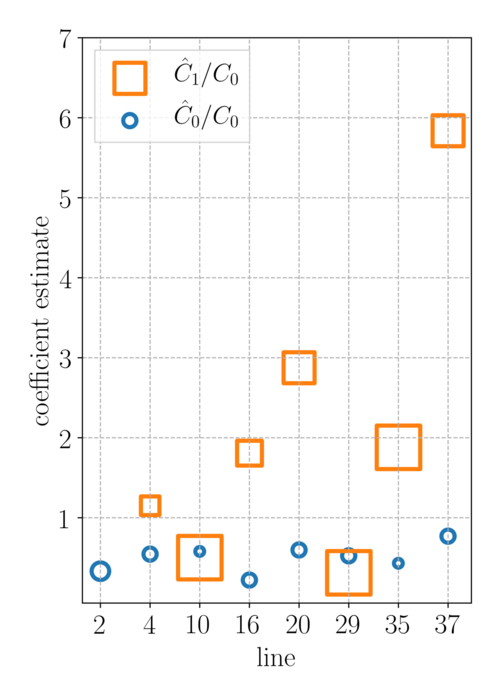}}
\end{minipage}
\begin{minipage}[t]{0.65\linewidth}
	\vspace{-8pt}
	\caption{\small{Mean absolute logarithmic error ($\bar{\delta} = N^{-1} \sum_{i=1}^N |\log(\lambda_{i} / \log \hat{\lambda}_{i})|$ across $N$ temperatures for simulated rate $\lambda$ and analytic rate $\hat{\lambda}$) \cite{tornqvist_how_1985} and standard deviation of absolute logarithmic error (sd) across all studied lines over all temperatures. The \texttt{aggregate} row is computed from the combined dataset.}}
	\label{tab:convergence-summary}
\end{minipage}
\hfill%
\begin{minipage}[t]{0.32\linewidth}
	\vspace{-8pt}
	\caption{\small{$\hat{C}_0/C_0$ and $\hat{C}_1/C_0$ relative estimates from data. Marker size corresponds to the number of data points used in the estimation.}}
	\label{fig:rate-modification-fit}
\end{minipage}
\end{figure}
\egroup

%

\subsubsection*{Summary, impact, and limitations}
Our numerical experiments provide evidence in support of the applicability of large deviation theory to the problem of unconditional line failure.
Specifically, our simulations validate the low-noise behavior predicted by both zeroth- and first-order theories for nonpathological lines.
Outside the low-noise regime, we find that the first-order approximation introduces additional robustness across a range of temperatures, and we observe an aggregate mean error of around 0.7 logarithmic units across all lines and temperatures studied.
Since regime classification is line dependent, this robustness is particularly important in ``relatively'' high-temperature cases where a line's energy difference may be smaller than the system's ambient temperature.
In these cases, the $\lambda \propto \tau^{1/2}$ behavior of the first order theory generally coincides with our simulations, although in a single case (line 37) the failure rate grew faster.
For practical uses, the large-noise limit is less of a concern, however; if a line is expected to fail in the near future, whether it fails in the next second or the next fraction of a second is of less importance.

We acknowledge a number of limitations in our framework.
First, the swing equation and dynamics are typically valid only for relatively small perturbations; we assume that the stochastic forcing is small enough that the dynamics are applicable.
Next, we were limited in exploring low-temperature (and, to a lesser extent, high-temperature) regimes because of the computational cost.
In addition, aside from potential modeling concerns such as appropriate system settings and network structure,\footnote{The IEEE 30-bus system is an approximation of a portion of the American Electrical Power System from the 1960s.} the failure rate is intimately connected to the system temperature, which may be difficult to estimate.\footnote{One approach, proposed in \cite{matthews_2018_simulating}, is to utilize the approximately ergodic properties of the system \cref{eq:stochastic-port-hamiltonian} to compute the mean magnitude of power grid frequency fluctuations over a long period of time.}

Most notably, however, the simulation methodology discussed above handles the tractable \textit{unconditional} failure problem, but the \textit{conditional} failure problem is of interest in practice.
Since conditional line $l$ failures must avoid certain regions of the state space (by definition, regions where other lines fail), computing the conditional failure rate requires solving a different set of Fokker-Planck equations~\cite{redner_2001_guide} to evaluate the corresponding conditional exit rate.
Solving the appropriate Fokker-Planck equation may be nontrivial, and for the remainder of this document we treat the unconditional failure rate as an approximation of the conditional failure rate.
In the next section, we present numerical evidence in support of the claim that the approximation is reasonable in the low-noise limit.


\section{Approximating the conditional failure rate}
\label{sec:adapt}
Here we take a detailed look at potential issues that may cause the unconditional failure rate to be a poor approximation of the conditional failure rate.
In the network dynamics \cref{eq:stochastic-port-hamiltonian}, the boundary of first failure (across any network line) is defined by the union of the individual line failure boundaries.
The conditional failure problem (\cref{def:unconditional-failure}) accounts for the presence of other line failure boundaries and allows for the decomposition of the joint boundary failure problem into separate failure problems that can be studied independently.
Since power systems are sparsely connected and designed to prevent simultaneous line failures, state variables that contribute to line $l$'s energy often do not significantly influence the line energies of other lines.
Accordingly, we anticipate that the regions of state space that correspond to line $l$ failure are in some sense isolated from regions of state space that may trigger other failures, especially at low loadings (see \cite{dobson_complex_2007}).
Based on these observations, we make the following hypothesis.
\begin{hypothesis}[Conditional failure approximation]\label{assn:cond-uncond}
	The conditional failure rate can be well approximated by the unconditional failure rate in the low-noise limit.
\end{hypothesis}
In the remainder of this section, we study the implications of this hypothesis and its validity by analyzing both static failure points and dynamic failure paths.

\subsection{Nested failure point analysis}
A direct indicator that the unconditional failure problem will be a poor approximation of the conditional failure problem is the presence of multiple failures associated with an unconditional failure point \cref{eq:xstar}.
For a line $l$ of interest, such a scenario indicates that the lowest energy failure point $x^{\star}$ is contained within the failure region of another line.
The lowest energy failure point such that \emph{only} the line of interest is tripped can be computed by the ``conditional'' NLP formulated as
\begin{subequations}
	\label{eq:cond-nlp}
	\begin{alignat}{3}
	\underset{x}{\min} & \quad && \mathcal{H}(x) \\
	\text{s.t.} && \quad & \Theta_l(x) = \Theta_l^{\max} \\
	&&& \Theta_k(x) \leq \Theta_k^{\max} \pm \epsilon, \quad \forall k \in \mathcal{A},
	\end{alignat}
\end{subequations}
where $\mathcal{A}$ is the set of acceptable lines (those which are still active and exclude line $l$) and $\epsilon$ is a small positive parameter that enforces strict/loose conformity.
When \cref{eq:cond-nlp} is infeasible, the line of interest will never fail by itself, meaning that the line's failure region is nested entirely within the region of another line and that the MFPT theory is not applicable.
When \cref{eq:cond-nlp} is feasible, however, the unconditional problem provides a low-noise asymptotic upper bound on the failure rate.
We summarize the scenario as follows.
\begin{definition}[Nested pathology]\label{def:nested}
	A transmission line has a \emph{conditionally nested} failure region if its conditional failure point problem \cref{eq:cond-nlp} is infeasible.
	Similarly, a transmission line is \emph{unconditionally nested} if the solution to its unconditional failure point problem \cref{eq:xstar} results in line failures besides the line of interest.
\end{definition}

To quantify the prevalence and impact of nestedness, we proceed in two steps.
First we determine prevalence by computing the proportion of infeasible conditional NLPs across all system lines (for conditional nestedness) and the proportion of NLP solutions \cref{eq:xstar} that trigger multiple lines (for unconditional nestedness), as summarized in \cref{fig:cond-uncond-diagnostics} panel (a).
We scan for nestedness in both the IEEE 30-bus and 118-bus systems \cite{matpower}, which we augment with physical parameters from \cref{fig:30bus-dynamics-parameters}.\footnote{The 118-bus case file does not contain line limits, so we set limits to ensure a standard measure of system security ($N$-$1$ security, defined and studied in \cite{vonmeier_electric_2006,alsac_optimal_1974}).}
Second we use \cref{eq:1-order-qst-final} to compare failure rates computed at both the conditional and unconditional failure point solutions, as summarized in \cref{fig:cond-uncond-diagnostics} panels (b)--(c).

In each problem instance, we observe that the majority of the system's lines are neither conditionally nor unconditionally nested.
Furthermore, we find (i) that nestedness becomes rarer in the larger system and (ii) that more vulnerable (lower $\Delta \mathcal{H}$) lines have a lower rate of nestedness than the overall system rate.
Three of forty lines in the 30-bus system exhibit conditional NLP infeasibility, but only three of 131 lines studied in the 118-bus system do (see \cref{fig:30-failrate-cond-uncond-sort,fig:118-failrate-cond-uncond-sort}).
Similarly, fifteen lines are unconditionally nested in the 30-bus system, but only eighteen lines are in the 118-bus system (see \cref{fig:30-trips-cond-uncond,fig:118-trips-cond-uncond}).
Additionally, we observe that the set of conditionally nested lines is a subset of the set of unconditionally nested lines, indicating that unconditional nestedness is a useful proxy for conditional nestedness.
As expected for non-nested lines, panels (b)--(c) of \cref{fig:cond-uncond-diagnostics} show that the relative error between failure rates \cref{eq:1-order-qst-final} computed from the conditional failure point and the unconditional failure point is small (below $10^{-5}$).
For unconditionally nested lines, larger errors are observed, indicating that such lines cannot be studied by the theory in \cref{subsec:rate-asymptotic}.
Across all lines, about 75\% of the lines in the 30-bus case have a relative error less than 1\%, and about 85\% of the lines in the 118-bus case have relative error less than $10^{-5}$.
We find these results encouraging, not only because nestedness is uncommon, but also because the low energy difference lines that are the limiting factors for system reliability exhibit lower rates of nestedness.
For larger systems, we expect that nestedness will become even less prevalent, and we use the unconditional failure point \cref{eq:xstar} when computing individual failure rates.
\newcommand{\CondUncondWidth}{0.325}
\begin{figure}[H]
	\begin{subfigure}[T]{\CondUncondWidth\linewidth}%
		\centering
		\resizebox{\linewidth}{!}{\includegraphics{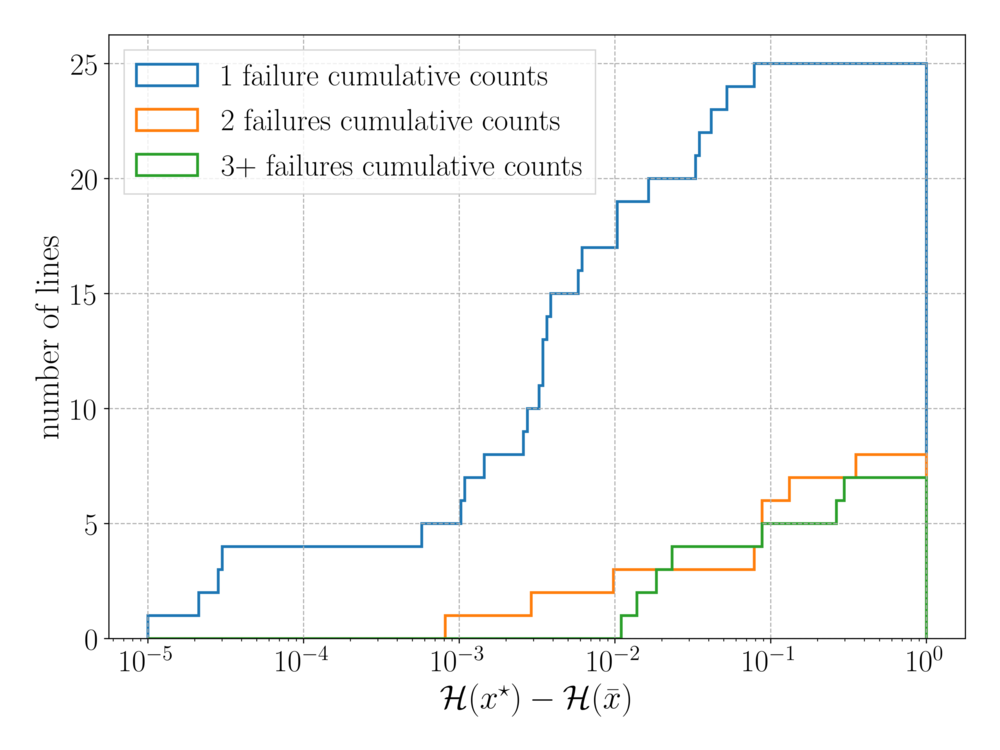}}%
	\end{subfigure}	
	\begin{subfigure}[T]{\CondUncondWidth\linewidth}%
		\centering
		\resizebox{\linewidth}{!}{\includegraphics{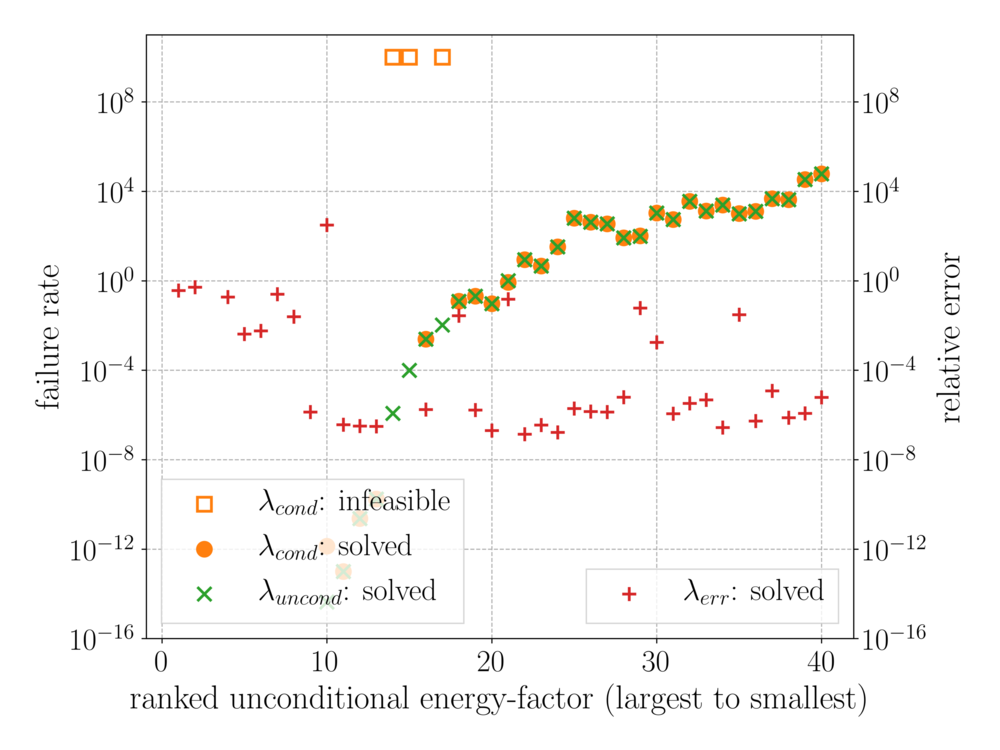}}%
	\end{subfigure}
	\begin{subfigure}[T]{\CondUncondWidth\linewidth}%
		\centering
		\resizebox{\linewidth}{!}{\includegraphics{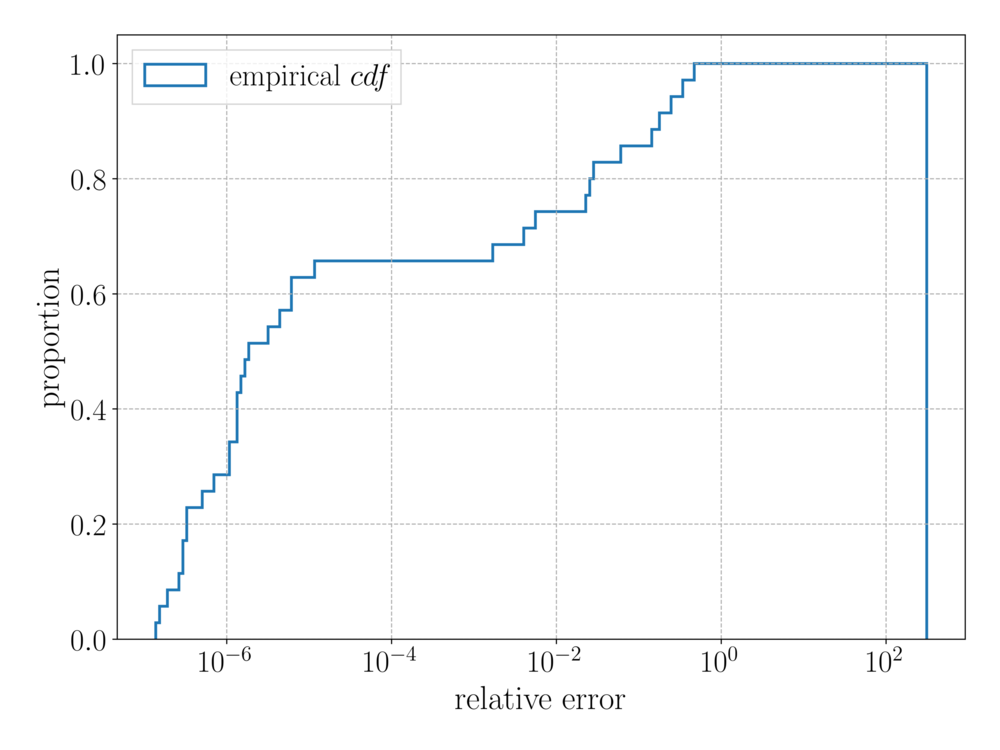}}%
	\end{subfigure}
	\begin{subfigure}[b]{\CondUncondWidth\linewidth}%
		\renewcommand\thesubfigure{I.\alph{subfigure}}%
		\vspace{-10pt}
		\caption{}
		\label{fig:30-trips-cond-uncond}
	\end{subfigure}
	\hfill%
	\begin{subfigure}[b]{\CondUncondWidth\linewidth}%
		\renewcommand\thesubfigure{I.\alph{subfigure}}%
		\vspace{-10pt}
		\caption{}
		\label{fig:30-failrate-cond-uncond-sort}
	\end{subfigure}
	\hfill%
	\begin{subfigure}[b]{\CondUncondWidth\linewidth}%
		\renewcommand\thesubfigure{I.\alph{subfigure}}%
		\vspace{-10pt}
		\caption{}
		\label{fig:30-failrate-cond-uncond-hist}
	\end{subfigure}
\end{figure}
\vspace{-1cm}
\begin{figure}[H]\ContinuedFloat
	\begin{subfigure}[T]{\CondUncondWidth\linewidth}%
		\centering
		\resizebox{\linewidth}{!}{\includegraphics{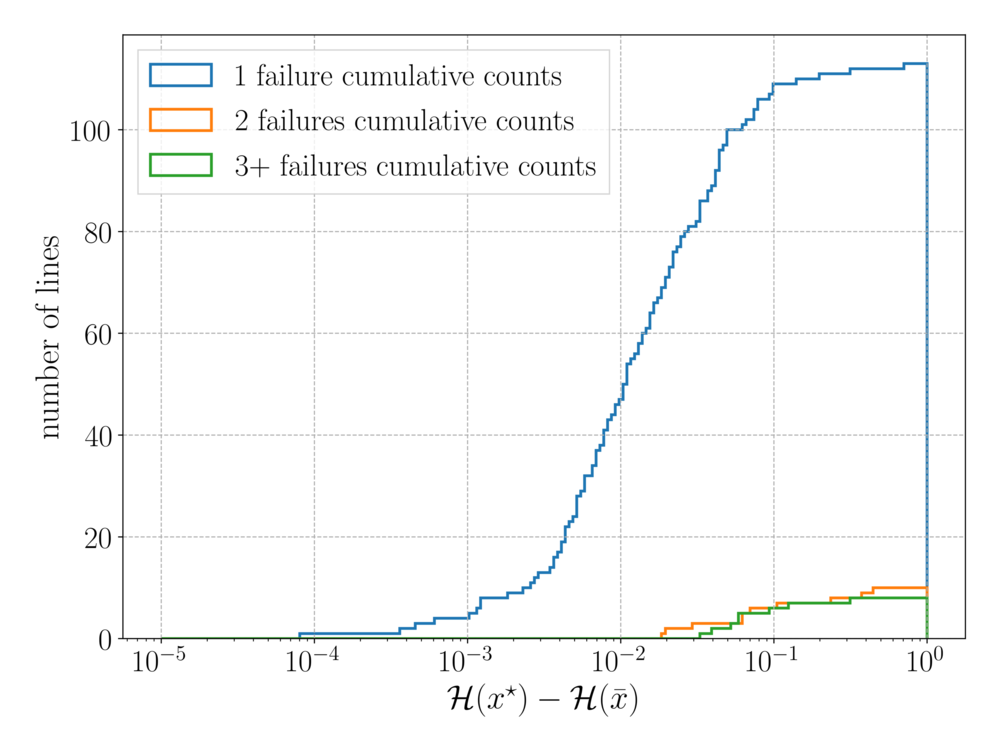}}%
	\end{subfigure}
	\begin{subfigure}[T]{\CondUncondWidth\linewidth}%
		\centering
		\resizebox{\linewidth}{!}{\includegraphics{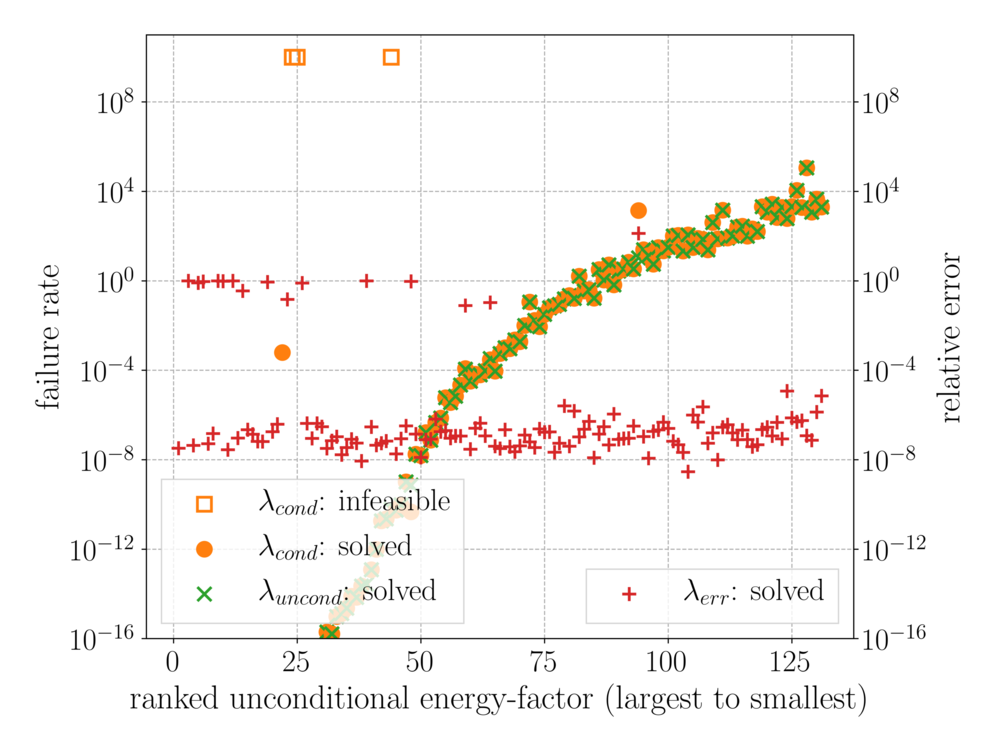}}%
	\end{subfigure}
	\begin{subfigure}[T]{\CondUncondWidth\linewidth}%
		\centering
		\resizebox{\linewidth}{!}{\includegraphics{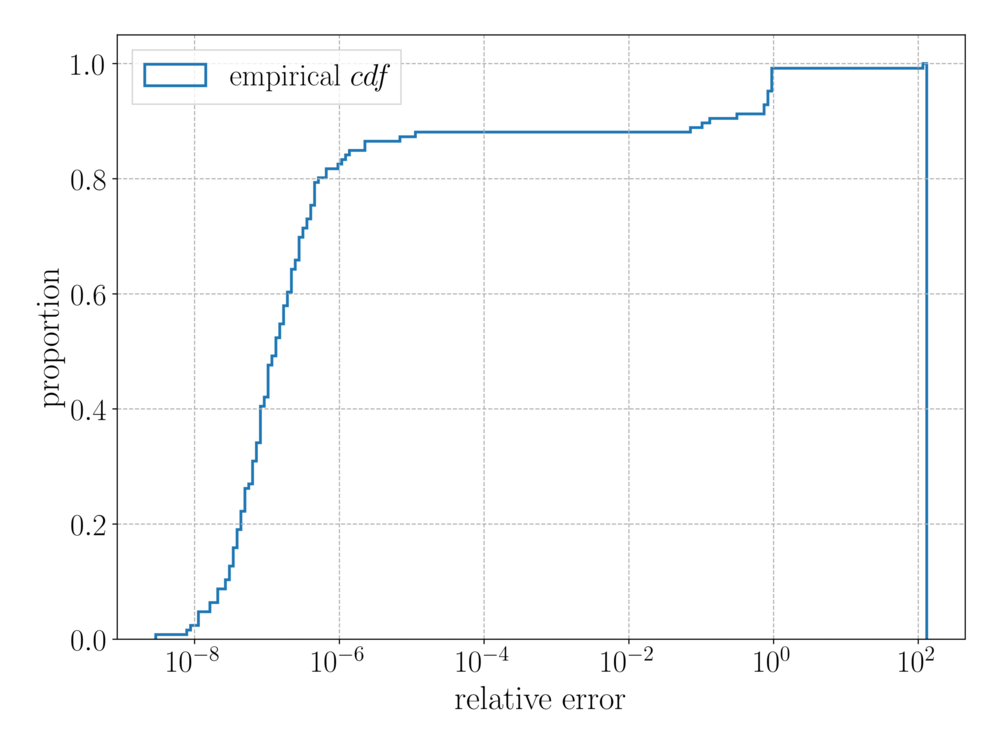}}%
	\end{subfigure}
	\addtocounter{subfigure}{-3}%
	\begin{subfigure}[b]{\CondUncondWidth\linewidth}%
		\renewcommand\thesubfigure{II.\alph{subfigure}}%
		\vspace{-10pt}
		\caption{}
		\label{fig:118-trips-cond-uncond}%
	\end{subfigure}
	\hfill%
	\begin{subfigure}[b]{\CondUncondWidth\linewidth}
		\renewcommand\thesubfigure{II.\alph{subfigure}}%
		\vspace{-10pt}
		\caption{}
		\label{fig:118-failrate-cond-uncond-sort}%
	\end{subfigure}
	\hfill%
	\begin{subfigure}[b]{\CondUncondWidth\linewidth}%
		\renewcommand\thesubfigure{II.\alph{subfigure}}%
		\vspace{-10pt}
		\caption{}
		\label{fig:118-failrate-cond-uncond-hist}%
	\end{subfigure}
\vspace{-0.4cm}
\caption{\small{Nested failure diagnostics for 30-bus (I) and 118-bus (II) IEEE systems.}
\small{(a) Empirical cumulative counts of total number of line failures associated with line $l$ failure as a function of (unconditional) energy difference $\Delta \mathcal{H}$. 1 line failure means that line $l$ is the only line to fail.}
\small{(b) Unconditional failure rate computed at unconditional failure rate and relative error (versus unconditional failure rate computed at conditional failure point) ordered from largest $\Delta \mathcal{H}$ to smallest $\Delta \mathcal{H}$.}
\small{(c) Empirical cdf of relative error.}
}
\label{fig:cond-uncond-diagnostics}
\end{figure}

\subsection{Nonisolated failure point analysis}
Even when considering failure boundaries in isolation of each other, the constrained energy minimization problems~\cref{eq:xstar,eq:cond-nlp} may have multiple local solutions, since they are nonconvex, nonlinear programs.
Solutions that are isolated from each other are straightforward to diagnose; in the limit of low noise, the minimal energy solution will dominate.
Conversely, lines whose constraint intersects a level set of the energy function cannot be handled by the asymptotic theory.
We collect such cases in the following definition.
\begin{definition}[Nonisolated pathology]\label{def:non-isolated}
	A line $l$ is pathological in the \emph{nonisolated} sense if solutions to the unconditional failure problem form a continuum in state space.
\end{definition}

We study isolatedness by solving the unconditional NLP from 100 randomly initialized seeds corresponding to perturbed solutions of the initial dispatch point, in addition to one seed at $\bar{x}$.
For those problems that converge, we round the failure point $x^{\star}$ to four digits of accuracy and obtain a lower bound for the number of isolated solutions.
We compute the failure rate at each isolated solution and compare the rates with unconditional failure rate (that is, the solution to the unconditional NLP initialized at the operating point $\bar{x}$).

\begin{figure}[H]
\begin{subfigure}[T]{0.495\linewidth}
	\centering
	\resizebox{\linewidth}{!}{\includegraphics{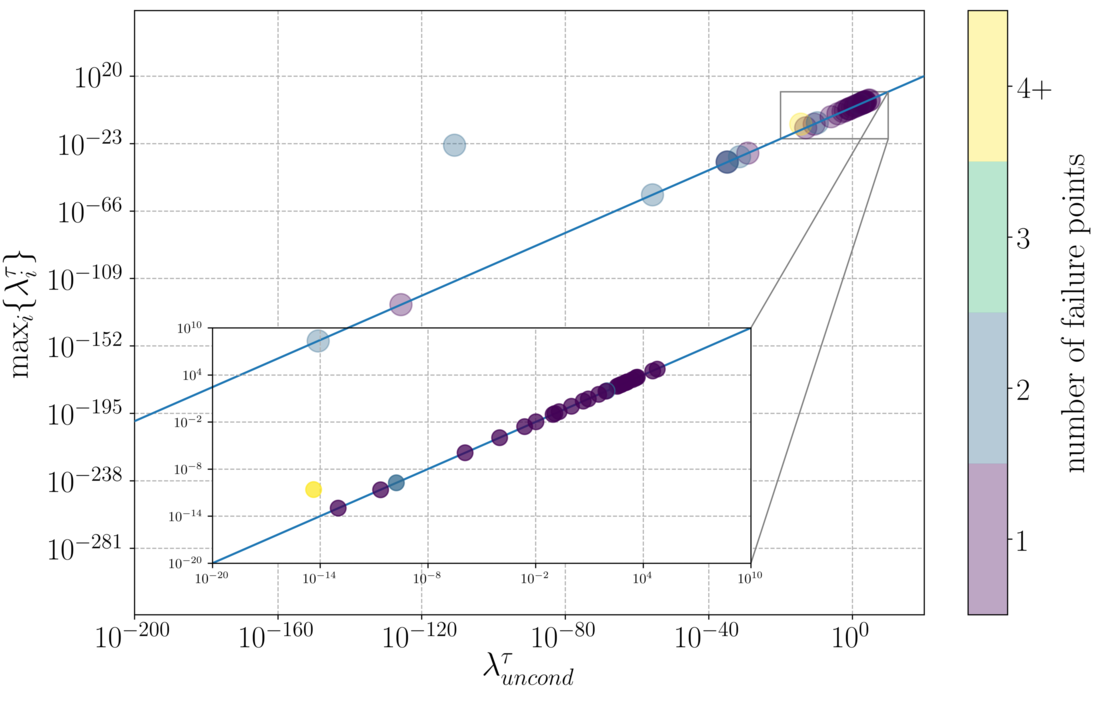}}
\end{subfigure}
\begin{subfigure}[T]{0.495\linewidth}
	\centering
	\resizebox{\linewidth}{!}{\includegraphics{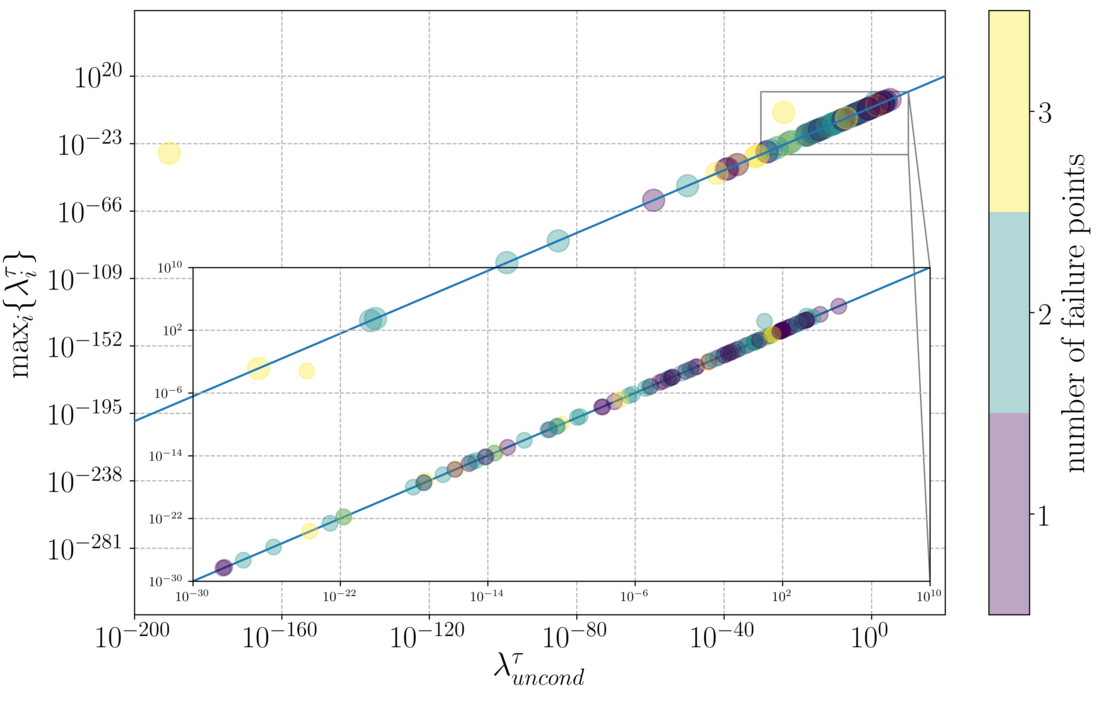}}
\end{subfigure}
\begin{subfigure}[b]{0.495\linewidth}
	\vspace{-8pt}
	\caption{}
	\label{fig:case30-multisolve-impact}
\end{subfigure}
\begin{subfigure}[b]{0.495\linewidth}
	\vspace{-8pt}
	\caption{}
	\label{fig:case118-multisolve-impact}
\end{subfigure}
\vspace{-0.75cm}
\caption{\small{Study of multiplicity of solutions to \cref{eq:xstar}. First order failure rates \cref{eq:1-order-qst-final} computed at $\tau = 10^{-3}$; $y$-axis shows the maximum first order failure rate computed across solutions to the randomly initialized unconditional NLP \cref{eq:xstar}, and $x$-axis shows the first order failure rate computed from the solution to \ref{eq:xstar} initialized at $\bar{x}$. Blue line indicates perfect agreement.}
	(\subref{fig:case30-multisolve-impact}) \small{30-bus system.}
	(\subref{fig:case118-multisolve-impact}) \small{118-bus system.}
}
\label{fig:multisolve}
\end{figure}

In the vast majority of lines studied numerically, the presence of multiple solutions does not pose a significant problem to failure rate estimation (see \cref{fig:multisolve}), and hence lines with nonisolated solutions are in some sense ``pathological.''
In \cref{fig:multisolve} we find good agreement between the maximum failure rate (across isolated solutions) and the $\bar{x}$-initialized failure rate across both test systems, especially for lines with higher unconditional failure rates.
In the larger system, we observe that increased NLP complexity contributed to an increased prevalence of local minima.
In the 30-bus system, across all non \emph{generator-generator} lines, 78\% of NLP solutions find a single failure point, 20\% find two, and 2\% (line 13, a unique case of an isolated load connected to the network through another load) find a continuum of failure points.
In the 118-bus system, across all non \emph{generator-generator} lines, 44\% of NLP solutions find a single failure point, 40\% find two, 8\% find three, and 7\% find four.
Ultimately, solutions initialized at $\bar{x}$ well approximate the lowest energy minimizer obtained from multiple restarts in both test cases.

We remark that violation of the applicability conditions \ref{assn:non-characteristic}--\ref{assn:noise} can be detected first by inspection (that is, by line type) and second by solving the unconditional NLP.
Optimality conditions require $\nabla \mathcal{H}(x^{\star}) = k \nabla \Theta_l(x)$; and if $k < 0$, we can conclude that conditions \ref{assn:non-characteristic} and \ref{assn:name-tbd} are not satisfied at $x^{\star}$.
In practice, we have observed numerical satisfaction of the applicability criteria near $x^{\star}$, although this is not consistent along the entire boundary $\partial D_l$ of the safety region \cref{eq:safety-region}.

\subsection{Failure path analysis}
Another indicator that the unconditional problem will poorly approximate the conditional problem is if the most likely exit path from the equilibrium to the constrained energy minimizer passes through a region of state space which triggers another line failure.
We collect such instances in the following definition.
\begin{definition}[Dynamically inaccessible pathology]\label{def:inaccessible}
	A line $l$ is pathological in the (dynamically) \emph{inaccessible} sense if the most likely exit path triggers line relays for lines $k \ne l$.
\end{definition}

\begin{figure}[H]
\begin{subfigure}[T]{0.325\linewidth}
	\centering
	\resizebox{\linewidth}{!}{\includegraphics{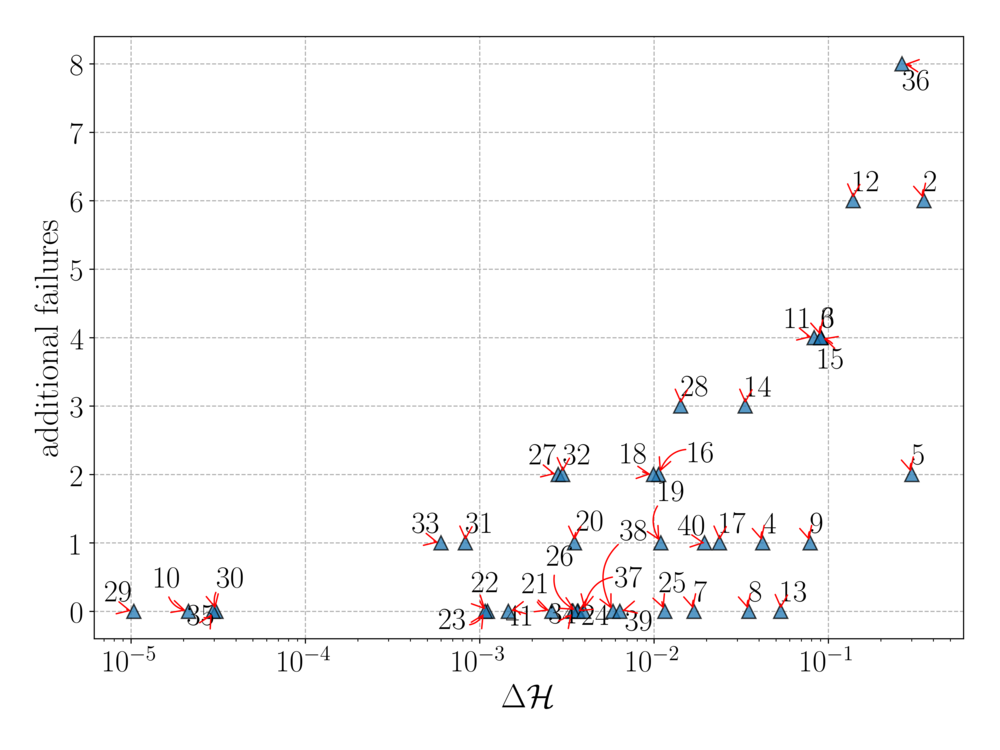}}
\end{subfigure}
\begin{subfigure}[T]{0.325\linewidth}
	\centering
	\resizebox{\linewidth}{!}{\includegraphics{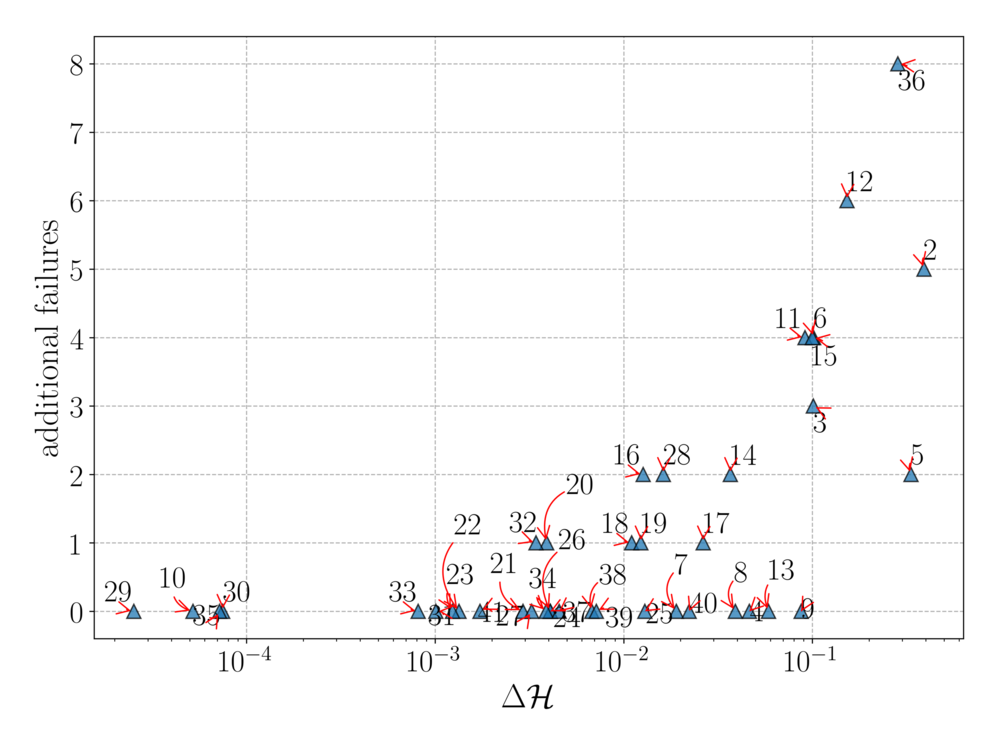}}
\end{subfigure}
\begin{subfigure}[T]{0.325\linewidth}
	\centering
	\resizebox{\linewidth}{!}{\includegraphics{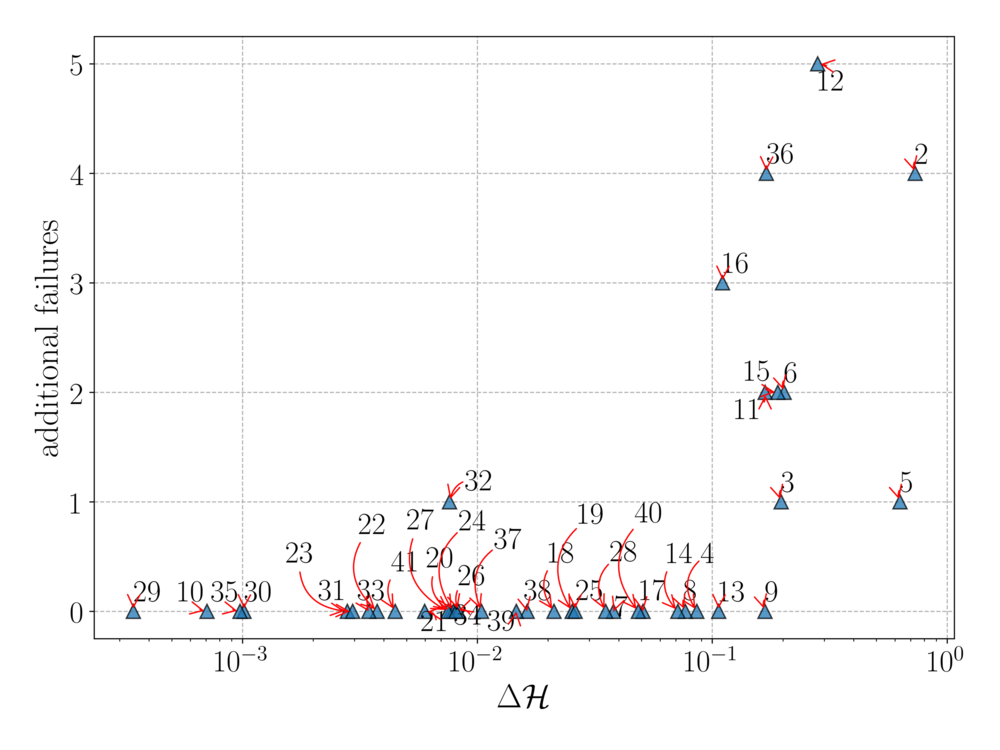}}
\end{subfigure}
\begin{subfigure}[b]{0.325\linewidth}
	\vspace{-8pt}
	\caption{}
	\label{fig:path-diagnostic-30}
\end{subfigure}
\begin{subfigure}[b]{0.325\linewidth}
	\vspace{-8pt}
	\caption{}
	\label{fig:path-diagnostic-30-1_1}
\end{subfigure}
\begin{subfigure}[b]{0.325\linewidth}
	\vspace{-8pt}
	\caption{}
	\label{fig:path-diagnostic-30-2_0}
\end{subfigure}
\vspace{-0.5cm}
\caption{\small{Study of fluctuation dynamics paths displaying number of unintended failures along the pathway between $x^{\star}$ and $\bar{x}$ for each line in the 30-bus system (see supplemental materials for more detail).}
	(\subref{fig:path-diagnostic-30}) \small{Standard (\texttt{Matpower}) line limits.}
	(\subref{fig:path-diagnostic-30-1_1}) \small{Line limits uniformly increased by 10\%.}
	(\subref{fig:path-diagnostic-30-2_0}) \small{Line limits uniformly increased by 100\%.}
}
\label{fig:path-diagnostic-30bus}
\end{figure}

We study inaccessibility by computing the most likely exit path of the dynamics through the boundary $\partial D_l$.
For the power system dynamics~\cref{eq:stochastic-port-hamiltonian}, the most likely exit path solves the backwards problem (see Section 2.1.1 in~\cite{bouchet_2016_generalisation} and Chapter 4, Theorem 3.1 in~\cite{wf_random_2012})
\begin{equation}
  \label{eq:fluctuation}
  \dot{\phi}^x_t = S \nabla \mathcal{H}(x) + J \nabla \mathcal{H}(x), \quad
  \phi^x_0 = x^{\star}, \quad t \leq 0.
\end{equation}
By numerically solving \cref{eq:fluctuation} backwards in time, we compute failure paths for each line in both the IEEE 30-bus and 118-bus networks and find that the majority do not trigger other line failures.
In particular, we remark that uniformly increasing the line limits generally separates failure pathways in state space, as shown in \cref{fig:path-diagnostic-30bus}.
Similarly, although not directly comparable because of network differences, a comparison of failure pathways under $N$-1 secure line limits in both 30-bus and 118-bus systems indicates that well-designed limits also increasingly tend to separate in higher dimensions.
In summary, the most likely exit path analysis provides evidence in favor of using the unconditional failure rate as an approximation of the conditional failure rate in realistic models at low temperatures.
%


\section{Chemical kinetics Markov model}
\label{sec:markov}
In this section, we introduce a chemical kinetics-inspired Markov model of cascading line failure under specified operating conditions.
Specifically, we interpret the unconditional failure rate \cref{eq:1-order-qst-final} as a chemical ``reaction rate'' in the sense of transition state theory (see \cite{bao_variational_2017} for a recent overview) and aggregate individual rates into a composite model derived from kinetic Monte Carlo (KMC, also known as dynamic Monte Carlo) methods.
KMC methods were developed to discretize a system's continuous dynamical behavior into a sequence of distinct events representing transitions between discrete states that occur at known rates \cite{voter_introduction_2007}.
An attractive feature of these models is that if the rates corresponding to mutually exclusive events are both correct and independent of the preceding states, then the event-based system evolves at the same time scale as the original system \cite{voter_introduction_2007,serebrinsky_2011_physical}.

Our model discretizes the dynamics of \cref{eq:stochastic-port-hamiltonian} by employing the first-order failure rate expression \cref{eq:1-order-qst-final} to catalog transition rates such that each event corresponds to (i) an individual line failure and (ii) an irreversible transition to an increasingly degraded operating point.
By storing an on-the-fly event list and rate catalog in a weighted directed acyclic graph (DAG), we can generate realizations of possible failure pathways by traversing the DAG rather than by integrating the dynamics of \cref{eq:stochastic-port-hamiltonian}.
The event-based approach can generate sequences of transmission line failures with computational effort that is independent of the temperature $\tau$.
As a result, the method allows for the study of cascades in settings that may otherwise be intractable because of the prohibitively high computational cost of direct numerical integration at low temperatures.
Below, we detail the model structure and simulation methodology (\cref{subsec:markov-algorithm}), its assumptions (\cref{subsec:markov-assns}), and our numerical experiments (\cref{subsec:markov-numerical}).


\subsection{DAG failure pathways and Markov simulation}
\label{subsec:markov-algorithm}
The organizing principle of the DAG is to enumerate all possible failure paths between the fully operational topological network state (equivalently, \textit{energy surface}) $S_0$ (where no lines have failed) and the fully degraded topological network state $S_{\infty}$ (where all lines have failed).
In this context, each directed edge hierarchically links two energy surfaces, $S$ and $S^{\prime}$ such that $S^{\prime} \backslash S$ contains the set of failed lines linking $S$ to $S^{\prime}$ and its weight is specified by the transition rate from $S$ to $S^{\prime}$.
For simplicity, we disregard simultaneous line failures so that edges connect only topological states that are separated by a single line failure.

At state $S$, under certain assumptions described in \cref{subsec:markov-assns}, the failure rate to any child state $\lambda_{S \to {S^{\prime}}}$ can be computed directly from the information contained in the parent state $S$ and hence defines a Markov process.
Specifically, the set of line failures in $S$ (in conjunction with the original dispatch data) defines an energy surface with equilibrium minimizer $\bar{x}_{S}$, and the computation of failure point ${x}^{\star}_{{S} \to {S^{\prime}}}$ proceeds as before \cref{eq:xstar}.
Since the failure rate model depends only on the combination of failed lines in a topological state $S$, the full DAG $\mathcal{G} = (\mathcal{V},\, \mathcal{D})$ composed of vertices $\mathcal{V}$ and directed edges $\mathcal{D}$ (depicted in \cref{apx:markov}) can be expressed with a number of vertices that scale exponentially rather than combinatorially. 
Once all such topological states have been enumerated, failure sequences can be generated by traversing the graph according to the relative weights of each edge.

In order to enumerate the DAG, three inputs are required: (i) an initial network topology, (ii) dispatch data (which defines an initial operating point), and (iii) a set of exogenous dynamics parameters including (temperature $\tau$ and physical parameters like damping and mass).
The first input determines each vertex in $\mathcal{V}$ as well as the hierarchical interconnections of each edge in $\mathcal{D}$, while the second two inputs determine the weights of each edge.
In this way, the entire DAG stems from the original dispatch point.
For nontrivial networks, enumerating every possible vertex in the graph \textit{a priori} is infeasible, so we employ a depth-first search of the failure pathway space to enumerate ``likely'' sequences of failures, as summarized in \cref{alg:kmc}.


\subsection{Markov model hypotheses and evidence}
\label{subsec:markov-assns}
The efficient exploration of the combinatorial statespace of likely failure pathways depends on identifying susceptible lines at each vertex.
We use the unconditional failure rate model \cref{eq:1-order-qst-final} as a proxy for susceptibility along each directed edge.

To compute directed edge weights for the Markov model using transition state theory, we rely on the notion of ``equilibration'' for the stochastic dynamics \cref{eq:stochastic-port-hamiltonian}:
\begin{definition}[Equilibration]\label{defn:markov-reeq}
	A stochastic process has \emph{equilibrated} if its law has converged to an appropriate quasistationary distribution.
\end{definition}
Following a line failure in the continuous power grid dynamics \cref{eq:stochastic-port-hamiltonian}, the system has equilibrated according to \cref{defn:markov-reeq} if (after updating the susceptance matrix $B$), an ensemble of trajectories from the parent state has relaxed to a quasistationary distribution over the child state.
In certain cases, however, the process may fail to equilibrate or may fail to do so in an amount of time that is negligible relative to the time of the next failure.
The theory in \cref{sec:rate} does not handle the computation of failure rates for processes that have not equilibrated, so we propose the following hypothesis:
\begin{hypothesis}[Rarity of nonequilibration]\label{assn:markov-reeq}
	Instances of nonequilibration are uncommon when modeling cascading line failure. 
\end{hypothesis}
Researchers have shown that there exist deterministic failure pathways between cascade topologies~\cite{yang-motter_2017_continuous}, but we assume that such cases are rare.
\cref{assn:markov-reeq} allows for the formulation of conditional and unconditional failure rate problems for individual lines.
Though the conditional failure problem gives the correct line failure rates at each level, we rely on \cref{assn:cond-uncond} and its encouraging validation results to justify computing edge weights via the unconditional failure rate \cref{eq:1-order-qst-final}.

\begin{figure}[H]
	\begin{minipage}[T]{0.65\linewidth}
		\begin{subfigure}{0.325\linewidth}
			\centering
			\resizebox{0.97\linewidth}{!}{\includegraphics[trim={0.5cm 0 3cm 1.25cm}, clip=true, keepaspectratio]{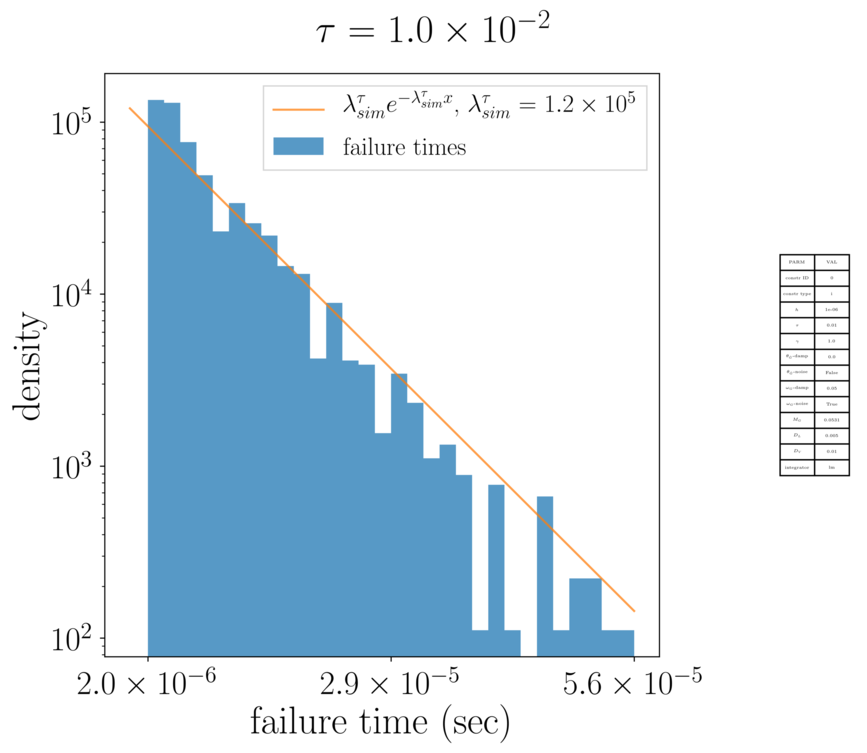}}
		\end{subfigure}%
		\begin{subfigure}{0.325\linewidth}
			\centering
			\resizebox{0.97\linewidth}{!}{\includegraphics[trim={0.5cm 0 3cm 1.25cm}, clip=true, keepaspectratio]{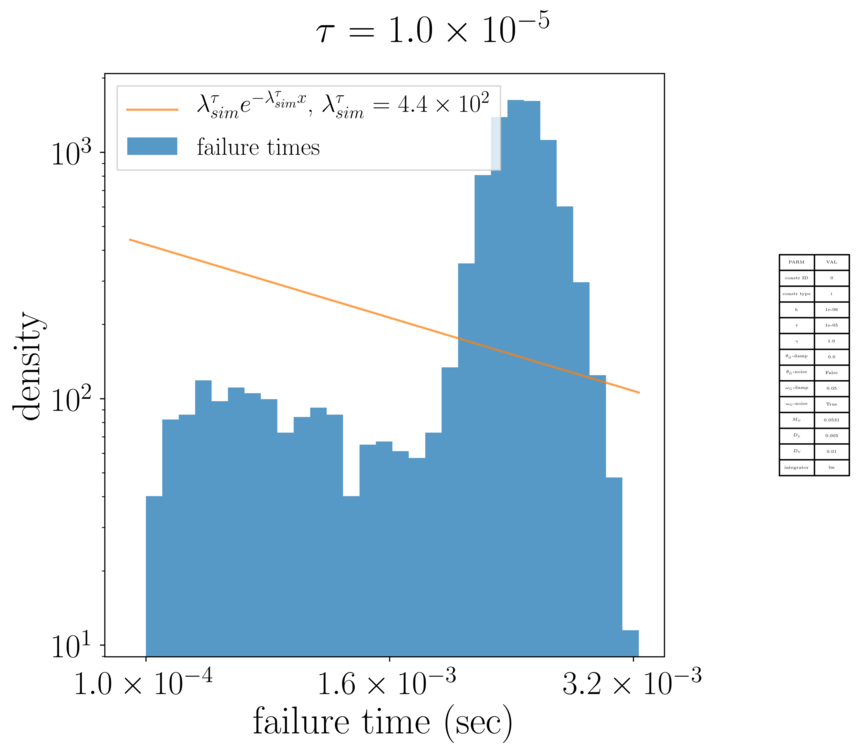}}
		\end{subfigure}%
		\begin{subfigure}{0.325\linewidth}
			\centering
			\resizebox{0.97\linewidth}{!}{\includegraphics[trim={0.5cm 0 3cm 1.25cm}, clip=true, keepaspectratio]{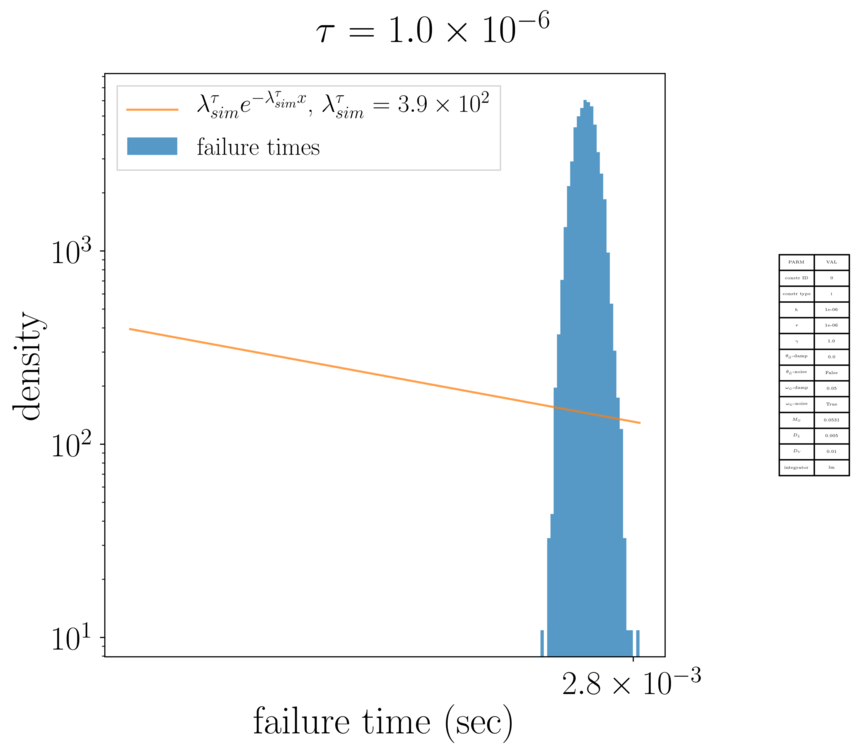}}
		\end{subfigure}%
		\hfill%
		\\
		\begin{subfigure}{0.325\linewidth}
			\vspace{-8pt}
			\caption{}
			\label{fig:reeq-a-lo}
		\end{subfigure}%
		\hfill%
		\begin{subfigure}{0.325\linewidth}
			\vspace{-8pt}
			\caption{}
			\label{fig:reeq-b-lo}
		\end{subfigure}%
		\hfill%
		\begin{subfigure}{0.325\linewidth}
			\vspace{-8pt}
			\caption{}
			\label{fig:reeq-c-lo}
		\end{subfigure}%
		\hfill%
		\\
		\begin{subfigure}{0.325\linewidth}
			\centering
			\resizebox{0.97\linewidth}{!}{\includegraphics[trim={0.5cm 0 3cm 1.25cm}, clip=true, keepaspectratio]{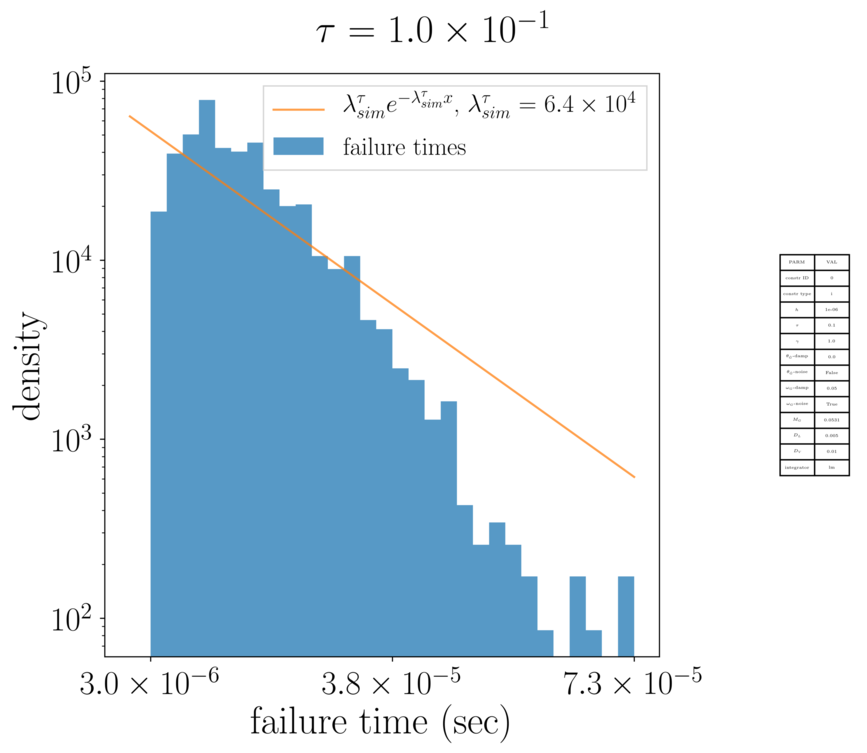}}
		\end{subfigure}%
		\begin{subfigure}{0.325\linewidth}
			\centering
			\resizebox{0.97\linewidth}{!}{\includegraphics[trim={0.5cm 0 3cm 1.25cm}, clip=true, keepaspectratio]{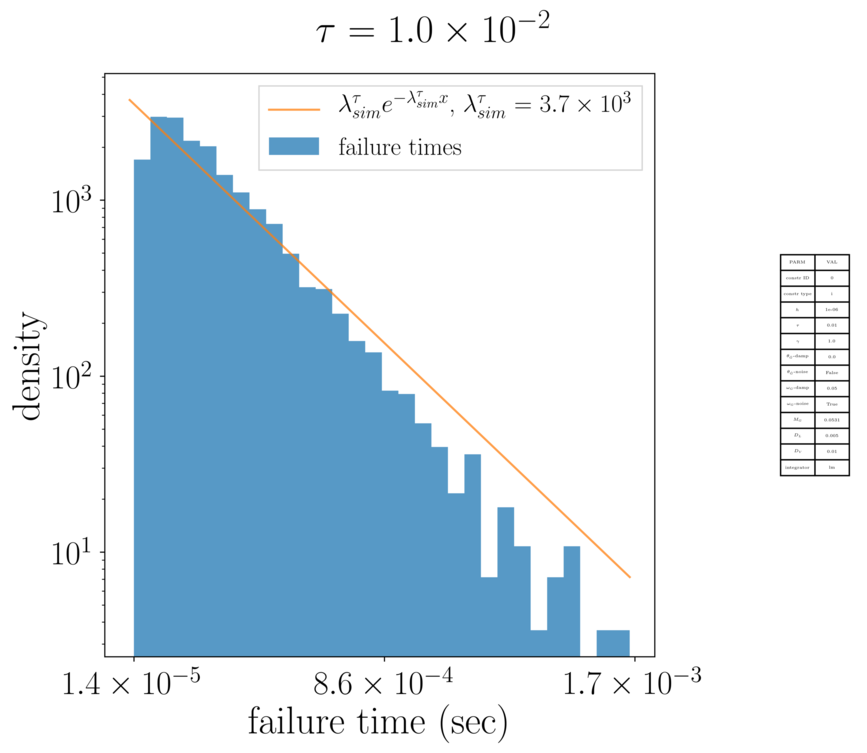}}
		\end{subfigure}%
		\begin{subfigure}{0.325\linewidth}
			\centering
			\resizebox{0.97\linewidth}{!}{\includegraphics[trim={0.5cm 0 3cm 1.25cm}, clip=true, keepaspectratio]{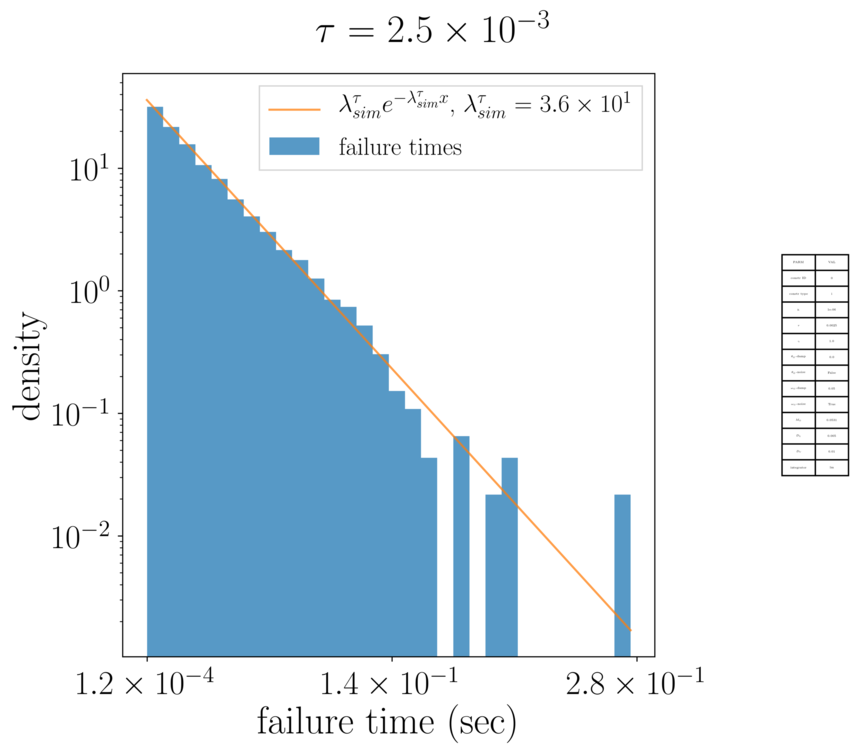}}
		\end{subfigure}%
		\hfill%
		\\
		\begin{subfigure}{0.325\linewidth}
			\vspace{-8pt}
			\caption{}
			\label{fig:reeq-a-hi}
		\end{subfigure}%
		\hfill%
		\begin{subfigure}{0.325\linewidth}
			\vspace{-8pt}
			\caption{}
			\label{fig:reeq-b-hi}
		\end{subfigure}%
		\hfill%
		\begin{subfigure}{0.325\linewidth}
			\vspace{-8pt}
			\caption{}
			\label{fig:reeq-c-hi}
		\end{subfigure}%
		\hfill%
	\end{minipage}%
	\begin{minipage}[T]{0.35\linewidth}
		\begin{subfigure}{\linewidth}
			\centering
			\resizebox{0.97\linewidth}{!}{\includegraphics{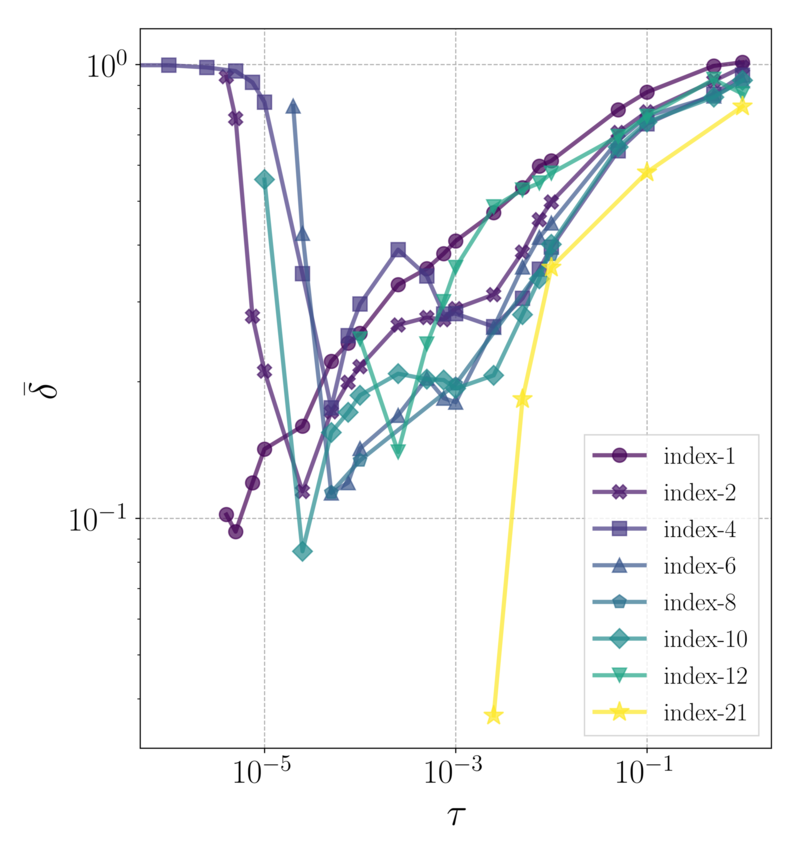}}
		\end{subfigure}
		\begin{subfigure}{\linewidth}
			\vspace{-8pt}
			\caption{}
			\label{fig:reeq-summary}
		\end{subfigure}%
	\end{minipage}%
\vspace{-0.4cm}
\caption{\small{Equilibration diagnostics. Panels (\subref{fig:reeq-a-lo})--(\subref{fig:reeq-c-hi}) show observed and (estimated) exponential failure time distributions.}
	(\subref{fig:reeq-a-lo})--(\subref{fig:reeq-c-lo}) \small{4 failures at $\tau = 10^{-2}, \, 10^{-5}, \, 10^{-6}$.}
	(\subref{fig:reeq-a-hi})--(\subref{fig:reeq-c-hi}) \small{21 failures at $\tau = 10^{-1}, \, 10^{-2}, \, 2.5 \times 10^{-3}$.}
	(\subref{fig:reeq-summary}) \small{$\bar{\delta}$ measures mean error between observed and (estimated) exponential failure time distribution across different temperatures and levels of network degradation (i.e., ``index'').}
}
\label{fig:reeq}
\end{figure}

Next we study the reasonableness of \cref{assn:markov-reeq}.
Symptoms of equilibration can be observed numerically since it is known that the distribution of first exit times for a Markov process that has equilibrated is approximately exponential~\cite{collet_2013_qst}.\footnote{In fact, a positive exponential moment of the failure time distribution is a necessary condition for the existence of a quasistationary distribution~\cite{collet_2013_qst}.}
While we cannot definitively verify equilibration via simulation, we can identify lack of equilibration.
To explore equilibration, we return to studying the dynamics \cref{eq:stochastic-port-hamiltonian} in the 30-bus system.
Specifically, we initialize an ensemble of trajectories at points of failure in degraded networks (according to failure sequences generated by the Markov model), randomize each trajectory's momentum variables (angular frequency), and simulate dynamics until the first overall line failure.
Equilibration experiments in \cref{fig:reeq} are performed on the longest sequence of failures generated by the Markov model and carried out at decreasing temperatures until computation time exceeds 2 hours.
Convergence in failure time distribution toward an exponential distribution is presented in \cref{fig:reeq}.

In \cref{fig:reeq}, we summarize convergence toward an exponential distribution (or lack thereof) in two ways, first visually and next quantitatively.
Panels (\subref{fig:reeq-a-lo})--(\subref{fig:reeq-c-lo}) visually demonstrate an example of lack of equilibration across decreasing temperatures (each at the same topology), indicating the presence of deterministic failure pathways as studied in~\cite{yang-motter_2017_continuous}.
Similarly, panels (\subref{fig:reeq-a-hi})--(\subref{fig:reeq-c-hi}) visually demonstrate an example of successful equilibration across decreasing temperatures (each at the same topology).
Together, panels (\subref{fig:reeq-a-lo})--(\subref{fig:reeq-c-hi}) offer a visual key for interpreting the quantitative summary of failure time distribution convergence summarized in panel (\subref{fig:reeq-summary}).
In particular, panel (\subref{fig:reeq-summary}) displays mean absolute deviation $\bar{\delta} \coloneqq \tfrac 1 N \sum_{i=1}^N | \log(\hat{q}_i / q_i) |$ across (i) temperature and (ii) number of initially failed lines (i.e., failure-``index'') where $\hat{q}_i$ denotes the $i$th quantile of the empirically observed failure time distribution and $q_i$ denotes the $i$th quantile of the exponential distribution fitted to the empirical data.

Along this particular line failure sequence, we find that in topologies with fewer line failures (i.e., a lower failure-index), the distribution of first failure times tends to approach a Dirac delta function in the zero noise limit, thus indicating lack of equilibration.
In fact, line failures in \cref{fig:reeq-c-lo} correspond to a deterministic failure pathways associated with a single dynamically inaccessible line (line 32).
Accordingly, when failure times do not follow an approximately exponential distribution, we believe that the reason is the presence of dynamically inaccessible lines that dominate the most likely equilibration pathways.
At higher-indexed topologies (those with a greater number of line failures), however, we observe that the system generally exhibits equilibration in the low noise limit.
This indicates that the higher-indexed topologies are relatively stable and that the dynamically inaccessible lines have previously failed.
At moderately higher temperatures, we observe approximate equilibration in the majority of cases, even at topologies with multiple dynamically inaccessible lines.
In such cases, noise appears to stabilize the equilibration pathway (see \cref{fig:reeq-a-lo}), even at unstable operating points that contain dynamically inaccessible lines.

In summary, results in \cref{fig:reeq-summary} indicate the existence of both ``stable'' and ``unstable'' equilibration pathways at various levels of degradation (i.e., failure-index), 
but this relationship is moderated by temperature such that there appears to be an intermediate regime ($10^{-2}$ to $10^{-4}$) where approximate equilibration is the predominant response.
These results provide encouraging evidence in support of the claim that at moderate temperatures, equilibration is not a strong assumption, even for unstable operating points.
\vspace{-0.1in}
%


\subsection{Experimental results}
\label{subsec:markov-numerical}
In this section, we explore the practical implications of the Markov model.
Validating the model against ``real'' data is difficult since cascades are rare events and line failure data is often private.
One point of reference, however, is the timestamped line status data provided by the Bonneville Power Administration (BPA, \cite{bpa_data}, part of the Western Interconnect with thousands of buses), which has been analyzed by Dobson and collaborators \cite{dobson_finding_2018,dobson_estimating_2012}.

The dataset contains times of line outage and reinstatement events, and following \cite{dobson_estimating_2012}, cascades can be constructed by appropriately grouping outages that occur in close succession to one another.
Specifically, Dobson separates successive outage events by event time according to two bandwidths: one that determines failures that are part of the same \textit{cascade} (failures occurring within 60 minutes of each other) and one that determines failures that are part of the same \textit{generation} (failures occurring within 1 minute of each other within the same cascade) \cite{ren+dobson_2008_using}.

His results indicate that the distribution of the number of cascades containing $g$ generations is approximately governed by a power law with exponent $-s$ \cite{dobson_finding_2018} and that the number of generations is a useful measure of cascade severity \cite{zhou_markovian_2019}.
Furthermore, $s$ (the negative of the Zipf distribution slope) can be interpreted as a measurement of the system's risk of cascading failure, called SEPSI (system event propagation slope index).

\begin{figure}[H]
	\begin{subfigure}[T]{0.325\linewidth}
		\centering
		\resizebox{\linewidth}{!}{\includegraphics[trim={0 0 3cm 0}, clip=true]{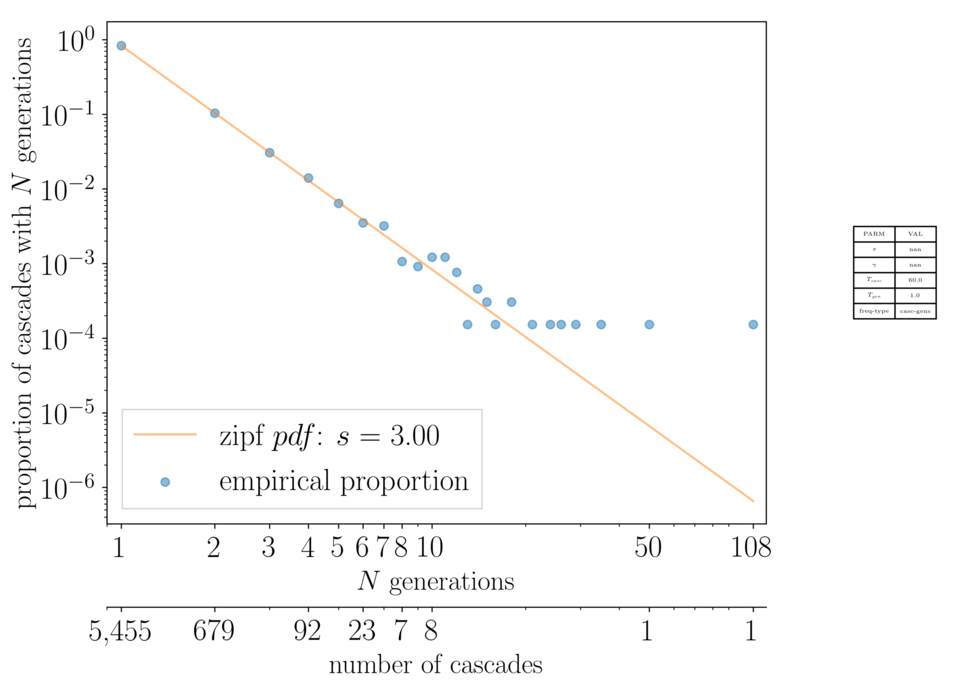}}
	\end{subfigure}
	\begin{subfigure}[T]{0.325\linewidth}
		\centering
		\resizebox{\linewidth}{!}{\includegraphics[trim={0 0 3cm 0}, clip=true]{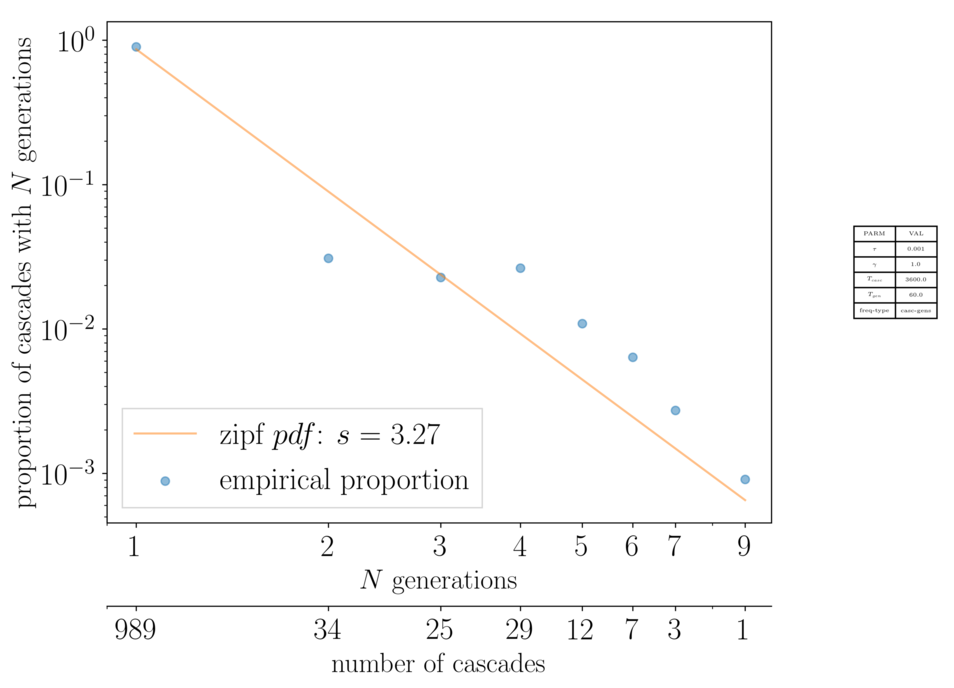}}
	\end{subfigure}
	\begin{subfigure}[T]{0.325\linewidth}
		\vspace{-12pt}
		\centering
		\resizebox{\linewidth}{!}{\includegraphics{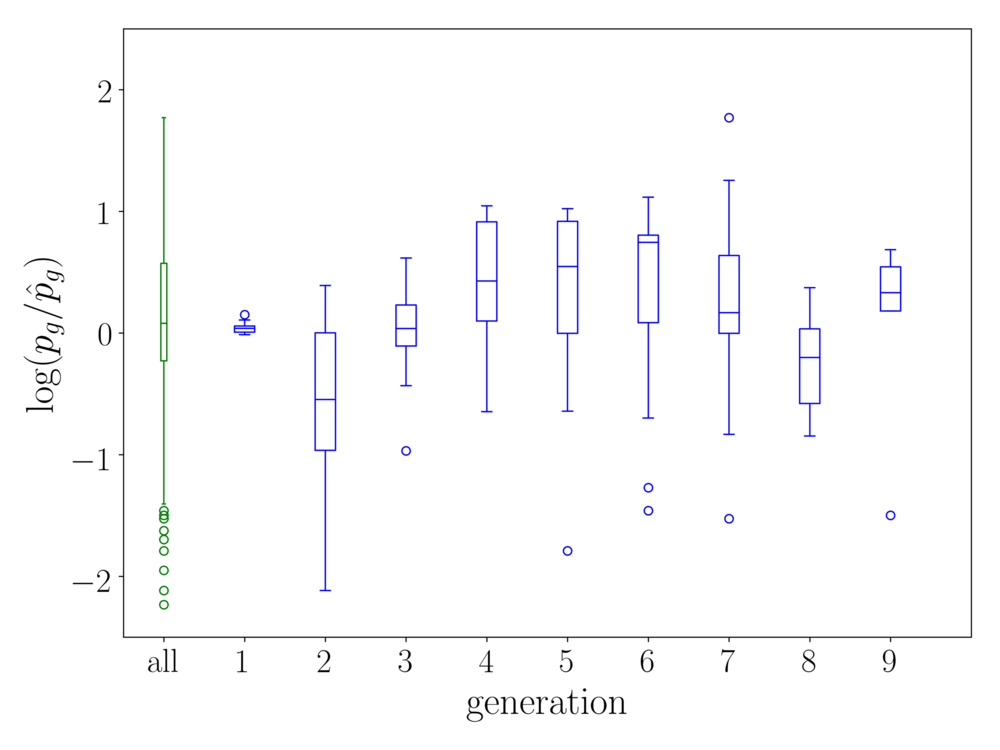}}
	\end{subfigure}
	\begin{subfigure}[T]{0.325\linewidth}
		\caption{}
		\label{fig:zipf-bpa}
	\end{subfigure}	
	\hfill%
	\begin{subfigure}[T]{0.325\linewidth}
		\caption{}
		\label{fig:zipf-kmc-n1}		
	\end{subfigure}
	\hfill%
	\begin{subfigure}[T]{0.325\linewidth}
		\caption{}
		\label{fig:zipf-resids-n1}
	\end{subfigure}
\vspace{-0.5cm}
\caption{\small{Empirical failure proportion versus Zipf fit (estimated from the first nine generations).}
(\subref{fig:zipf-bpa}) \small{SEPSI computed from 6,563 cascades from BPA outage data.}
(\subref{fig:zipf-kmc-n1}) \small{SEPSI computed from 1,100 cascades from the Markov model at $\tau = 10^{-3}$.}
(\subref{fig:zipf-resids-n1}) \small{Aggregated Zipf residuals from Markov model at twelve temperatures between $10^{-4}$--$10^{-2}$.} 
}
\label{fig:118bus-bpa-kmc-zipf}
\end{figure}

From observed BPA outage frequency data, we estimate SEPSI by maximum likelihood based on the first nine generations.
In \cref{fig:zipf-bpa}, Dobson computes SEPSI of 3.02 for similar data, and the choice of nine generations best aligns with the SEPSI calculated in \cite{dobson_finding_2018}.
Dobson finds that SEPSI ranges from 2.2 to 3.2 under various grid operating conditions \cite{dobson_finding_2018}.

SEPSI provides a tractable metric for comparing observed failure data with simulated failure data, and we study the Markov model's capability of producing ``realistic'' cascades through estimates of SEPSI.
To generate raw failure data for cascade-processing, we repeat 100 independent experiments of \cref{alg:kmc} in the IEEE 118-bus system.
Each repetition generates a sequence of timestamped line failures which continues until either the entire network collapses or the cumulative sequence time exceeds $10^{10}$ seconds.\footnote{We choose a long stopping time ($10^{10}$ seconds) to introduce variability in the original network topology to better reflect the heterogeneous conditions across which the BPA data were collected.}
We then aggregate failures to count the frequencies of cascades and generations under 1-hour and 1-minute bandwidth assumptions \cite{dobson_finding_2018,dobson_estimating_2012}, assuming that the timescale of the dynamics is in seconds.
In this way, each repetition can generate multiple sets of cascades from topologies that are related to the original topology through the appropriate number of line failures.
Hence it also allows for exploring network stability after likely configurations of initial lines have failed.

Findings in \cref{fig:118bus-bpa-kmc-zipf} indicate that the Markov model is capable of reproducing sequences of cascading line failures with approximate power law behavior.
The quality of the power law relationship is moderated by network configuration, system temperature, and the cascade-splitting mechansim; but across a range of parameter configurations, we have observed variations in SEPSI that fall within the realistic range.

Across a range of system temperatures, we find that observations exhibit oscillatory behavior around the power law (\cref{fig:zipf-resids-n1}).
An analysis of logarithmic residuals (defined as $\log (p_g / \hat{p}_g)$ where $p_g$ is empirical proportion of cascades with $g$ generations and $\hat{p}_g$ is the Zipf prediction) indicates that the Zipf fit overestimates the number of cascades with two generations (though we do not expect this amount to be of practical significance in realistic situations).
The moderate over- and underestimation varies with generation and may be attributable to the network's size.
In larger networks, we expect our preliminary results to stabilize.

We next discuss some numerical properties of the Markov model.
First, the computational effort of generating cascading failure sequences is significantly more stable than that of direct numerical integration.
At higher temperatures, direct integration may outperform the Markov model, but at temperatures where numerical integration may by prohibitively expensive, the computational cost of the Markov model is fixed and involves only solving a sequence of NLPs.
Such examples may occur in intermediate stages of the cascade where the system settles into a relatively stable configuration.
Second, solution of individual NLP problems to determine the set of outgoing failure rates at each failure level can be parallelized and therefore should have positive scaling properties for larger networks.
Third, once the majority of dominant Markov pathways have been computed, cascade generation amounts to scanning a lookup table and generating exponential random variables.

\subsubsection*{Limitations}
The Markov model can reproduce approximate power law behavior for cascades with $N$ generations under certain parameter settings, but it is limited by its complex dependence on tunable parameters, including system temperature and cascade-generation time scale.
The choices of system temperature and cascade-generation bandwidths affect the intracascade failure time properties of the Markov model, and it is not clear how best to determine these without numerical experimentation.
The faithfulness with which the Markov simulator can reproduce failure sequences corresponding to numerical integration of the network depends on the validity of the re-equilibration assumption (\cref{assn:markov-reeq}).
Our discussion from \cref{subsec:markov-assns} leads us to believe that this is not a significant effect.
We also note that the BPA data is not directly comparable to the failure data generated by the Markov model, since the BPA data allows for line reconnection whereas the Markov model has no such mechanism.
A continuous mechanism for line reinstatement and analytic adjustments to the system's energy function is introduced in \cite{zheng_2010_bistable}, but we do not address this extension.


\section{Conclusions}
\label{sec:conclusions}
In this article, we described a transition rate theory approach to modeling both individual and cascading transmission line failure in power systems.
We introduced the interpretation of individual line failures as large-deviation events on an energy surface defined by a first-principles model of the power system, and we derived expressions for the asymptotic failure rates.
Although the power grid dynamics \cref{eq:stochastic-port-hamiltonian} were modeled as an irreversible SDE, their close relation to a gradient system (by virtue of the transverse decomposition of the drift vector field) provided a framework for analytical treatment of the problem of first escape through a dividing surface.
In the limit of small noise, validation results indicated that our analytic approximations were well suited for determining the unconditional failure rate of any applicable network line.
Further, a first-order expansion of the prefactor suggested a modification that increased the approximation's applicability to higher temperature regimes as well.
In the first-order model of line failure \cref{eq:1-order-qst-final}, we observed an average (absolute) logarithmic error of $0.7$ across all experiments and inferred that the model is capable of describing the failure mechanism of isolated transmission lines in a restricted network.

Next we studied the applicability of the unconditional failure model in the context of realistic networks, relaxing the assumption of studying line failure as an isolated phenomenon.
We described various network pathologies (\cref{def:non-isolated,def:nested,def:inaccessible}) and their potential effects on the validity of the unconditional failure model as an approximation to the conditional failure model.
From numerical experiments, we inferred that the prevalence and impact of these pathologies in realistic networks is relatively limited.
In addition, such pathologies are related to network parameters (e.g., line limits) that can be adjusted.

We incorporated individual models of line failure into a finite-temperature, KMC-based Markov model to simulate cascades.
The Markov model generates cascades by traversing a directed graph containing edges with weights specified by the individual line failure model.
Computation within a cascade sequence can be parallelized; and once dominant pathways have been computed, simulation amounts to finding values in a lookup table (albeit an exponentially large one) and generating exponential random variables.
Our numerical experiments show that under certain parameter settings, the Markov model can approximate the power law behavior of empirically observed cascades.
Tuning the Markov model parameters is currently based on experimentation; however, we anticipate that future research will address this issue, as well as the following potential directions:
\begin{enumerate}
	\item Further explore the relationship between temperature, cascade splitting parameters, and cascade severity for describing cascade behavior.
	\item Analyze the qualitative features of failure time sequences of cascades.
	\item Explore alternative characterizations of cascade severity beyond SEPSI (including the number of outages in a cascade or generation and the amount of load lost in a cascade).
	\item Introduce mechanisms for line reconnection.
	\item Consider more complex stochastic drivers.
	\item Relax the lossless power flow assumption.
\end{enumerate}


\appendix
\section{Transverse decomposition of the port-Hamiltonian dynamics}
\label{apx:rate-transverse}
Bouchet and Reygner \cite{bouchet_2016_generalisation} note that a stochastic differential equation of the form
\begin{equation}
\label{eq:generic-transverse}
\mathrm{d} x^{\tau}_t = -K \nabla U \left (x^{\tau}_t \right ) \mathrm{d} t + \sqrt{2 \tau} \sigma \mathrm{d} W_t
\end{equation}
for potential $U$ and any appropriately sized (constant) matrix $K$ with symmetric part $\tfrac 12 (K + K^{\top}) = \sigma \sigma^{\top}$ has two salient properties:
\begin{enumerate}
	\item Its generator preserves stationarity of the Gibbs measure.
	\item It possesses a natural transverse decomposition of the drift.
\end{enumerate}
Denoting symmetric and antisymmetric components by $S = \tfrac 12 (K + K^{\top})$ and $J = \tfrac 12 (K - K^{\top})$, respectively, we can express our model \cref{eq:stochastic-port-hamiltonian} in the form of \cref{eq:generic-transverse}.
Specifically, we can perform a transverse decomposition of the port-Hamiltonian drift term into
\begin{equation}\label{drift}
b(x) \coloneqq (J - S) \nabla \mathcal{H}(x) = -S \nabla \mathcal{H}(x) + \ell(x)
\end{equation}
with $\ell(x) \coloneqq J \nabla \mathcal{H}(x)$.
Since $J$ is skew-symmetric, we have that $\langle \nabla \mathcal{H}(x), \, \ell(x) \rangle = 0$.
By the same property, we also have $\mathrm{div}(\ell(x)) = \mathrm{div}(J \nabla \mathcal{H}(x)) = \mathrm{tr}(J\nabla^2 \mathcal{H)}(x) = 0$ (using the cycle property of the trace operator and symmetry of the Hessian).
Hence, despite being irreversible, the dynamics \cref{eq:stochastic-port-hamiltonian} preserve the Gibbs measure and do not introduce any ``non-Gibbsianness'' into estimates involving the stationary density.

\section{Derivation of the quasistationary exit rate}
\label{apx:rate-qst-exit-rate-derivation}
The surface integral in \cref{eq:qst-exit-rate-pf-ef} can be evaluated by using the Laplace method.
Specifically, we are interested in evaluating the asymptotic leading term of integrals of the form
\begin{equation*}
  \int_{x \in \partial D_l} v^{\top}(x) n(x) e^{-f(x) / \tau} \, \mathrm{d} S(x), \quad \tau \to 0,
\end{equation*}
where the minima of $f(x)$ along the surface $\partial D_l$ is unique and is not a critical point of $f(x)$.
In the limit $\tau \to 0$, the integral is dominated by the contribution due to $e^{-f(x) / \tau}$ around $x^{\star} = \argmin_{x \in \partial D_l} f(x)$.

To evaluate the leading term, we parametrically represent $\partial D_l \equiv \{ x \colon \Theta_l(x) = \Theta^{\max}_l \}$ in the vicinity of $x^{\star}$ by
\begin{equation*}
  x = x(u), \quad u = (u_1, \dots, u_{d - 1})^{\top} \in \partial D_l,
\end{equation*}
where $u$ denotes orthogonal curvilinear coordinates, so that $x^{\star} = x(u^{\star})$.
Formula~(5.20) of~\cite{wong_2001_asymptotic}, Chapter 9, provides the leading term, which reads
\begin{equation}
  \label{eq:wong-laplace-surface}
  \int v^{\top}(x) n(x) e^{-f(x) / \tau} \, \mathrm{d} S(x) \sim (v^{\top} n)(x^\star) D(u^\star) (\det \mathcal{B})^{- 1 / 2} (2 \pi \tau)^{(d - 1) / 2} e^{-f(x^{\star}) / \tau},
\end{equation}
where
\begin{equation*}
  \mathcal{B} = \Hess_u f(x(u^{\star})), \quad D(u) = \sqrt{\det \mathcal{J}^{\top}(u) \mathcal{J}(u)}, \quad \mathcal{J}_{ij}(u) = \frac{\partial x_i}{\partial u_j}, \quad \mathcal{J} \in \mathbb{R}^{d \times (d - 1)}.
\end{equation*}
To write $\det \mathcal{B}$ in terms of $f(x)$ and $\Theta_l(x)$, we employ Formula (3.24) of~\cite{wong_2001_asymptotic}, which reads
\begin{equation}
  \label{eq:wong-det-curvilinear}
  \det \mathcal{B} = \frac{D^2(u^{\star})}{|\nabla f(x^{\star})|^2} B^{\star},
\end{equation}
where
\begin{equation}
  \label{eq:wong-B}
  B^{\star} = \sum^d_{p = 1} \sum^d_{q = 1} \frac{\partial f(x^{\star})}{\partial x_p} \frac{\partial f(x^{\star})}{\partial x_q} \cof_{pq} \left [ \Hess \mathcal{H}(x^{\star}) - k \Hess \Theta_l(x^{\star}) \right ],
\end{equation}
and $k$ is a constant such that $\nabla f(x^{\star}) = k \nabla \Theta_l(x^{\star})$.
In the NLPs \ref{eq:xstar} and \ref{eq:cond-nlp}, $k$ corresponds to the Lagrange multiplier on the line energy constraint.

We now evaluate these expressions for the stochastic port-Hamiltonian model.
First, we have the equivalencies $f(x) = V(\bar{x}, x) = \mathcal{H}(x) - \mathcal{H}(\bar{x})$, and $v(x) = (J + S) \nabla \mathcal{H}(x)$.
Second, we have
\begin{equation*}
  \begin{aligned}
    (v^{\top} n)(x^\star) D(u^\star) (\det \mathcal{B})^{-1 / 2} &= (v^{\top} n)  D(u^\star) \frac{|\nabla f(x^\star)|}{D(u^\star) \sqrt{B^{\star}}} = \frac{(v^{\top} \nabla f)(x^\star)}{\sqrt{B^{\star}}}\\
    &= \frac{1}{\sqrt{B^{\star}}} \nabla^{\top} \mathcal{H}(x^\star) (J + S)^{\top} \nabla \mathcal{H}(x^\star)\\
    &= \frac{1}{\sqrt{B^{\star}}} \nabla^{\top} \mathcal{H}(x^\star) S \nabla \mathcal{H}(x^\star),
  \end{aligned}
\end{equation*}
because $n = \nabla f(x^\star) / |\nabla f(x^\star)|$, given that $f(x^\star)$ over $\partial D_l$ has its minimum on $x^\star$, and $\nabla^{\top} \mathcal{H}(x^\star) J \nabla^{\top} \mathcal{H}(x^\star) = 0$.
Substituting this expression above into~\cref{eq:wong-laplace-surface} leads to~\cref{eq:qst-exit-rate-final}.

We also note that $\nabla \Theta_l(x)$ points outwards of $D_l$.
Furthermore, given that $f(x)$ over $\partial D$ is minimal at $x^\star$, $\nabla f(x)$ is parallel and in the direction of $n(x)$.
Therefore, $\nabla f(x^\star)$ and $\nabla \Theta_l(x^\star)$ point in the same direction, so that $k = |\nabla f(x^\star)| / |\nabla \Theta_l(x^\star)| > 0$.
Substituting into~\cref{eq:wong-det-curvilinear} results in~\cref{eq:B-star}.

\section{Higher-order expansion of the exponent and the prefactor of the stationary density}
\label{apx:rate:higher}

We use the same notation as the zero-order analysis for the stationary density that can be found in \cite{bouchet_2016_generalisation}.
For the first order of the expansion in the exponent (using up to $j=1$ in Eq. \cref{eq:n-order-stationary-wkb}), we assume
\begin{equation}\label{first_order_exponent}
\pre{p}{x} = \frac{\pre{C}{x}}{\tau^{1/2}}\exp \left[ -\frac{\mathcal{H}(x)}{\tau} + \mathcal{H}_1(x) \right],
\end{equation}
where $\mathcal{H}_1(x)$ is an unknown function of $x.$
We plug \cref{first_order_exponent} for the density in the stationary Fokker-Planck equation
\begin{equation}\label{fp_stationary}
\tau \sum_{i,j=1}^d \partial_{ij} \left [ S_{ij}(x) \pre{p}{x} \right ] = \sum_{i=1}^d \partial_i \left [ \, b(x) \pre{p}{x} \right ]
\end{equation}
where $b(x)$ is given by \cref{drift}.
We now collect terms in orders of $\tau$.
We follow Section 3 in \cite{bouchet_2016_generalisation} and use the properties of the power grid model that (i) the matrix $S$ is constant, (ii) $F(x) \coloneqq {\operatorname{div}} J \nabla \mathcal{H}(x) + \langle A(x), S\nabla \mathcal{H}(x) \rangle=0,$ and (iii) $A_i(x)=\sum_{j=1}^d \partial_{ij} S_{ij}(x).$
The equation of order $\tau^{-1}$ ({\it eikonal}) is the Hamilton-Jacobi equation for the quasipotential.
Using the eikonal equation and the aforementioned, the equation of order $\tau^0$ gives
\begin{equation}\label{transport_exponent}
\langle \nabla (\log \pre{C}{x}+ \mathcal{H}_1(x)), S \nabla \mathcal{H}(x) + J \nabla \mathcal{H}(x) \rangle=0.
\end{equation}
Through the method of characteristics we obtain
\begin{equation}\label{transport_exponent_2}
\pre {C}{x} = \pre {C}{\bar{x}} \exp \bigl[ -\mathcal{H}_1(x) + \mathcal{H}_1(\bar{x})  \bigl]
\end{equation}
and for the stationary density
\begin{equation}\label{first_order_exponent_final}
\pre{p}{x} = \frac{\pre{C}{\bar{x}}}{\tau^{1/2}}\exp \left[ -\frac{\mathcal{H}(x)}{\tau} + \mathcal{H}_1(\bar{x}) \right].
\end{equation}
Thus, the contribution from the first order term $\mathcal{H}_1(x)$ in the exponent is a constant that can be absorbed in the prefactor.
We note that we would arrive at the same result even if $F(x) \neq 0.$
However, starting from the second order and onwards, the condition $F(x)=0$ is required in order to show the contribution of only constants from the higher order terms in the exponent.

Using a similar analysis (albeit with more involved algebra), one can show that the second order approximation (using up to $j=2$ in Eq. \cref{eq:n-order-stationary-wkb})
\begin{equation}\label{second_order_exponent}
\pre{p}{x} = \frac{\pre{C}{x}}{\tau^{1/2}}\exp \left[ -\frac{\mathcal{H}(x)}{\tau} + \mathcal{H}_1(x)+ \tau \mathcal{H}_2(x) \right],
\end{equation}
becomes
\begin{equation}\label{second_order_exponent_final}
\pre{p}{x} = \frac{\pre{C}{\bar{x}}}{\tau^{1/2}}\exp \left[ -\frac{\mathcal{H}(x)}{\tau} + \mathcal{H}_1(\bar{x})+ \tau \mathcal{H}_2(\bar{x}) \right],
\end{equation}
and thus the contributions of the first- and second-order terms can be absorbed in the prefactor.
Similarly, one can show that all the higher order terms in the exponent end up contributing constants.
We omit the details.

For the expansion of the prefactor we use the expression \cref{eq:n-order-prefactor-prefinal}.
The analysis to show that all terms contribute a constant (in $x$) is performed in a similar way as for the expansion of the exponent. For example, for the first order ($n=1$ in \cref{eq:n-order-prefactor-prefinal}) we have
\begin{equation}\label{first_order_prefactor}
\pre{p}{x} = \frac{C_0(x)+\tau C_1(x)}{\tau^{1/2}}\exp \left[ -\frac{\mathcal{H}(x)}{\tau} \right].
\end{equation}
We plug \cref{first_order_prefactor} in Eq. \cref{fp_stationary} and collect terms in orders of $\tau.$
From the equations at order $\tau^{-1}, \tau^{0}$, and $\tau^1$ and using properties (i) and (ii) above for the power grid model, we find $C_0(x)=C_0(\bar{x})$ and
\begin{equation}\label{transport_prefactor}
\langle \nabla C_1(x), b(x) + 2 S \nabla \mathcal{H}(x) \rangle=\langle \nabla C_1(x), S \nabla \mathcal{H}(x) + J \nabla \mathcal{H}(x) \rangle=0.
\end{equation}
Using the method of characteristics, we get $C_1(x)=C_1(\bar{x})$, and so
\begin{equation}\label{first_order_prefactor_final}
\pre{p}{x} = \frac{C_0(\bar{x})+\tau C_1(\bar{x})}{\tau^{1/2}}\exp \left[ -\frac{\mathcal{H}(x)}{\tau} \right].
\end{equation}
The analysis for the higher-order terms in the prefactor expansion proceeds in a similar way.
We omit the details.

\section{Failure rate dynamics algorithm}
\label{apx:rate:simulators}
The procedure for unconditional failure rate simulation is summarized below.
\begin{algorithm}[H]
\bgroup
\scriptsize
\caption{Line $l$ unconditional failure}
\label{alg:line-l-uncond-failure}
\begin{algorithmic}
	\Require Equilibrium point $\bar{x}$, system parameters $y$, temperature $\tau$, step size $\Delta t$, and line limits $\Theta_l^{\max}$
	\State {Construct $J$ and $S$ per \cref{eq:J-S-matrices}}
	\State $x \gets \bar{x}$
	\State $\Theta_l \gets \Theta_l(x)$
	\State $t \gets 0$
	\While {$\Theta_l < \Theta_l^{\max}$}
		\State $x \gets x + (J - S) \cdot \nabla_x\mathcal{H}^y(x) \Delta t + \sqrt{ 2 \tau}\sqrt{S} \, \Delta W$
		\State $t \gets t + \Delta t$
		\State $\Theta_l \gets \Theta_l(x)$
	\EndWhile
\end{algorithmic}
\egroup
\end{algorithm}%
%


\section{Markov model}
\label{apx:markov}
The rejectionless, on-the-fly Markov simulation procedure is described below.
\begin{algorithm}[H]
\bgroup
\scriptsize
\begin{algorithmic}
  \Require Network line data $\mathcal{N}_L$, network state $\mathcal{N}_S$, equilibrium point $\bar{x}$, system parameters $y$, temperature $\tau$, and stopping time $t_{\max}$
  \State Initialize topological network state: $S \leftarrow \{ \varnothing \}$%
  \State Initialize DAG: $\mathcal{G} \gets (\varnothing,\, \varnothing)$
  \State Initialize unfailed lines: $U \gets \{ \mathcal{N}_{L} \}$
  \State Initialize state and time: $S \gets \mathcal{N}_S$, $t \gets 0$
  \While{$t \geq t_{\max}$ or $S \equiv S_{\infty}$}%
  \State Compute $\bar{x}_S$ (more detail in supplemental material) and update $U$ for any failed lines 
  \For {$l \in U$}
  \State Look up or compute $x^{\star}_{S \to \{S \, \cup \, l\}}$ (more detail in supplemental material) 
  \State Compute $\lambda^{\tau}_{S \to \{S \, \cup \, l\}}$ from $\bar{x}_S$ and $x^{\star}_{S \to \{S \, \cup \, l\}}$%
  \State $\mathcal{V} \gets \mathcal{V} \cup \{ S \cup l \}$%
  \State $\mathcal{A} \gets \mathcal{A} \cup \lambda^{\tau}_{S \to \{S \, \cup \, l\}}$%
  \EndFor
  \State Compute aggregate rate: $\lambda_{S \to {S}^{\prime}} = \sum_l \lambda_{S \to \{ S \, \cup \, l \}}$%
  \State Sample failure time: draw $\Delta t \sim \operatorname{Exp} \left( \lambda_{S \to {S}^{\prime}} \right)$%
  \State Sample failure line: draw ${l}^{\prime}$ proportionally from aggregate rate $\lambda_{S \to {S}^{\prime}}$
  \State $t \leftarrow t + \Delta t$
  \State $S \leftarrow {S}^{\prime} \equiv \{ S \cup {l}^{\prime} \}$
  \State $U \leftarrow U \backslash l^{\prime}$
  \EndWhile
\end{algorithmic}
\egroup
\caption{Rejectionless (Markov) KMC Model}
\label{alg:kmc}
\end{algorithm}
The entire discrete event space for a four-line model is enumerated below.
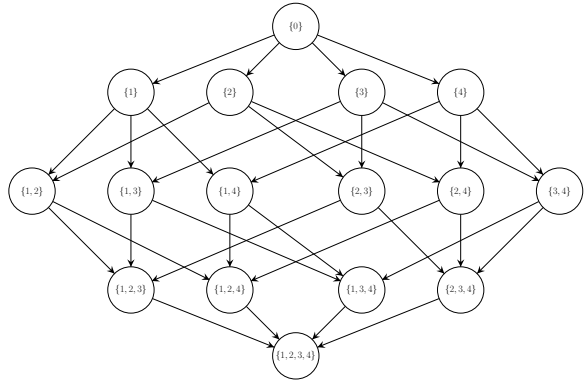
\begin{figure}[H]
\centering
\rotatebox{0}{ \resizebox{!}{5cm}{\begin{tikzpicture}
\tikzset{vertex/.style = {shape=circle,draw,minimum size=100pt}}
\tikzset{edge/.style = {->,> = latex', scale=3, ultra thick}}
\tikzset{edge/.style={decoration={markings,mark=at position 1 with %
			{\arrow[scale=5,>=stealth]{>}}},postaction={decorate}}}

\node[vertex, rotate=0] (0) at  (0,10) {\huge$\{0\}$};

\node[vertex, rotate=0] (1) at (-12.5,5) {\huge$\{1\}$};
\node[vertex, rotate=0] (2) at (-5,5) {\huge$\{2\}$};
\node[vertex, rotate=0] (3) at (5,5) {\huge$\{3\}$};
\node[vertex, rotate=0] (4) at (12.5,5) {\huge$\{4\}$};

\node[vertex, rotate=0] (12) at (-20,-2.5) {\huge$\{1,2\}$};
\node[vertex, rotate=0] (13) at (-12.5,-2.5) {\huge$\{1,3\}$};
\node[vertex, rotate=0] (14) at (-5,-2.5) {\huge$\{1,4\}$};
\node[vertex, rotate=0] (23) at (5,-2.5) {\huge$\{2,3\}$};
\node[vertex, rotate=0] (24) at (12.5,-2.5) {\huge$\{2,4\}$};
\node[vertex, rotate=0] (34) at (20,-2.5) {\huge$\{3,4\}$};

\node[vertex, rotate=0] (123) at (-12.5,-10) {\huge$\{1,2,3\}$};
\node[vertex, rotate=0] (124) at (-5,-10) {\huge$\{1,2,4\}$};
\node[vertex, rotate=0] (134) at (5,-10) {\huge$\{1,3,4\}$};
\node[vertex, rotate=0] (234) at (12.5,-10) {\huge$\{2,3,4\}$};

\node[vertex, rotate=0] (1234) at (0,-15) {\huge$\{1,2,3,4\}$};

\draw[edge] (0) to (1);
\draw[edge] (0) to (2);
\draw[edge] (0) to (3);
\draw[edge] (0) to (4);

\draw[edge] (1) to (12);
\draw[edge] (1) to (13);
\draw[edge] (1) to (14);
\draw[edge] (2) to (12);
\draw[edge] (2) to (23);
\draw[edge] (2) to (24);
\draw[edge] (3) to (13);
\draw[edge] (3) to (23);
\draw[edge] (3) to (34);
\draw[edge] (4) to (14);
\draw[edge] (4) to (24);
\draw[edge] (4) to (34);

\draw[edge] (12) to (123);
\draw[edge] (12) to (124);
\draw[edge] (13) to (123);
\draw[edge] (13) to (134);
\draw[edge] (14) to (124);
\draw[edge] (14) to (134);
\draw[edge] (23) to (123);
\draw[edge] (23) to (234);
\draw[edge] (24) to (124);
\draw[edge] (24) to (234);
\draw[edge] (34) to (134);
\draw[edge] (34) to (234);

\draw[edge] (123) to (1234);
\draw[edge] (124) to (1234);
\draw[edge] (134) to (1234);
\draw[edge] (234) to (1234);
\end{tikzpicture}} }
\caption{
	\small{(Unweighted) Hasse diagram of all possible failure states for a four-line (labeled 1, 2, 3, and 4) model.
	Each node contains a topological state $S$, and the (directed) edges represent transitions to further degraded surfaces $S^{\prime}$ (from left to right).
	Each directed edge represents an additional line failure $l$ (where $l = S^{\prime} \backslash S$) and is weighted by $\lambda_{S}^{S^{\prime}}$, the analytic transition rate between $S$ and $S^{\prime}$.}
}
\label{fig:4line-tree}
\end{figure}

Each edge of the Markov model relates the rate of failure from an equilibrium point $\bar{x}$ to a failure of a given line $l$ through a failure point $x^{\star}$.
In degraded surfaces, the computation of $\bar{x}$ may result in (i) a nonphysical solution (for example, one with low voltage at a given bus, bus-$i$) or (ii) direct infeasibility.
We encounter both cases in the Markov model and adjust the solution procedure by shedding load \cite{mollah_automatic_2012}; see the supplemental material for further detail.


\section*{Acknowledgments}
This material was based upon work supported by the U.S. Department of Energy, Office of Science, Office of Advanced Scientific Computing Research (ASCR) under Contract DE-AC02-06CH11347.
We acknowledge partial NSF funding through awards FP061151-01-PR and CNS-1545046 to MA.
We also thank Christopher DeMarco, Ian Dobson, Charles Matthews, Adrian Maldonado, and Anirudh Subramanyam for insightful discussions.


\bibliographystyle{siam}
\bibliography{_article}

\begin{flushright}
	\scriptsize \framebox{\parbox{2.5in}{Government License: The
			submitted manuscript has been created by UChicago Argonne,
			LLC, Operator of Argonne National Laboratory
			(``Argonne’’).  Argonne, a U.S. Department of Energy
			Office of Science laboratory, is operated under Contract
			No. DE-AC02-06CH11357.  The U.S. Government retains for
			itself, and others acting on its behalf, a paid-up
			nonexclusive, irrevocable worldwide license in said
			article to reproduce, prepare derivative works, distribute
			copies to the public, and perform publicly and display
			publicly, by or on behalf of the Government. The
			Department of Energy will provide public access to these
			results of federally sponsored research in accordance with
			the DOE Public Access
			Plan. http://energy.gov/downloads/doe-public-access-plan. }}
	\normalsize
\end{flushright}
\end{document}


\maketitle

\section{Port-Hamiltonian power transmission model and energy function}
\label{sm:dynamics-model}

For the following presentation we will employ the following notation: For vector quantities in $\mathbb{R}^N$ or $\mathbb{C}^N$, we denote by $[ \cdot ]_X$ the subvector corresponding to the subset of indices $X \subseteq [1, N]$.
For square matrices of dimension $N$, we denote by $[ \cdot]_X$ the diagonal block corresponding to the subset $X \subseteq [1, N]$ of row and column indices.
Next, we denote the identity matrix of dimension $|X|$ by $\eye_X$, the vector of ones of dimension $|X|$ by $\one_X$, and rectangular zero matrices of $|X|$ rows and $|Y|$ columns by $\zero_{X \times Y}$.
Finally, we denote by $\odot$ the Hadamard product.

\subsection*{AC power flow equations}
The state of the $i$th bus at time $t$ is specified by its net power phasor, $[S_t]_i \coloneqq [P_t]_i + j [Q_t]_i$, and its voltage phasor, $[v_t]_i \coloneqq [V_t]_i \exp({[\delta_t]_i})$, where $[P_t]_i$ and $[Q_t]_i$ denote the net active and reactive powers, respectively, and $[V_t]_i$ and $[\delta_t]_i$ denote the voltage magnitude and phase angle, respectively.
The power and voltage phasors are related via the AC power flow equations
\begin{equation*}
  [S_t]_i = \left[ v_t \odot {(Y v_t)}^* \right]_i, \quad i \in \mathcal{B}
\end{equation*}
where the complex matrix $Y \coloneqq G + j B$ denotes the network admittance matrix.
Assuming that the network is lossless, we have that $Y$ is purely imaginary, with $Y = jB$.
The real and imaginary components of the AC power flow equations read
\begin{subequations}
  \label{eq:acpf}
  \begin{align}
    \label{eq:acpf-active}
    [P_t]_i &= \sum_{k \in \mathcal{B}} [V_t]_i [V_t]_k B_{ik} \sin \left ( [\delta_t]_i - [\delta_t]_k \right ), \\
    \label{eq:acpf-reactive}
    [Q_t]_i &= -\sum_{k \in \mathcal{B}} [V_t]_i [V_t]_k B_{ik} \cos \left ( [\delta_t]_i - [\delta_t]_k \right ), \quad i \in \mathcal{B}.
  \end{align}
\end{subequations}
Here, we adopt the convention that $[P_t]_i \coloneqq [P_t^g]_i - [P_t^d]_i$ and $[Q_t]_i \coloneqq [Q_t^d]_i - [Q_t^g]_i$.
Hence, $[P_t]_i > 0$ ($[Q_t]_i > 0$) indicates that net active (reactive) power is being injected at the $i$th bus, while $[P_t]_i < 0$ ($[Q_t]_i < 0$) indicates that net active (reactive) power is being consumed at the $i$th bus.

We assume that at $t = 0$ the system is at equilibrium and the generators are synchronized.
Generator buses are modeled as PV buses, so that their equilibrium values of net real power and voltage magnitudes, $[P_0]_{\mathcal{G}}$ and $[V_0]_{\mathcal{G}}$, respectively, are prescribed.
For the reference bus, on the other hand, the voltage magnitude and angle, $[V_0]_{\mathcal{S}}$ and $[\delta_0]_{\mathcal{S}}$, are specified.
Finally, for load buses, the equilibrium values of the net real and reactive powers, $[P_0]_{\mathcal{L}}$ and $[Q_0]_{\mathcal{L}}$, respectively, are prescribed.
The AC power flow equations~\eqref{eq:acpf-active} and~\eqref{eq:acpf-reactive} can be solved together with these prescribed quantities to calculate all other equilibrium quantities.


\subsection*{Classical machine model}
Generators are modeled using the classical model.
The deterministic dynamics of the generators are then governed by the swing equations
%
\begin{subequations}
  \begin{align}
    \label{eq:swing-eq-velocity}
    [M^g]_{ii} [\dot{\omega}_t]_i &= -[D^g]_{ii} [\omega_t]_i - [P_t]_i + [P_0]_i, \\
    \label{eq:swing-eq-angle}
    [\dot{\delta}_t]_i &= [\omega_t]_i, \quad i \in \mathcal{S} \cup \mathcal{G},
  \end{align}
\end{subequations}
where $[\omega_t]_i$ denotes the generator angular velocity of the $i$th bus, and $M^g$ and $D^g$ denote the diagonal matrices of generator mass and damping constants.

For generator buses we assume that their voltage magnitudes remain constant throughout the evolution of the system, so that $[V_t]_{\mathcal{G}} \equiv [V_0]_{\mathcal{G}}$.
Introducing the auxiliary angular variables
\begin{equation}
  \label{eq:phase-angles-diff}
  [\theta_t]_i \coloneqq [\delta_t]_i - [\delta_t]_{\mathcal{S}}, \quad i \in \mathcal{G} \cup \mathcal{L},
\end{equation}
we can rewrite the AC power flow equations~\eqref{eq:acpf-active} and~\eqref{eq:acpf-reactive} for $i \in \mathcal{\mathcal{G}} \cup \mathcal{L}$ in terms of the auxiliary functions $f_i$ and $g_i$ defined as
\begin{subequations}
  \begin{align}
    \label{eq:f-function}
    [P_t]_i = f_i([\theta_t]_{\mathcal{G} \cup \mathcal{L}}, [V_t]_{\mathcal{L}}) &\coloneqq \sum_{k \in \mathcal{G} \cup \mathcal{L}} [V_t]_i [V_t]_k B_{ik} \sin \left ([\theta_t]_i - [\theta_t]_k \right ) + B_{iS} [V_t]_i [V_t]_S \sin \left ([\theta_t]_i \right ), \\
    \label{eq:g-function}
    [Q_t]_i = g_i([\theta_t]_{\mathcal{G} \cup \mathcal{L}}, [V_t]_{\mathcal{L}}) &\coloneqq -\sum_{k \in \mathcal{G} \cup \mathcal{L}} [V_t]_k [V_t]_k B_{ik} \cos \left ([\theta_t]_i - [\theta_t]_k \right ) - B_{iS} [V_t]_i [V_t]_S \cos \left ([\theta_t]_i \right ), \\ \quad i \in \mathcal{G} \cup \mathcal{L} \nonumber.
  \end{align}
\end{subequations}
Finally, we note that (i) as the system is lossless, the net real power at all buses is zero and (ii) by setting a reference angle, we have the implicit constraint $\sum_j \theta_j = 0$.
From this, it follows that
\begin{equation*}
  [P_t]_{\mathcal{S}} - [P_0]_{\mathcal{S}} = - \sum_{i \in \mathcal{G} \cup \mathcal{L}} \left \{ [f]_i - [P_0]_i \right \}.
\end{equation*}
We can then rewrite the swing equations~\eqref{eq:swing-eq-velocity} and~\eqref{eq:swing-eq-angle} as
\begin{subequations}
  \label{eq:swing-eq-SG}
  \begin{align}
    \label{eq:swing-eq-vel-S}
    [M^g]_{\mathcal{S}} [\dot{\omega}_t]_{\mathcal{S}} &= - [D^g]_{\mathcal{S}} [\omega_t]_{\mathcal{S}} + \sum_{i \in \mathcal{G} \cup \mathcal{L}} \left \{ [f]_i - [P_0]_i \right \},\\
    \label{eq:swing-eq-vel-G}
    [M^g]_{\mathcal{G}} [\dot{\omega}_t]_{\mathcal{G}} &= - [D^g]_{\mathcal{G}} [\omega_t]_{\mathcal{G}} - [f]_{\mathcal{G}} + [P_0]_{\mathcal{G}},\\
    \label{eq:swing-eq-theta-G}
    [\dot{\theta}_t]_{\mathcal{G}} &= -\one_{\mathcal{G}} [\omega_t]_{\mathcal{S}} + [\omega_t]_{\mathcal{G}}.
  \end{align}
\end{subequations}
\subsection*{Deterministic load model}
For load buses, we model the real power as linear functions of the bus angular velocity, and we assume that the reactive power remains constant throughout the simulation:
%
\begin{subequations}
  \label{eq:load-model}
  \begin{align}
    \label{eq:active-load-model}
    [P_t]_i &\coloneqq  [P_0]_i - [D^d]_{ii} [\dot{\delta}_t]_i,\\
    \label{eq:reactive-load-model}
    [Q_t]_i &\coloneqq [Q_0]_i, \quad i \in \mathcal{L}.
  \end{align}
\end{subequations}
where $D^d$ is a diagonal matrix of load damping constants.
In terms of the relative phase angles $\theta$, \eqref{eq:active-load-model} can be rewritten as
\begin{equation}
  \label{eq:active-load-model-theta}
  [\dot{\theta}_t]_{\mathcal{L}} = -[D^d]^{-1}_{\mathcal{L}} \left \{ [f]_{\mathcal{L}} - [P_0]_{\mathcal{L}} \right \} - [\dot{\omega}_t]_{\mathcal{S}}.
\end{equation}

\subsection*{DAE system}
From \eqref{eq:f-function}--\eqref{eq:active-load-model-theta} it can be seen that the dynamics of the network can be described by $[\omega_t]_{\mathcal{S} \cup \mathcal{G}}$, $[\theta_t]_{\mathcal{G}}$, and $[V_t]_{\mathcal{L}}$.
We collect these state variables into the state vector $x^{\top}_t \coloneqq \left ([\omega_t]^{\top}_{\mathcal{S} \cup \mathcal{G}}, [\theta_t]^{\top}_{\mathcal{G}}, [V_t]^{\top}_{\mathcal{L}} \right )$.
To assemble a system of DAEs governing the dynamics of $x_t$, we introduce the auxiliary functions
\begin{subequations}
  \label{eq:H-function-dae}
  \begin{align}
    H_1(x_t) &\coloneqq M^g [\omega_t]_{\mathcal{S} \cup \mathcal{G}}, \\
    H_2(x_t) &\coloneqq [f(x_t)]_{\mathcal{G} \cup \mathcal{L}} - [P_0]_{\mathcal{G} \cup \mathcal{L}}.
  \end{align}
\end{subequations}
and the matrices
\begin{equation*}
  T_1 =
  \begin{bmatrix}
    -\one_{\mathcal{G}} & \eye_{\mathcal{G}}\\
    -\one_{\mathcal{L}} & \zero_{\mathcal{L} \times \mathcal{G}}
  \end{bmatrix}, \quad
  T_2 =
  \begin{bmatrix}
    \zero_{\mathcal{G} \times \mathcal{L}} \\ \eye_{\mathcal{L}}
  \end{bmatrix},
\end{equation*}
so that the governing equations~\eqref{eq:reactive-load-model}, \eqref{eq:swing-eq-SG} and \eqref{eq:active-load-model-theta} can be rewritten in DAE form as
\begin{subequations}
  \label{eq:dae-system}
  \begin{align}
    [\dot{\omega}_t]_{\mathcal{S} \cup \mathcal{G}} &= -(M^g)^{-1} D^g (M^g)^{-1} H_1 - (M^g)^{-1} T^{\top}_1 H_2, \\
    [\dot{\theta}_t]_{\mathcal{G} \cup \mathcal{L}} &= T_1 (M^g)^{-1} H_1 - T_2 (D^d)^{-1} T^{\top}_2 H_2, \\
    \label{eq:dae-system-algebraic}
    0 &= [V_t]_{\mathcal{L}}^{-1} \odot \left \{ [g]_{\mathcal{L}} - [Q_0]_{\mathcal{L}} \right \}
  \end{align}
\end{subequations}
where the set of algebraic equations \eqref{eq:dae-system-algebraic} corresponds to the reactive load model~\eqref{eq:reactive-load-model}, which constrains the dynamics to the power flow manifold.

\subsection*{Singularly-perturbed ODE system}
The DAE system \eqref{eq:dae-system} is not globally well posed.
Following \cite{demarco_1987_security}, we consider a singularly-perturbed version of the algebraic constraint with perturbation parameter $D^{\epsilon}$ such that in the limit $D^{\epsilon} \to 0^{+}$, the singularly-perturbed model approximates the DAE system arbitrarily well.
Introducing the additional auxiliary function
\begin{equation}
  \label{eq:H3-function}
  H_3(x_t) \coloneqq [V_t]_{\mathcal{L}}^{-1} \odot \left\{ [g]_{\mathcal{L}} - [Q_0]_{\mathcal{L}} \right \}
\end{equation}
we can recast the DAE system~\eqref{eq:dae-system} into the ODE system
\begin{subequations}
  \label{eq:ode}
  \begin{align}
    [\dot{\omega}_t]_{\mathcal{S} \cup \mathcal{G}} &= -(M^g)^{-1} D^g (M^g)^{-1} H_1 - (M^g)^{-1} T^{\top}_1 H_2, \\
    [\dot{\theta}_t]_{\mathcal{G} \cup \mathcal{L}} &= T_1 (M^g)^{-1} H_1 - T_2 (D^d)^{-1} T^{\top}_2 H_2,\\
    [\dot{V}_t]_{\mathcal{L}} &= -(D^{\epsilon})^{-1} H_3,
  \end{align}
\end{subequations}
or in compact form,
\begin{equation}
  \label{eq:ode-compact}
  \dot{x}_t = -K H(x_t),
\end{equation}
where $H^{\top} \coloneqq (H^{\top}_1, H^{\top}_2, H^{\top}_3)$, and $K \coloneqq S- J$, with the matrices $J$ and $S$ given in the in the main article. 

\subsection*{Energy function derivation}
We proceed to obtain a candidate Lyapunov function.
Defining $\nabla \mathcal{H}(x_t) \coloneqq H(x_t)$, we can obtain $\mathcal{H}(x_t)$ by integrating $H(x)$ along a path from $x_0$ to $x_t$, that is,
\begin{equation}
  \mathcal{H}(x_t) \coloneqq
  \int_{(0,\, [\theta_0]_{\mathcal{G} \cup \mathcal{L}},\, [V_0]_{\mathcal{L}})}^{([\omega_t]_{\mathcal{S} \cup \mathcal{G}},\, [\theta_t]_{\mathcal{G} \cup \mathcal{L}},\, [V_t]_{\mathcal{L}})}  \left \langle H(y), \mathrm{d} y \right \rangle.
\end{equation}
Disregarding the constant of integration, we obtain the scalar potential \cite{narasimhamurthi_1984_generalized,demarco_1987_security,pai_energy_1989,zheng_2010_bistable,zheng_2016_new}
\begin{align*}
  \begin{split}
    \mathcal{H}(x_t) &= \frac{1}{2} [\omega_t]_{\mathcal{S} \cup \mathcal{G}}^{\top} M^g\, [\omega_t]_{\mathcal{S} \cup \mathcal{G}} + \frac{1}{2} \sum_{i, j \in \mathcal{B}} B_{ij} [V_t]_i [V_t]_j \cos([\theta_t]_i - [\theta_T]_j) \\
    &+ \left \langle  [P_0]_{\mathcal{G} \cup \mathcal{L}}, \, [\theta_t]_{\mathcal{G} \cup \mathcal{S}} \right \rangle + \left \langle [Q_0]_{\mathcal{L}}, \, \log \left( [V_t]_{\mathcal{L}} \right) \right \rangle + C \\
    &= \frac{1}{2} ([\omega_t]_{\mathcal{S} \cup \mathcal{G}})^{\top} M^g\, [\omega_T]_{\mathcal{S} \cup \mathcal{G}} + \frac{1}{2} v_{t}^{\herm} B\, v_{t} + [P_0]_{\mathcal{G} \cup \mathcal{L}}^{\top} [\theta_t]_{\mathcal{G} \cup \mathcal{L}} + [Q_0]_{\mathcal{L}}^{\top} \log \left ( [V_t]_{\mathcal{L}} \right ),
  \end{split}
\end{align*}
thus recovering the energy function. 
Substituting this energy function into the ODE system \cref{eq:ode} leads to the port-Hamiltonian form of the system dynamics.
Also, it can be verified that the condition $\nabla \mathcal{H}(x) = 0$ is equivalent to the power flow equations~\cref{eq:acpf}.
Finally, we note that the equilibrium position $\bar{x}$ is asymptotically stable~\cite{bergen_1984_application}.

\subsection*{Noise model}
Following DeMarco \cite{demarco_1987_security}, we introduce scaled, additive noise to the deterministic dynamics \cref{eq:ode} in such a way as to satisfy fluctuation-dissipation dynamics.
Specifically, this means adding independent Wiener processes with covariance (diffusion) matrix $S/2$.
Since the net $P_0$ and $Q_0$ vectors enter the gradient term $\mathcal{H}$ linearly, we can interpret the stochastic fluctuation as direct perturbations to elements of active and reactive power (either demand or generation) with magnitude $Z_t^P$ and $Z_t^Q$, respectively.
Componentwise, we now make this relationship explicit by replacing $P_0$ with $P_0 + Z_t^P$ and $Q_0$ with $Q_0 + Z_t^Q$.
We then determine the appropriate scaling coefficients of $Z_t^P$ and $Z_t^Q$ in terms of the components of $S$.

Consider the stochastic dynamics
\begin{equation}
\begin{aligned}
\label{eq:ode+noise}
\dot{\omega}_{\mathcal{S}} &= -(M^g)^{-1} D^g \omega_{\mathcal{S}} - \one_{\mathcal{G} \cup \mathcal{L}}^{\top} [f - (P_0 + Z_t^P)]_{\mathcal{G} \cup \mathcal{L}} \\
\dot{\omega}_{\mathcal{G}} &= -(M^g)^{-1} D^g \omega_{\mathcal{G}} + (M^g)^{-1} \eye_{\mathcal{G}} [f - (P_0 + Z_t^P)]_{\mathcal{G}} \\
\dot{\theta}_{\mathcal{G}} &= -\one_{\mathcal{G}} \omega_{\mathcal{S}} - \omega_{\mathcal{G}} \\
\dot{\theta}_{\mathcal{L}} &= -\one_{\mathcal{L}} \omega_{\mathcal{S}} + (D^d)^{-1} [f - (P_0 + Z_t^P)]_{\mathcal{L}} \\
\dot{V}_{\mathcal{L}} &= V_{\mathcal{L}}^{-1} \odot [g - (Q_0 + Z_t^Q)]_{\mathcal{L}}
\end{aligned}
\end{equation}
and define
\begin{equation}
\begin{aligned}
\label{eq:w-p-noise}
[Z_t^P]_i &\coloneqq (M^g)^{-1} D^g (M^g)^{-1} [\mathrm{d} W_t]_i , \quad &i &\in \mathcal{S} \\
[Z_t^P]_i &\coloneqq D^g (M^g)^{-1} [\mathrm{d} W_t]_i , \quad &i &\in \mathcal{G} \\
[Z_t^P]_i &\coloneqq [\mathrm{d} W_t]_i , \quad &i &\in \mathcal{L}
\end{aligned}
\end{equation}
and
\begin{equation}
\begin{aligned}
\label{eq:w-q-noise}
[Z_t^Q]_i &\coloneqq 0 , \quad &i &\in \mathcal{S} \\
[Z_t^Q]_i &\coloneqq 0 , \quad &i &\in \mathcal{G} \\
[Z_t^Q]_i &\coloneqq V_{\mathcal{L}} [\mathrm{d} W_t]_i , \quad &i &\in \mathcal{L}.
\end{aligned}
\end{equation}
We remark that \cref{eq:ode+noise} yields the same dynamics as $-K\nabla \mathcal{H}(x) + \sqrt{2S} \mathrm{d}W_t$ if we assume that at the reference (slack) bus, the aggregate frequency dynamics are computed on power balance differences with respect to the unperturbed $P_0$ (following \cite{demarco_1987_security}) but are perturbed according to an independent forcing term with magnitude $(M^g)^{-1} D^g (M^g)^{-1}$.
In \cite{demarco_1987_security}, it is assumed that there is no additional independent forcing term; in our model, this amounts to removing damping from the generator equation.
For consistency with the other generators, we include slack generator damping (and its corresponding scaled noise).
Finally, we control the magnitude of the additive noise by $\tau$.
Thus, the noise model introduces stochastic disturbances to the (i) active power at both load and generator buses, and (ii) reactive power at the load buses.

\section{Algorithm and data details}
We provide additional detail on the KMC algorithm and data for a small test system.
\label{sm:alg-data}
\subsection{Line limit calculation}
\label{sm:line-limits}
\noindent For the adjusted DC method, current magnitude limits were set as the square root of DC limit data.
DC limit data was computed through a system feedback loop where limits were iteratively increased if an overflow was observed in the network under OPA simulation (a cascading failure model developed at Oak Ridge National Laboratory used in \cite{dobson_complex_2007}).
The simulation was run 100 times until the network became reasonably stable and flow limits were such that failures were due to overloads due to initial failures rather than normal operating fluctuations.
The DC limit data did not contain limits for each line, and we computed limits for the few remaining lines as the geometric average of the line's OPF flow and the maximum flow over all lines at the OPF point.
\begin{algorithm}[H]
\bgroup
\scriptsize
\begin{algorithmic}
	\Require DC (real power) line limit data $D$, Network topology $\mathcal{N}$
	\State $\bar{x}_0 \gets $ ACOPF($\mathcal{N}$)
	\State $\Theta_l = \Theta_l(\bar{x}_0)$
	\For {$l \in \mathcal{N}.\mathcal{L}$}
		\If {$l \in D$}
			\State $\Theta^{\max}_l = \sqrt{D_l}$
		\Else
			\State $\Theta^{\max}_l = \sqrt{\max_k \{\Theta_k\} \times \Theta_l}$
		\EndIf
	\EndFor
	\State \Return $\Theta^{\max}$
\end{algorithmic}
\egroup
\caption{Adjusted DC Approximation}
\label{alg:adj-dc}
\end{algorithm}

\noindent For the $N$-$1$ procedure, line limits were set to ensure $N$-$1$ feasibility under all non-islanding line failure contingencies (that is, line failures which do not separate a bus from the slack) in the heavily congested network settings.
The method involves iteratively solving security-constrained AC optimal power flow problems (see \cite{frank_primer_2012}) at different loading levels.
\begin{algorithm}[H]
\bgroup
\scriptsize
\begin{algorithmic}
	\Require Network $\mathcal{N}$, non-islanding contingencies $\mathcal{C}$, loading factor $a$, step factor $b$, power factor $c$, rating buffer $d$
	\While {\textit{true}}
		\State $\Theta^{\max} \gets |Y_{ft}|^2 \times \left( (V_f^{\max})^2 + (V_t^{\max})^2 + 2V_f^{\max}V_t^{\max} \right)$ $\forall (f,t) \in \mathcal{N}.\mathcal{L}$ \Comment{physically maximal current}
		\State Adjust $\mathcal{N}.P_d$ and $\mathcal{N}.Q_d$ network data to $a$\% of generation total capacity, maintaining $c$
		\State $\bar{x} \gets$ SC-ACOPF($\mathcal{N}, \mathcal{C}$) \Comment{run security constrained ACOPF}
		\For {$c \in \mathcal{C}$}
			\State $\bar{x}_c \gets$ ACPF($\mathcal{N}_c$) \Comment{run power flow to compute operating point $\bar{x}_c$ under contingency $c$}
			\State $\Theta_c \gets \Theta(\bar{x}_c)$ \Comment{compute flows for the operating point $\bar{x}_c$}
		\EndFor
		\State $\Theta^{\max} \gets \max_c \Theta_c \times d$ \Comment{update the vector of limits according to the maximum flow across contingencies}
		\State $\bar{x} \gets$ SC-ACOPF($\mathcal{N}$) for non-islanding line outages with new limits
		\If {$\bar{x}$ is feasible (SC-ACOPF is solved optimally)}
			\State \Return $\Theta^{\max}$
		\Else
			\State $a \gets a - b$
		\EndIf
	\EndWhile
\end{algorithmic}
\egroup
\caption{Loading-dependent $N$-$1$ Criteria}
\label{alg:n1-criteria}
\end{algorithm}
\subsection{Markov model optimization routines}
Occasionally, the Markov line failure computations may encounter either non-physical solutions or infeasibility.
In these cases, as outlined in \cref{alg:feasible-equilibrium-point}, we remove all lines associated with bus-$i$, shed any load at bus $i$, and re-solve for a new equilibrium point until the resulting point $\bar{x}$ has voltages that exceed 0.1 p.u.
Similarly, the constrained failure problem may not find an appropriate failure point on the first attempt.
To address these cases, we proceed by restarting the solver from a feasible point, and if that fails, we simply take a point which satisfies the line-limit constraint.
The failure point methodology is outlined in \cref{alg:feasible-failure-point}.
%
%
%

%
\begin{algorithm}[H]
\bgroup
\scriptsize
\begin{algorithmic}
	\Require Network topology $\mathcal{N}$, dispatch data $\mathcal{D}$, and dynamics parameters $\mathcal{P}$
	\State Attempt to solve unconditional NLP for $\hat{x}$ 
	\While {unconditional NLPis infeasible} 
		\State $\mathcal{I} \gets i$ such that $V_i < 0.8$ or $V_i > 1.2$ (non-supportable bus voltage)
		\State Shed load at buses $i \in \mathcal{I}$
		\State Remove lines connected to each bus in $\mathcal{I}$ 
		\State Attempt to solve unconditional NLP for $\hat{x}$ 
	\EndWhile
	\State $\bar{x} \gets \hat{x}$
	\State \Return $\bar{x}$
\end{algorithmic}
\egroup
\caption{Compute Equilibrium Point (with load shed for feasibility)}
\label{alg:feasible-equilibrium-point}
\end{algorithm}
\begin{algorithm}[H]
\bgroup
\scriptsize
\begin{algorithmic}
	\Require Network topology $\mathcal{N}$, dispatch data $\mathcal{D}$, and dynamics parameters $\mathcal{P}$, line $l$
	\State Attempt to solve unconditional NLP for $\hat{x}$ 
	\If {unconditional NLP is infeasible}
		\State Compute a feasible failure point $x_f$ such that $\Theta_l(x_f) = \Theta_l^{\max}$
		\State Attempt to solve relaxed unconditional NLP (with $\Theta_l(x) \geq \Theta_l^{\max}$) for $\hat{x}$ warm-started at $x_f$ 
		\If {Relaxed unconditional NLP is infeasible} 
			\State Solve for feasible failure point $x_f$ such that $\Theta_l(x_f) = \Theta_l^{\max}$
			\State $x^{\star} \gets x_f$
		\Else
			\State $x^{\star} \gets \hat{x}$
		\EndIf	
	\EndIf
	\State \Return $x^{\star}$
\end{algorithmic}
\egroup
\caption{Compute Failure Point}
\label{alg:feasible-failure-point}
\end{algorithm}

\subsection{3-bus detailed system}
\label{sm:3-bus-detailed}
The below model describes the 3bus system used to generate the energy surface contour plots and is largely based on the model in \cite{zheng_2010_bistable}. 
\begin{figure}[H]
	\resizebox{\textwidth}{!}{\tikzset{
	vertex/.style={circle,draw,minimum size=1.5em},
	edge/.style={-}
}
\begin{tikzpicture}
\node [vertex,draw] (1slack) at (0,0) {1-ref};
\draw ($(1slack)+(-0.5,0.5)$) node[left] {$x_1$: ($\omega_1$), $y_1$: $(V_1=1.02, \theta_1=0)$};
\node [vertex,draw] (2gen) at ($(1slack)+(0:3.5)$) {2-pv};
\draw ($(2gen)+(0.5,0.5)$) node[right] {$x_2$: ($\omega_2,\theta_2$), $y_2$: $(V_2=1.05, P_g^2=2)$};
\node [vertex,draw] (3load) at ($(1slack)+(-60:3.5)$) {3-pq};
\draw ($(3load)+(-0.0,-0.5)$) node[below] {$x_3$: ($\theta_3, V_3$), $y_3$: $(P_d^3=3, Q_d^3=0.1)$};
\path [-] (1slack) edge node[above] {line-$1$, $B_{12}=10$} (2gen);
\path [-] (1slack) edge node[below] {$\Theta_1^{\max}=5.8$} (2gen);
\path [-] (3load) edge node[left] {line-$2$, $B_{13}=10$, $\Theta_2^{\max}=5.8$} (1slack);
\path [-] (2gen) edge node[right] {line-$3$, $B_{23}=10$, $\Theta_3^{\max}=5.8$} (3load);
\end{tikzpicture}}
\end{figure}

\section{Additional experiments}
\label{sm:additional-experiments}
We present additional numerical results for failure path experiments, unconditional failure rate experiments, equilibration experiments, and cascade simulation experiments.

\subsection{Failure path experiments}
\label{sm:rate-path}
\begin{figure}[H]	
\begin{subfigure}[T]{0.325\linewidth}
	\centering
	\resizebox{\linewidth}{!}{\includegraphics{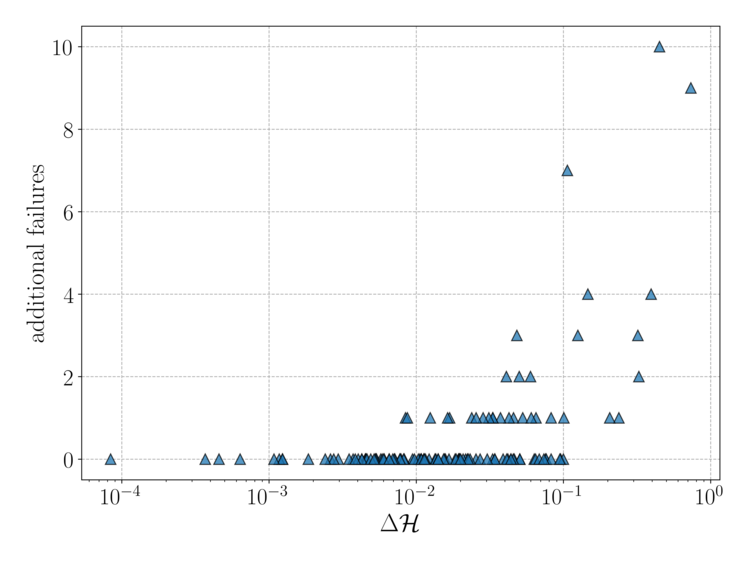}}
\end{subfigure}
\begin{subfigure}[T]{0.325\linewidth}
	\centering
	\resizebox{\linewidth}{!}{\includegraphics{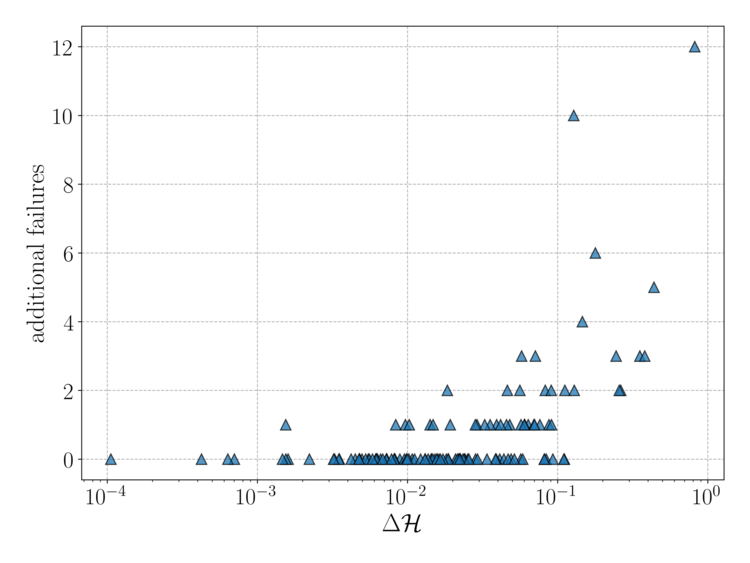}}
\end{subfigure}
\begin{subfigure}[T]{0.325\linewidth}
	\centering
	\resizebox{\linewidth}{!}{\includegraphics{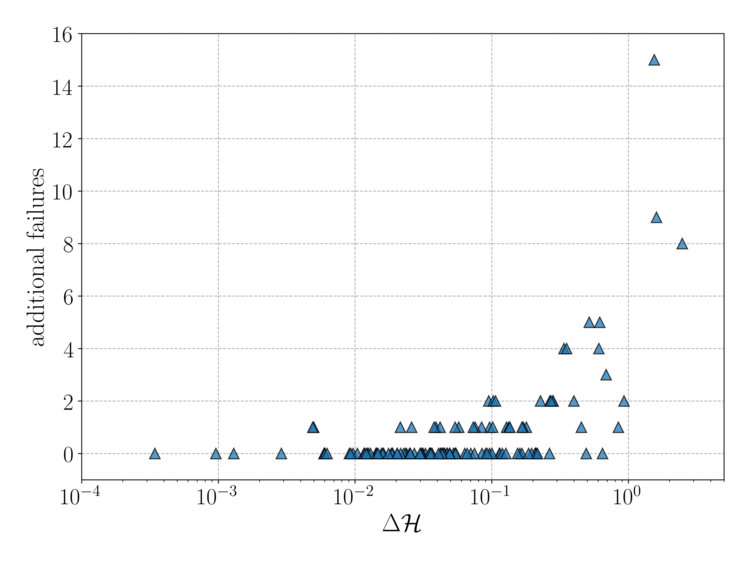}}
\end{subfigure}
\begin{subfigure}[b]{0.325\linewidth}
	\vspace{-8pt}
	\caption{}
	\label{fig:path-diagnostic-118}
\end{subfigure}
\begin{subfigure}[b]{0.325\linewidth}
	\vspace{-8pt}
	\caption{}
	\label{fig:path-diagnostic-118-1_1}
\end{subfigure}
\begin{subfigure}[b]{0.325\linewidth}
	\vspace{-8pt}
	\caption{}
	\label{fig:path-diagnostic-118-2_0}
\end{subfigure}
\vspace{-0.5cm}
\caption{\small{Study of approximate minimum energy failure path displaying number of unintended failures along the pathway from $x^{\star}$ to $\bar{x}$ for a particular line in the 118-bus system.}
	(\subref{fig:path-diagnostic-118}) \small{$N$-1 line limits.}
	(\subref{fig:path-diagnostic-118-1_1}) \small{$N$-1 line limits uniformly increased by 10\%.}
	(\subref{fig:path-diagnostic-118-2_0}) \small{$N$-1 line limits uniformly increased by 100\%.}
}
\label{fig:path-diagnostic-118bus}
\end{figure}

In this section, we present supplementary experiments for failure path experiments corresponding to the 118-bus system with $N$-1 secure line limits.
Note that in panels (\subref{fig:path-diagnostic-118-1_1})--(\subref{fig:path-diagnostic-118-2_0}) of \cref{fig:path-diagnostic-118bus}, line 133 is the lowest $\Delta \mathcal{H}$ line that exhibits an additional line failure along the most likely failure path; it is a corner case in the system.
For line 133's pathway, the additional failure corresponds to line 134, which has the lowest rating in the entire network.
Further, the low rating assigned to line 134 is a numerical artifact of the $N$-1 procedure (\cref{alg:n1-criteria}).
When setting line limits, the $N$-1 algorithm ignores ``islanding'' contingencies, and disconnection of line 133 automatically separates line 134 from the rest of the network; \cref{alg:n1-criteria} did not adjust line 134's rating based on possible contingencies associated with line 133.
However, increasing line 134's rating by just 6.5\% would prevent pathway failure.

We also note that in panel (\subref{fig:path-diagnostic-118-2_0}), line 58's failure pathway is not shown.
This line has a very large energy difference ($\Delta \mathcal{H} = 32.7$) and is associated with 127 additional line failures.
We treat it as an outlier and do not plot this point.

\newpage
\subsection{Unconditional failure rate experiments}
\label{sm:rate-uncond}
In this section, we present supplementary experiments for simulation experiments corresponding to the unconditional line failure rate.
\newcommand{\failurewidth}{0.245}

\newcommand{\LineA}{10}
\newcommand{\LineAFailrateComparison}{rate-30bus-uncond-line-10-failrate_comparison}
\newcommand{\LineAFailratioComparison}{rate-30bus-uncond-line-10-failratio_comparison}
\newcommand{\LineAFailtimeHistogram}{rate-30bus-uncond-line-10-failtime_histogram}
\newcommand{\LineAStatespaceHistogram}{rate-30bus-uncond-line-10-statespace_histogram_bus-8}
\newcommand{\LineANsims}{5,000}
\newcommand{\LineB}{20}
\newcommand{\LineBFailrateComparison}{rate-30bus-uncond-line-20-failrate_comparison}
\newcommand{\LineBFailratioComparison}{rate-30bus-uncond-line-20-failratio_comparison}
\newcommand{\LineBFailtimeHistogram}{rate-30bus-uncond-line-20-failtime_histogram}
\newcommand{\LineBStatespaceHistogram}{rate-30bus-uncond-line-20-statespace_histogram_bus-14}
\newcommand{\LineBNsims}{5,000}
\newcommand{\LineC}{4}
\newcommand{\LineCFailrateComparison}{rate-30bus-uncond-line-4-failrate_comparison}
\newcommand{\LineCFailratioComparison}{rate-30bus-uncond-line-4-failratio_comparison}
\newcommand{\LineCFailtimeHistogram}{rate-30bus-uncond-line-4-failtime_histogram}
\newcommand{\LineCStatespaceHistogram}{rate-30bus-uncond-line-4-statespace_histogram_bus-3}
\newcommand{\LineCNsims}{5,000}

\newcommand{\LineD}{29}
\newcommand{\LineDFailrateComparison}{rate-30bus-uncond-line-29-failrate_comparison}
\newcommand{\LineDFailratioComparison}{rate-30bus-uncond-line-29-failratio_comparison}
\newcommand{\LineDFailtimeHistogram}{rate-30bus-uncond-line-29-failtime_histogram}
\newcommand{\LineDStatespaceHistogram}{rate-30bus-uncond-line-29-statespace_histogram_bus-21}
\newcommand{\LineDNsims}{5,000}
\newcommand{\LineE}{35}
\newcommand{\LineEFailrateComparison}{rate-30bus-uncond-line-35-failrate_comparison}
\newcommand{\LineEFailratioComparison}{rate-30bus-uncond-line-35-failratio_comparison}
\newcommand{\LineEFailtimeHistogram}{rate-30bus-uncond-line-35-failtime_histogram}
\newcommand{\LineEStatespaceHistogram}{rate-30bus-uncond-line-35-statespace_histogram_bus-25}
\newcommand{\LineENsims}{5,000}
\newcommand{\LineF}{37}
\newcommand{\LineFFailrateComparison}{rate-30bus-uncond-line-37-failrate_comparison}
\newcommand{\LineFFailratioComparison}{rate-30bus-uncond-line-37-failratio_comparison}
\newcommand{\LineFFailtimeHistogram}{rate-30bus-uncond-line-37-failtime_histogram}
\newcommand{\LineFStatespaceHistogram}{rate-30bus-uncond-line-37-statespace_histogram_bus-29}
\newcommand{\LineFNsims}{500}
\newcommand{\LineG}{16}
\newcommand{\LineGFailrateComparison}{rate-30bus-uncond-line-16-failrate_comparison}
\newcommand{\LineGFailratioComparison}{rate-30bus-uncond-line-16-failratio_comparison}
\newcommand{\LineGFailtimeHistogram}{rate-30bus-uncond-line-16-failtime_histogram}
\newcommand{\LineGStatespaceHistogram}{rate-30bus-uncond-line-16-statespace_histogram_bus-12}
\newcommand{\LineGNsims}{500}

\newcommand{\LineH}{2}
\newcommand{\LineHFailrateComparison}{rate-30bus-uncond-line-2-failrate_comparison}
\newcommand{\LineHFailratioComparison}{rate-30bus-uncond-line-2-failratio_comparison}
\newcommand{\LineHFailtimeHistogram}{rate-30bus-uncond-line-2-failtime_histogram}
\newcommand{\LineHStatespaceHistogram}{rate-30bus-uncond-line-2-statespace_histogram_bus-3}
\newcommand{\LineHNsims}{5,000}

\begin{figure}[H]
\begin{minipage}{\linewidth}
\begin{subfigure}[t]{\failurewidth\linewidth}%
	\centering
	\resizebox{\linewidth}{!}{\includegraphics[trim={0 0 3cm 1.5cm}, clip=true]{\LineAFailrateComparison}}%
\end{subfigure}
\begin{subfigure}[t]{\failurewidth\linewidth}%
	\centering
	\resizebox{\linewidth}{!}{\includegraphics[trim={0 0 3cm 1.5cm}, clip=true]{\LineAFailratioComparison}}%
\end{subfigure}
\begin{subfigure}[t]{\failurewidth\linewidth}%
	\centering
	\resizebox{\linewidth}{!}{\includegraphics[trim={0 0 3cm 1.5cm}, clip=true]{\LineAFailtimeHistogram}}%
\end{subfigure}
\begin{subfigure}[t]{\failurewidth\linewidth}%
	\centering
	\resizebox{\linewidth}{!}{\includegraphics[trim={0 0 3cm 1.5cm}, clip=true]{\LineAStatespaceHistogram}}
\end{subfigure}
\begin{subfigure}[t]{\failurewidth\linewidth}
	\vspace{\failurecaption}%
	\renewcommand\thesubfigure{I.\alph{subfigure}}
	\caption{}
\end{subfigure}%
\hfill%
\begin{subfigure}[t]{\failurewidth\linewidth}
	\vspace{\failurecaption}%
	\renewcommand\thesubfigure{I.\alph{subfigure}}
	\caption{}
\end{subfigure}%
\hfill%
\begin{subfigure}[t]{\failurewidth\linewidth}
	\vspace{\failurecaption}%
	\renewcommand\thesubfigure{I.\alph{subfigure}}
	\caption{}
\end{subfigure}%
\hfill%
\begin{subfigure}[t]{\failurewidth\linewidth}
	\vspace{\failurecaption}%
	\renewcommand\thesubfigure{I.\alph{subfigure}}
	\caption{}
\end{subfigure}%
\hfill%
\label{fig:rate-line-A}
\end{minipage}
\begin{minipage}{\linewidth}
	\begin{subfigure}[t]{\failurewidth\linewidth}%
		\centering
		\resizebox{\linewidth}{!}{\includegraphics[trim={0 0 3cm 1.5cm}, clip=true]{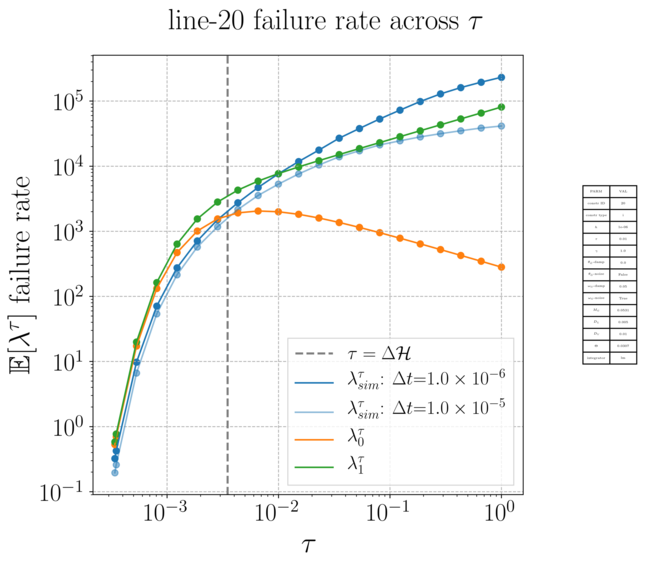}}%
	\end{subfigure}
	\begin{subfigure}[t]{\failurewidth\linewidth}%
		\centering
		\resizebox{\linewidth}{!}{\includegraphics[trim={0 0 3cm 1.5cm}, clip=true]{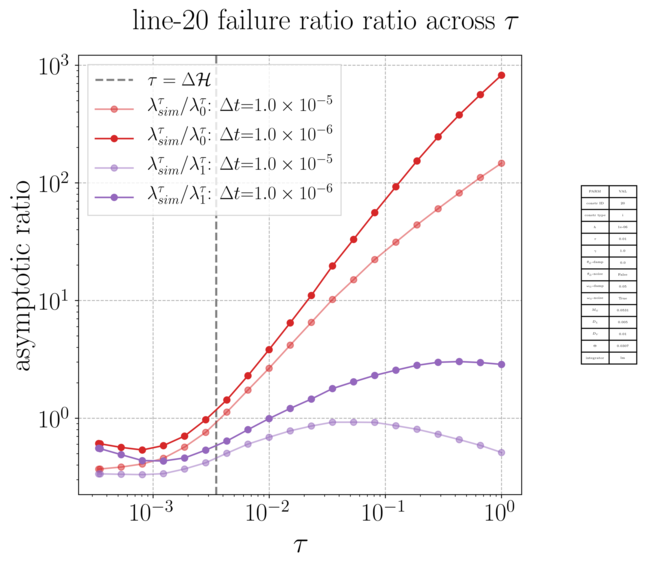}}%
	\end{subfigure}
	\begin{subfigure}[t]{\failurewidth\linewidth}%
		\centering
		\resizebox{\linewidth}{!}{\includegraphics[trim={0 0 3cm 1.5cm}, clip=true]{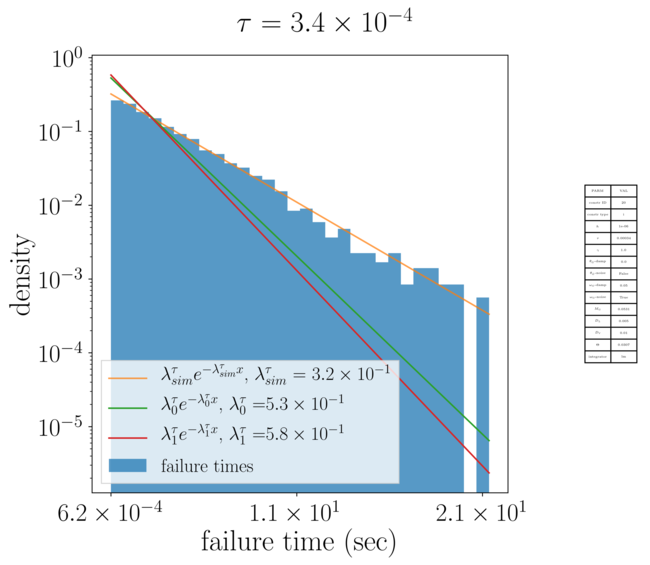}}%
	\end{subfigure}
	\begin{subfigure}[t]{\failurewidth\linewidth}%
		\centering
		\resizebox{\linewidth}{!}{\includegraphics[trim={0 0 3cm 1.5cm}, clip=true]{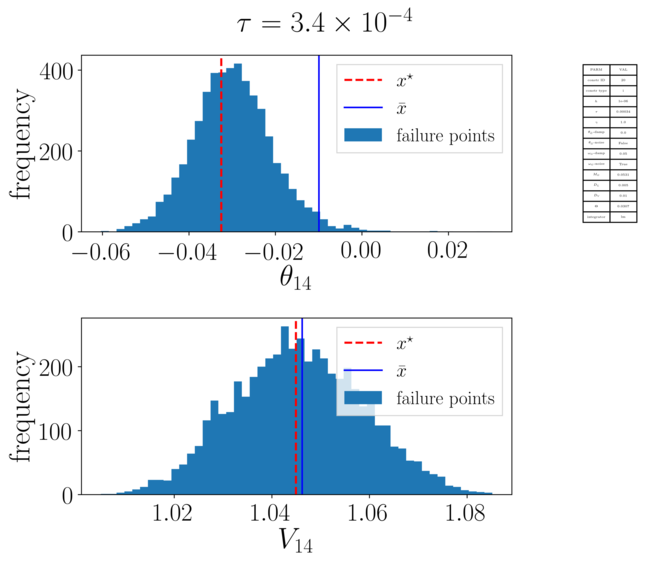}}
	\end{subfigure}
	\addtocounter{subfigure}{-4}
	\begin{subfigure}[t]{\failurewidth\linewidth}
		\vspace{\failurecaption}%
		\renewcommand\thesubfigure{II.\alph{subfigure}}
		\caption{}
	\end{subfigure}%
	\hfill%
	\begin{subfigure}[t]{\failurewidth\linewidth}
		\vspace{\failurecaption}%
		\renewcommand\thesubfigure{II.\alph{subfigure}}
		\caption{}
	\end{subfigure}%
	\hfill%
	\begin{subfigure}[t]{\failurewidth\linewidth}
		\vspace{\failurecaption}%
		\renewcommand\thesubfigure{II.\alph{subfigure}}
		\caption{}
	\end{subfigure}%
	\hfill%
	\begin{subfigure}[t]{\failurewidth\linewidth}
		\vspace{\failurecaption}%
		\renewcommand\thesubfigure{II.\alph{subfigure}}
		\caption{}
	\end{subfigure}%
	\hfill%
	\label{fig:rate-line-B}
\end{minipage}
\begin{minipage}{\linewidth}
	\begin{subfigure}[t]{\failurewidth\linewidth}%
		\centering
		\resizebox{\linewidth}{!}{\includegraphics[trim={0 0 3cm 1.5cm}, clip=true]{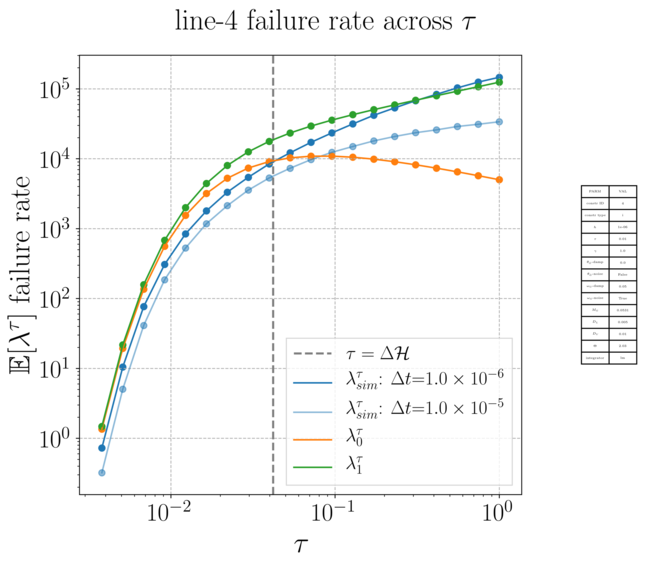}}%
	\end{subfigure}
	\begin{subfigure}[t]{\failurewidth\linewidth}%
		\centering
		\resizebox{\linewidth}{!}{\includegraphics[trim={0 0 3cm 1.5cm}, clip=true]{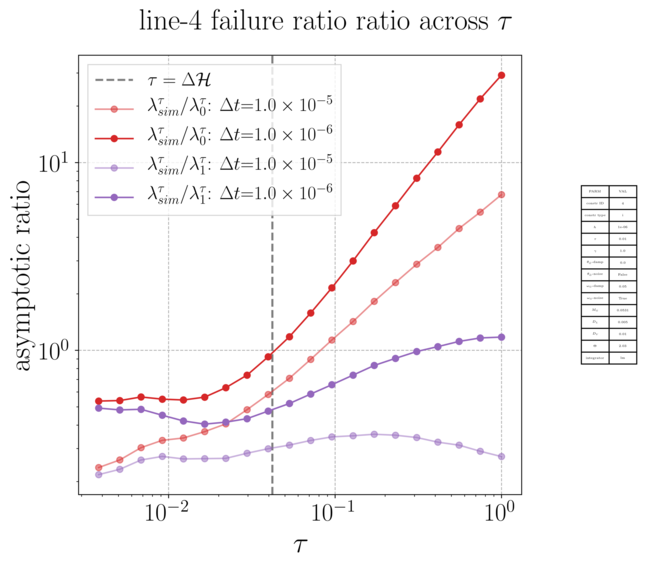}}%
	\end{subfigure}
	\begin{subfigure}[t]{\failurewidth\linewidth}%
		\centering
		\resizebox{\linewidth}{!}{\includegraphics[trim={0 0 3cm 1.5cm}, clip=true]{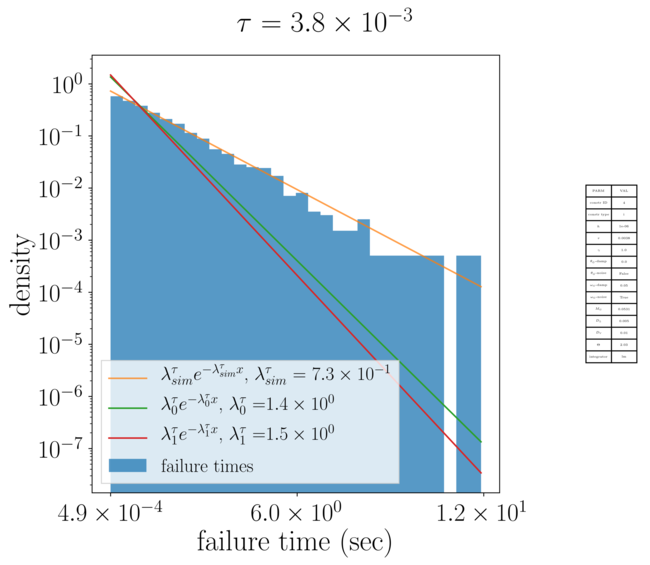}}%
	\end{subfigure}
	\begin{subfigure}[t]{\failurewidth\linewidth}%
		\centering
		\resizebox{\linewidth}{!}{\includegraphics[trim={0 0 3cm 1.5cm}, clip=true]{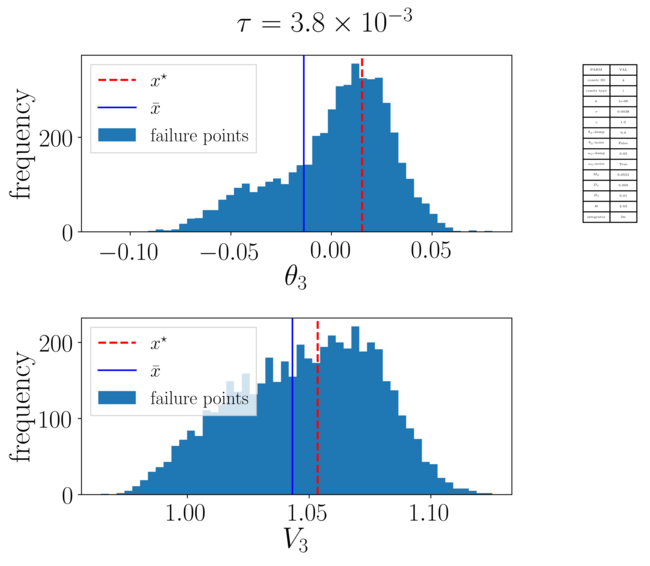}}
	\end{subfigure}
	\addtocounter{subfigure}{-4}
	\begin{subfigure}[t]{\failurewidth\linewidth}
		\vspace{\failurecaption}%
		\renewcommand\thesubfigure{III.\alph{subfigure}}
		\caption{}
	\end{subfigure}%
	\hfill%
	\begin{subfigure}[t]{\failurewidth\linewidth}
		\vspace{\failurecaption}%
		\renewcommand\thesubfigure{III.\alph{subfigure}}
		\caption{}
	\end{subfigure}%
	\hfill%
	\begin{subfigure}[t]{\failurewidth\linewidth}
		\vspace{\failurecaption}%
		\renewcommand\thesubfigure{III.\alph{subfigure}}
		\caption{}
	\end{subfigure}%
	\hfill%
	\begin{subfigure}[t]{\failurewidth\linewidth}
		\vspace{\failurecaption}%
		\renewcommand\thesubfigure{III.\alph{subfigure}}
		\caption{}
	\end{subfigure}%
	\hfill%
	\label{fig:rate-line-C}
\end{minipage}
\vspace{\failurecaption}
\caption{\small{Unconditional line failure diagnostics for load-load lines: line 10 (I), line 20 (II), and line 4 (III).}
\small{(a) Failure rate comparison across temperature.}
\small{(b) Asymptotic ratio across temperature.}
\small{(c) (Scaled) histogram of unconditional failure times at lowest temperature tested.}
\small{(d) Histogram of $x_i^{\star}$ where $i$ corresponds to the bus index of (one of) the line's connecting buses; shown at lowest temperature tested.}
}
\label{fig:rate-line-abc}
\end{figure}

\begin{figure}[H]
	\begin{minipage}{\linewidth}
		\begin{subfigure}[t]{\failurewidth\linewidth}%
			\centering
			\resizebox{\linewidth}{!}{\includegraphics[trim={0 0 3cm 1.5cm}, clip=true]{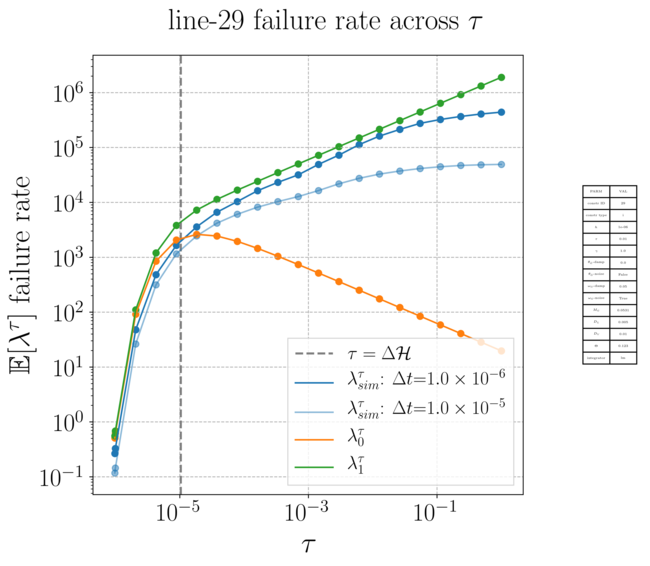}}%
		\end{subfigure}
		\begin{subfigure}[t]{\failurewidth\linewidth}%
			\centering
			\resizebox{\linewidth}{!}{\includegraphics[trim={0 0 3cm 1.5cm}, clip=true]{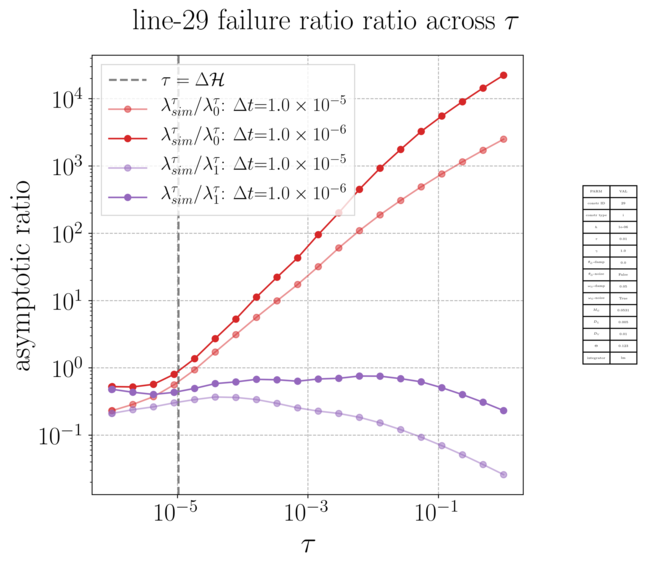}}%
		\end{subfigure}
		\begin{subfigure}[t]{\failurewidth\linewidth}%
			\centering
			\resizebox{\linewidth}{!}{\includegraphics[trim={0 0 3cm 1.5cm}, clip=true]{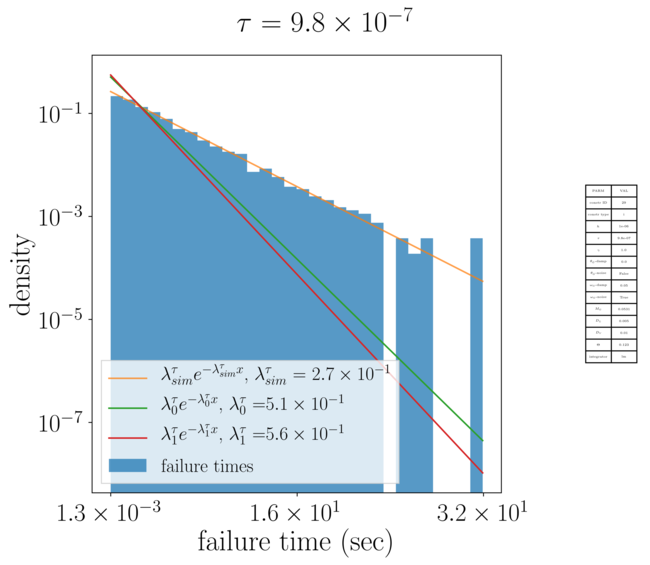}}%
		\end{subfigure}
		\begin{subfigure}[t]{\failurewidth\linewidth}%
			\centering
			\resizebox{\linewidth}{!}{\includegraphics[trim={0 0 3cm 1.5cm}, clip=true]{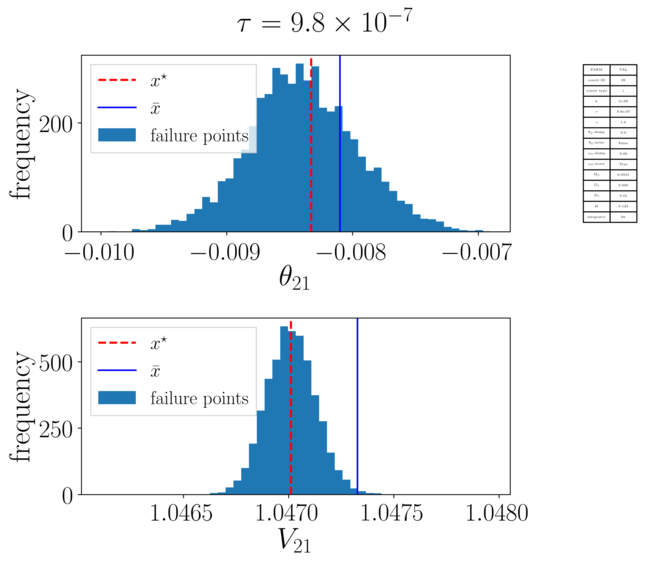}}
		\end{subfigure}
		\begin{subfigure}[t]{\failurewidth\linewidth}
			\vspace{\failurecaption}%
			\renewcommand\thesubfigure{I.\alph{subfigure}}
			\caption{}
		\end{subfigure}%
		\hfill%
		\begin{subfigure}[t]{\failurewidth\linewidth}
			\vspace{\failurecaption}%
			\renewcommand\thesubfigure{I.\alph{subfigure}}
			\caption{}
		\end{subfigure}%
		\hfill%
		\begin{subfigure}[t]{\failurewidth\linewidth}
			\vspace{\failurecaption}%
			\renewcommand\thesubfigure{I.\alph{subfigure}}
			\caption{}
		\end{subfigure}%
		\hfill%
		\begin{subfigure}[t]{\failurewidth\linewidth}
			\vspace{\failurecaption}%
			\renewcommand\thesubfigure{I.\alph{subfigure}}
			\caption{}
		\end{subfigure}%
		\hfill%
		\label{fig:rate-line-D}
	\end{minipage}
	\begin{minipage}{\linewidth}
		\begin{subfigure}[t]{\failurewidth\linewidth}%
			\centering
			\resizebox{\linewidth}{!}{\includegraphics[trim={0 0 3cm 1.5cm}, clip=true]{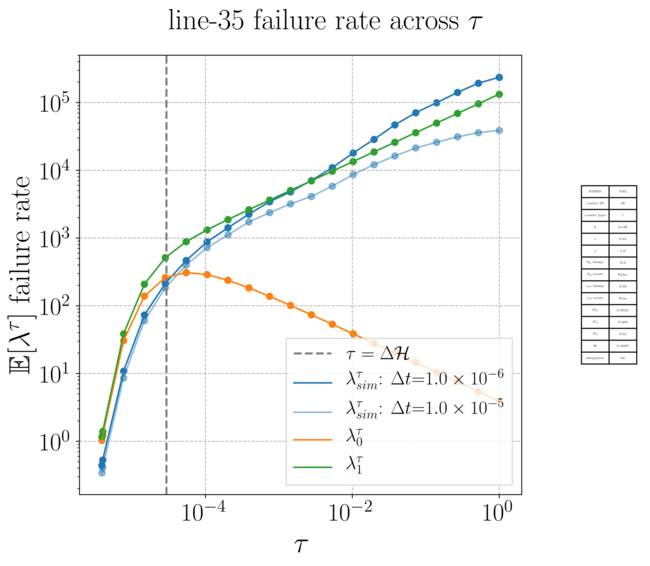}}%
		\end{subfigure}
		\begin{subfigure}[t]{\failurewidth\linewidth}%
			\centering
			\resizebox{\linewidth}{!}{\includegraphics[trim={0 0 3cm 1.5cm}, clip=true]{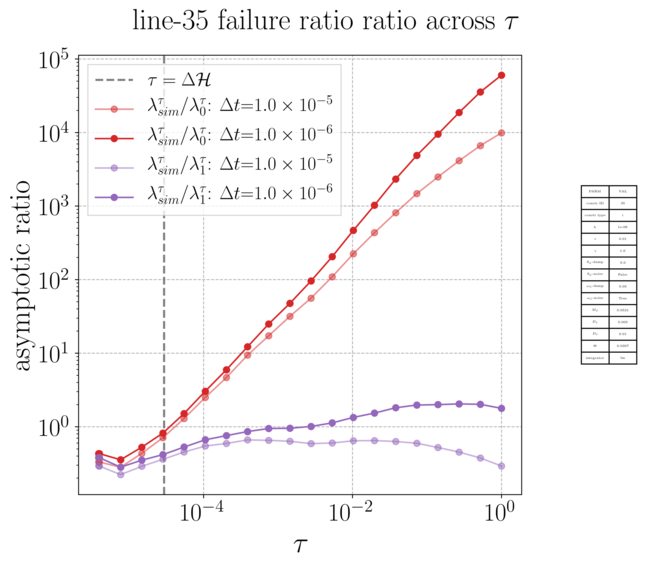}}%
		\end{subfigure}
		\begin{subfigure}[t]{\failurewidth\linewidth}%
			\centering
			\resizebox{\linewidth}{!}{\includegraphics[trim={0 0 3cm 1.5cm}, clip=true]{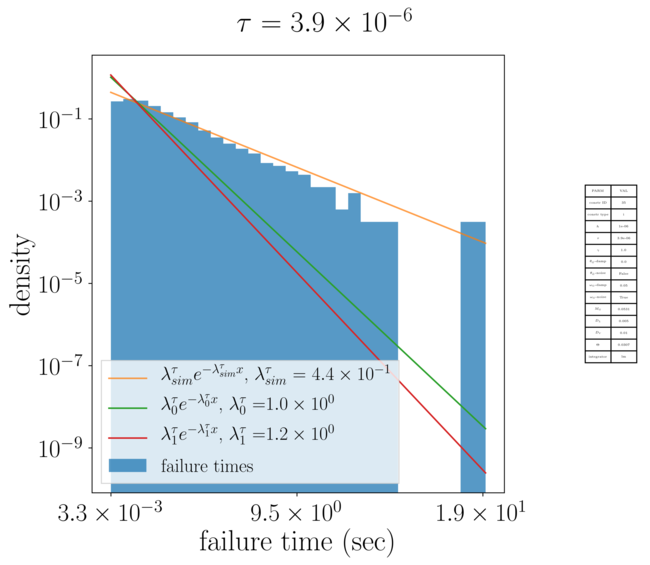}}%
		\end{subfigure}
		\begin{subfigure}[t]{\failurewidth\linewidth}%
			\centering
			\resizebox{\linewidth}{!}{\includegraphics[trim={0 0 3cm 1.5cm}, clip=true]{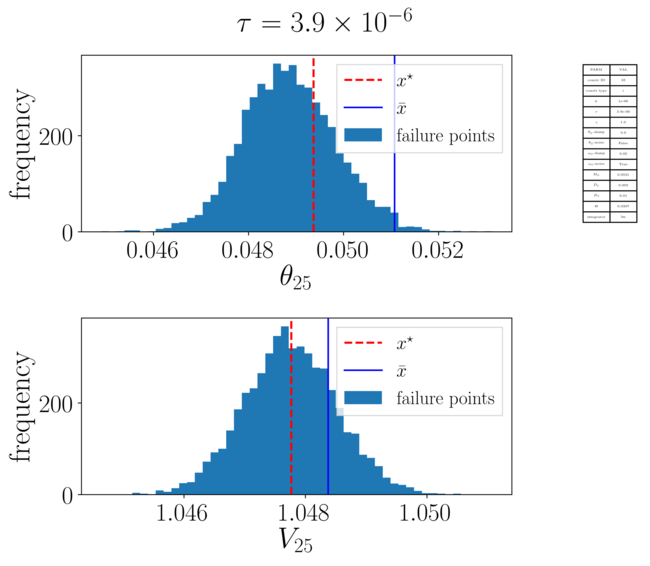}}
		\end{subfigure}
		\addtocounter{subfigure}{-4}
		\begin{subfigure}[t]{\failurewidth\linewidth}
			\vspace{\failurecaption}%
			\renewcommand\thesubfigure{II.\alph{subfigure}}
			\caption{}
		\end{subfigure}%
		\hfill%
		\begin{subfigure}[t]{\failurewidth\linewidth}
			\vspace{\failurecaption}%
			\renewcommand\thesubfigure{II.\alph{subfigure}}
			\caption{}
		\end{subfigure}%
		\hfill%
		\begin{subfigure}[t]{\failurewidth\linewidth}
			\vspace{\failurecaption}%
			\renewcommand\thesubfigure{II.\alph{subfigure}}
			\caption{}
		\end{subfigure}%
		\hfill%
		\begin{subfigure}[t]{\failurewidth\linewidth}
			\vspace{\failurecaption}%
			\renewcommand\thesubfigure{II.\alph{subfigure}}
			\caption{}
		\end{subfigure}%
		\hfill%
		\label{fig:rate-line-E}
	\end{minipage}
	\begin{minipage}{\linewidth}
		\begin{subfigure}[t]{\failurewidth\linewidth}%
			\centering
			\resizebox{\linewidth}{!}{\includegraphics[trim={0 0 3cm 1.5cm}, clip=true]{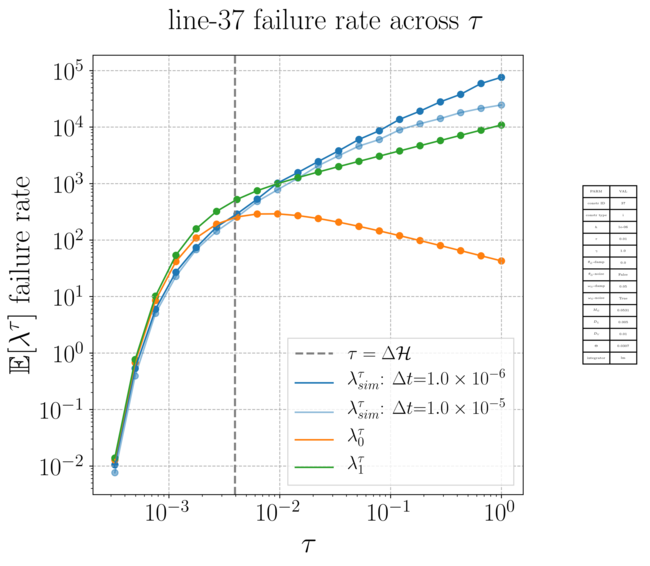}}%
		\end{subfigure}
		\begin{subfigure}[t]{\failurewidth\linewidth}%
			\centering
			\resizebox{\linewidth}{!}{\includegraphics[trim={0 0 3cm 1.5cm}, clip=true]{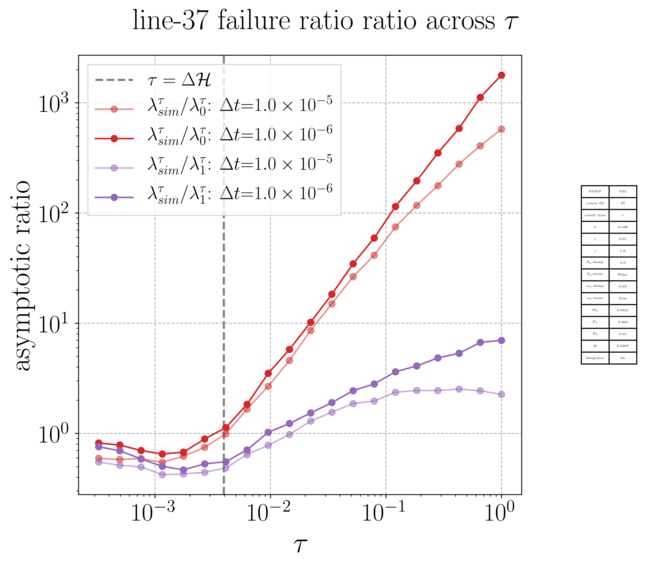}}%
		\end{subfigure}
		\begin{subfigure}[t]{\failurewidth\linewidth}%
			\centering
			\resizebox{\linewidth}{!}{\includegraphics[trim={0 0 3cm 1.5cm}, clip=true]{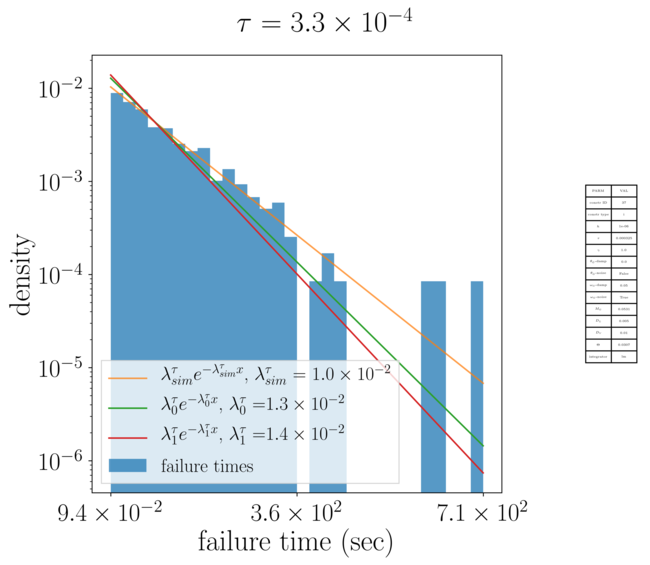}}%
		\end{subfigure}
		\begin{subfigure}[t]{\failurewidth\linewidth}%
			\centering
			\resizebox{\linewidth}{!}{\includegraphics[trim={0 0 3cm 1.5cm}, clip=true]{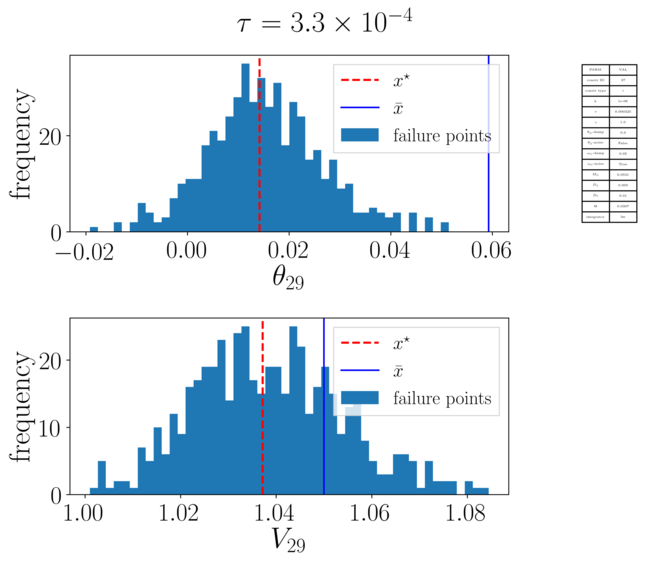}}
		\end{subfigure}
		\addtocounter{subfigure}{-4}
		\begin{subfigure}[t]{\failurewidth\linewidth}
			\vspace{\failurecaption}%
			\renewcommand\thesubfigure{III.\alph{subfigure}}
			\caption{}
		\end{subfigure}%
		\hfill%
		\begin{subfigure}[t]{\failurewidth\linewidth}
			\vspace{\failurecaption}%
			\renewcommand\thesubfigure{III.\alph{subfigure}}
			\caption{}
		\end{subfigure}%
		\hfill%
		\begin{subfigure}[t]{\failurewidth\linewidth}
			\vspace{\failurecaption}%
			\renewcommand\thesubfigure{III.\alph{subfigure}}
			\caption{}
		\end{subfigure}%
		\hfill%
		\begin{subfigure}[t]{\failurewidth\linewidth}
			\vspace{\failurecaption}%
			\renewcommand\thesubfigure{III.\alph{subfigure}}
			\caption{}
		\end{subfigure}%
		\hfill%
		\label{fig:rate-line-F}
	\end{minipage}
	\begin{minipage}{\linewidth}
	\begin{subfigure}[t]{\failurewidth\linewidth}%
		\centering
		\resizebox{\linewidth}{!}{\includegraphics[trim={0 0 3cm 1.5cm}, clip=true]{\LineGFailrateComparison}}%
	\end{subfigure}
	\begin{subfigure}[t]{\failurewidth\linewidth}%
		\centering
		\resizebox{\linewidth}{!}{\includegraphics[trim={0 0 3cm 1.5cm}, clip=true]{\LineGFailratioComparison}}%
	\end{subfigure}
	\begin{subfigure}[t]{\failurewidth\linewidth}%
		\centering
		\resizebox{\linewidth}{!}{\includegraphics[trim={0 0 3cm 1.5cm}, clip=true]{\LineGFailtimeHistogram}}%
	\end{subfigure}
	\begin{subfigure}[t]{\failurewidth\linewidth}%
		\centering
		\resizebox{\linewidth}{!}{\includegraphics[trim={0 0 3cm 1.5cm}, clip=true]{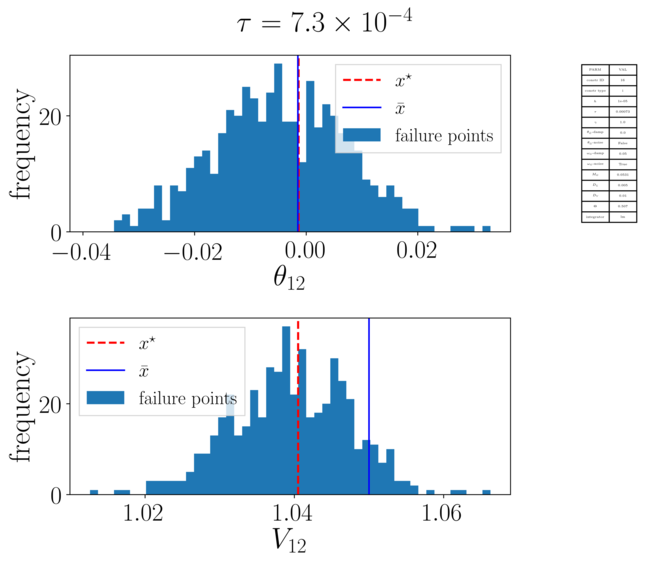}}
	\end{subfigure}
	\addtocounter{subfigure}{-4}
	\begin{subfigure}[t]{\failurewidth\linewidth}
		\vspace{\failurecaption}%
		\renewcommand\thesubfigure{IV.\alph{subfigure}}
		\caption{}
	\end{subfigure}%
	\hfill%
	\begin{subfigure}[t]{\failurewidth\linewidth}
		\vspace{\failurecaption}%
		\renewcommand\thesubfigure{IV.\alph{subfigure}}
		\caption{}
	\end{subfigure}%
	\hfill%
	\begin{subfigure}[t]{\failurewidth\linewidth}
		\vspace{\failurecaption}%
		\renewcommand\thesubfigure{IV.\alph{subfigure}}
		\caption{}
	\end{subfigure}%
	\hfill%
	\begin{subfigure}[t]{\failurewidth\linewidth}
		\vspace{\failurecaption}%
		\renewcommand\thesubfigure{IV.\alph{subfigure}}
		\caption{}
	\end{subfigure}%
	\hfill%
	\label{fig:rate-line-G}
\end{minipage}
\begin{minipage}{\linewidth}
	\begin{subfigure}[t]{\failurewidth\linewidth}%
		\centering
		\resizebox{\linewidth}{!}{\includegraphics[trim={0 0 3cm 1.5cm}, clip=true]{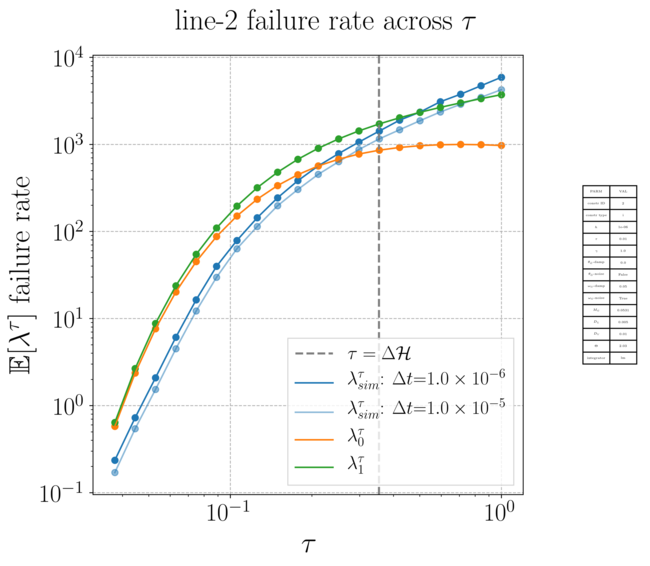}}%
	\end{subfigure}
	\begin{subfigure}[t]{\failurewidth\linewidth}%
		\centering
		\resizebox{\linewidth}{!}{\includegraphics[trim={0 0 3cm 1.5cm}, clip=true]{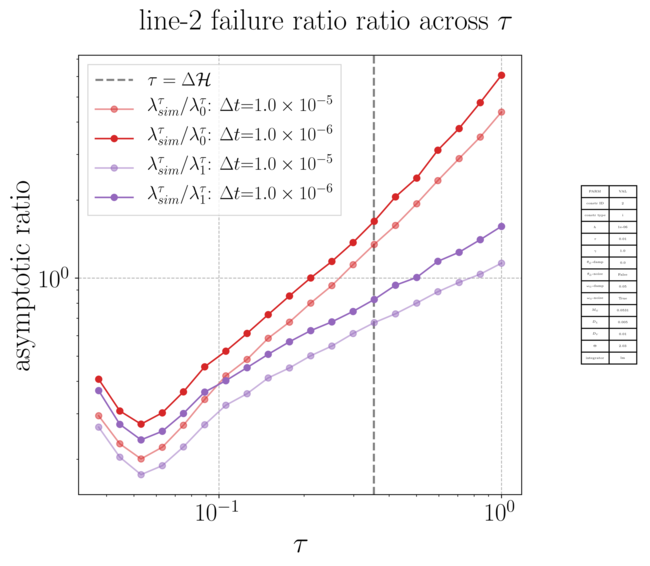}}%
	\end{subfigure}
	\begin{subfigure}[t]{\failurewidth\linewidth}%
		\centering
		\resizebox{\linewidth}{!}{\includegraphics[trim={0 0 3cm 1.5cm}, clip=true]{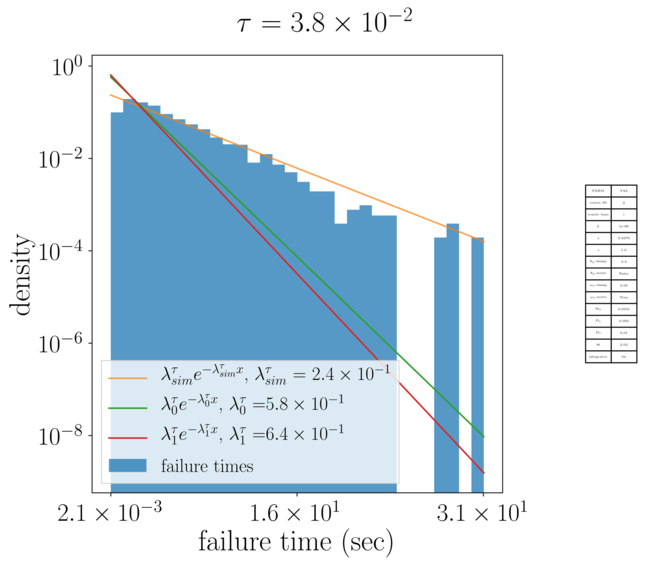}}%
	\end{subfigure}
	\begin{subfigure}[t]{\failurewidth\linewidth}%
		\centering
		\resizebox{\linewidth}{!}{\includegraphics[trim={0 0 3cm 1.5cm}, clip=true]{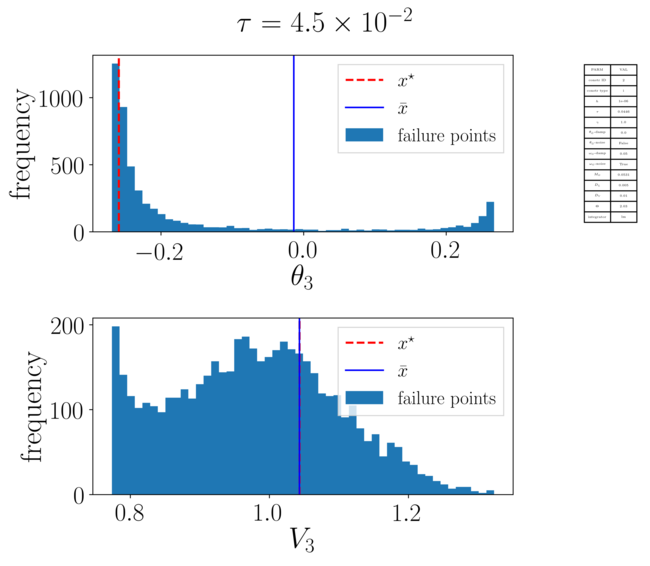}}
	\end{subfigure}
	\addtocounter{subfigure}{-4}
	\begin{subfigure}[t]{\failurewidth\linewidth}
		\vspace{\failurecaption}%
		\renewcommand\thesubfigure{V.\alph{subfigure}}
		\caption{}
	\end{subfigure}%
	\hfill%
	\begin{subfigure}[t]{\failurewidth\linewidth}
		\vspace{\failurecaption}%
		\renewcommand\thesubfigure{V.\alph{subfigure}}
		\caption{}
	\end{subfigure}%
	\hfill%
	\begin{subfigure}[t]{\failurewidth\linewidth}
		\vspace{\failurecaption}%
		\renewcommand\thesubfigure{V.\alph{subfigure}}
		\caption{}
	\end{subfigure}%
	\hfill%
	\begin{subfigure}[t]{\failurewidth\linewidth}
		\vspace{\failurecaption}%
		\renewcommand\thesubfigure{V.\alph{subfigure}}
		\caption{}
	\end{subfigure}%
	\hfill%
	\label{fig:rate-line-H}
\end{minipage}
	\vspace{\failurecaption}
	\caption{\small{Unconditional line failure diagnostics for generator-load lines: line 29 (I), line 25 (II), line 37 (III), line 16, and slack-load line 2 (V).}
		\small{(a) Failure rate comparison across temperature.}
		\small{(b) Asymptotic ratio across temperature.}
		\small{(c) (Scaled) histogram of unconditional failure times at lowest temperature tested.}
		\small{(d) Histogram of $x_i^{\star}$ where $i$ corresponds to the bus index of (one of) the line's connecting buses; shown at lowest temperature tested.}
	}
	\label{fig:rate-line-defgh}
\end{figure}

\subsection{Equilibration experiments}
\label{sm:kmc-reeq}
In this section, we present supplementary experiments for our study of the dynamical equilibration between Markov states.
%
\newcommand{\DepthA}{1}
\newcommand{\DepthAFirst}{kmc-reeq-0-0_01}
\newcommand{\DepthASecond}{kmc-reeq-0-0_0075}
\newcommand{\DepthAThird}{kmc-reeq-0-0_001}
\newcommand{\DepthAFourth}{kmc-reeq-0-0_00075}
\newcommand{\DepthANsims}{5,000}
\newcommand{\DepthB}{2}
\newcommand{\DepthBFirst}{kmc-reeq-1-0_01}
\newcommand{\DepthBSecond}{kmc-reeq-1-0_001}
\newcommand{\DepthBThird}{kmc-reeq-1-2_5e-5}
\newcommand{\DepthBFourth}{kmc-reeq-1-4_0e-6}
\newcommand{\DepthBNsims}{5,000}
\newcommand{\DepthC}{4}
\newcommand{\DepthCFirst}{kmc-reeq-3-0_01}
\newcommand{\DepthCSecond}{kmc-reeq-3-0_0025}
\newcommand{\DepthCThird}{kmc-reeq-3-1_0e-5}
\newcommand{\DepthCFourth}{kmc-reeq-3-1_0e-6}
\newcommand{\DepthCNsims}{5,000}
\newcommand{\DepthD}{6}
\newcommand{\DepthDFirst}{kmc-reeq-5-0_01}
\newcommand{\DepthDSecond}{kmc-reeq-5-0_001}
\newcommand{\DepthDThird}{kmc-reeq-5-5_0e-5}
\newcommand{\DepthDFourth}{kmc-reeq-5-2_5e-5}
\newcommand{\DepthDNsims}{5,000}
\newcommand{\DepthEE}{8}
\newcommand{\DepthEEFirst}{kmc-reeq-7-0_01}
\newcommand{\DepthEESecond}{kmc-reeq-7-0_001}
\newcommand{\DepthEEThird}{kmc-reeq-7-0_0001}
\newcommand{\DepthEEFourth}{kmc-reeq-7-5_0e-5}
\newcommand{\DepthEENsims}{5,000}
\newcommand{\DepthE}{10}
\newcommand{\DepthEFirst}{kmc-reeq-9-0_01}
\newcommand{\DepthESecond}{kmc-reeq-9-0_001}
\newcommand{\DepthEThird}{kmc-reeq-9-0_00075}
\newcommand{\DepthEFourth}{kmc-reeq-9-0_0005}
\newcommand{\DepthENsims}{5,000}
\newcommand{\DepthF}{1}
\newcommand{\DepthFFirst}{kmc-reeq-11-0_01}
\newcommand{\DepthFSecond}{kmc-reeq-11-0_001}
\newcommand{\DepthFThird}{kmc-reeq-11-0_00025}
\newcommand{\DepthFFourth}{kmc-reeq-11-0_0001}
\newcommand{\DepthFNsims}{5,000}
\newcommand{\DepthG}{1}
\newcommand{\DepthGFirst}{kmc-reeq-20-0_1}
\newcommand{\DepthGSecond}{kmc-reeq-20-0_01}
\newcommand{\DepthGThird}{kmc-reeq-20-0_005}
\newcommand{\DepthGFourth}{kmc-reeq-20-0_0025}
\newcommand{\DepthGNsims}{5,000}

\begin{figure}[H]
\begin{minipage}{\linewidth}
\begin{subfigure}[t]{\failurewidth\linewidth}%
	\centering
	\resizebox{\linewidth}{!}{\includegraphics[trim={0 0 3cm 0}, clip=true]{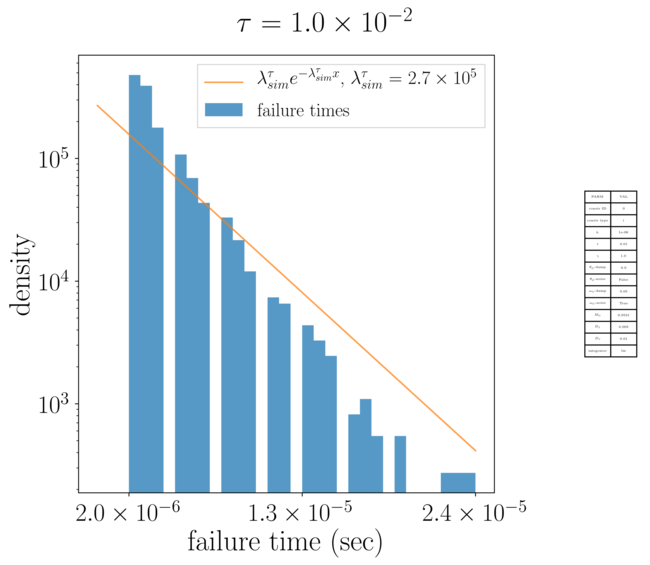}}%
\end{subfigure}
\begin{subfigure}[t]{\failurewidth\linewidth}%
	\centering
	\resizebox{\linewidth}{!}{\includegraphics[trim={0 0 3cm 0}, clip=true]{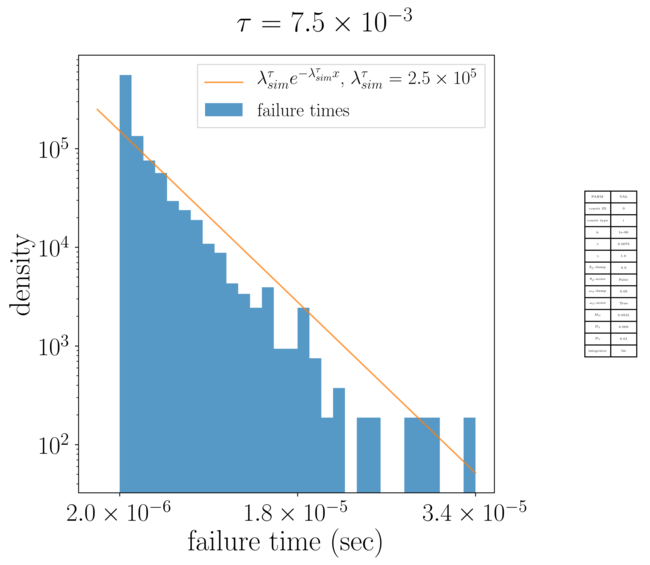}}%
\end{subfigure}
\begin{subfigure}[t]{\failurewidth\linewidth}%
	\centering
	\resizebox{\linewidth}{!}{\includegraphics[trim={0 0 3cm 0}, clip=true]{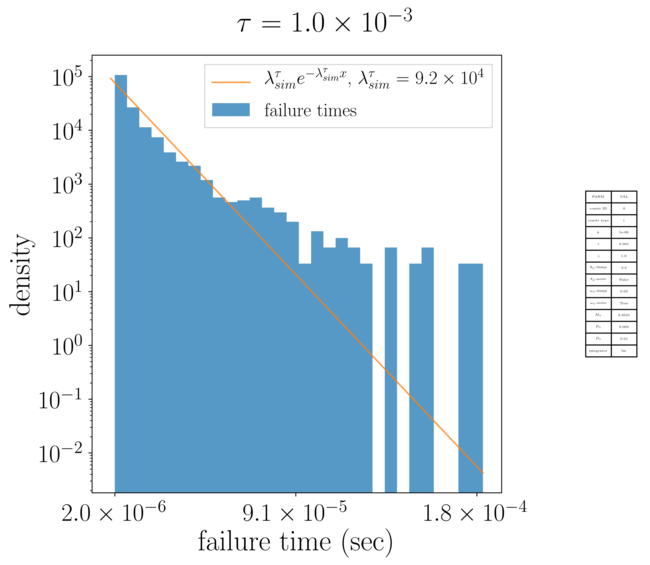}}%
\end{subfigure}
\begin{subfigure}[t]{\failurewidth\linewidth}%
	\centering
		\resizebox{\linewidth}{!}{\includegraphics[trim={0 0 3cm 0}, clip=true]{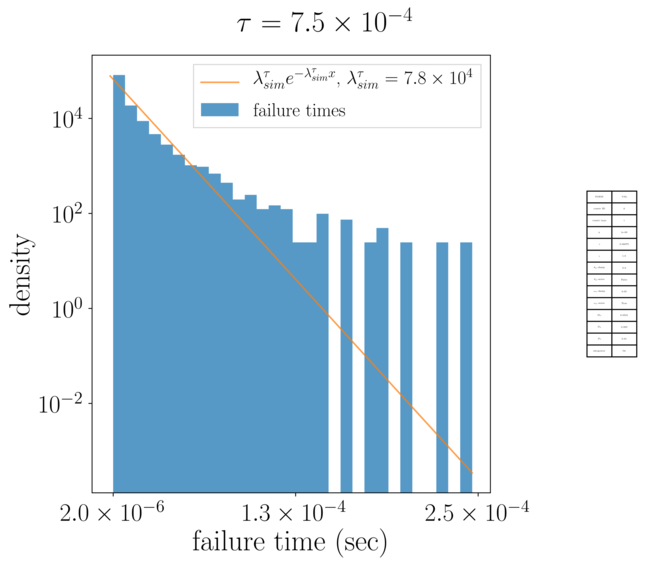}}%
\end{subfigure}
\begin{subfigure}[t]{\failurewidth\linewidth}
	\vspace{\failurecaption}%
	\renewcommand\thesubfigure{I.\alph{subfigure}}
	\caption{}
\end{subfigure}%
\hfill%
\begin{subfigure}[t]{\failurewidth\linewidth}
	\vspace{\failurecaption}%
	\renewcommand\thesubfigure{I.\alph{subfigure}}
	\caption{}
\end{subfigure}%
\hfill%
\begin{subfigure}[t]{\failurewidth\linewidth}
	\vspace{\failurecaption}%
	\renewcommand\thesubfigure{I.\alph{subfigure}}
	\caption{}
\end{subfigure}%
\hfill%
\begin{subfigure}[t]{\failurewidth\linewidth}
	\vspace{\failurecaption}%
	\renewcommand\thesubfigure{I.\alph{subfigure}}
	\caption{}
\end{subfigure}%
\hfill%
\label{fig:kmc-depth-A}
\end{minipage}
\begin{minipage}{\linewidth}
	\begin{subfigure}[t]{\failurewidth\linewidth}%
		\centering
		\resizebox{\linewidth}{!}{\includegraphics[trim={0 0 3cm 0}, clip=true]{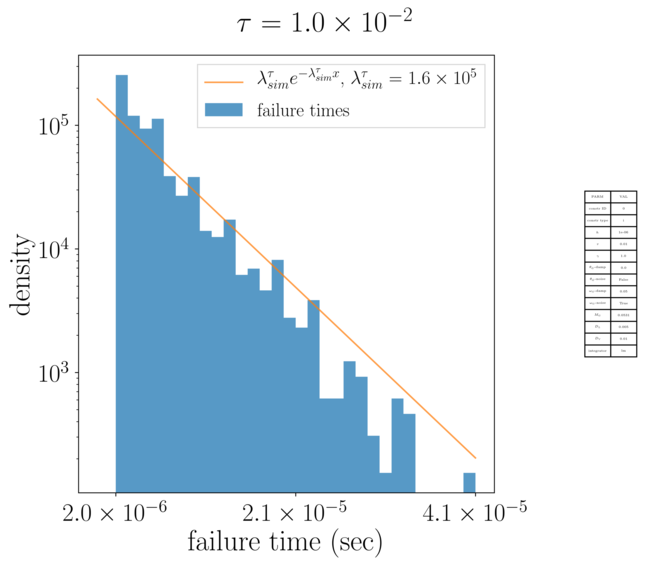}}%
	\end{subfigure}
	\begin{subfigure}[t]{\failurewidth\linewidth}%
		\centering
		\resizebox{\linewidth}{!}{\includegraphics[trim={0 0 3cm 0}, clip=true]{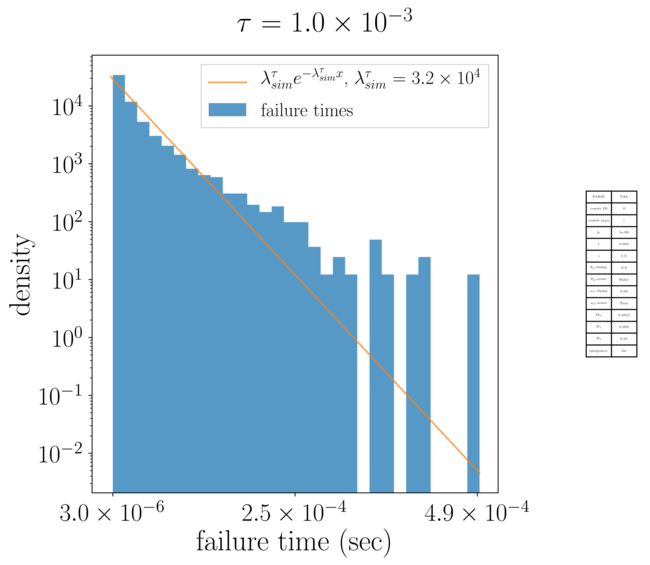}}%
	\end{subfigure}
	\begin{subfigure}[t]{\failurewidth\linewidth}%
		\centering
		\resizebox{\linewidth}{!}{\includegraphics[trim={0 0 3cm 0}, clip=true]{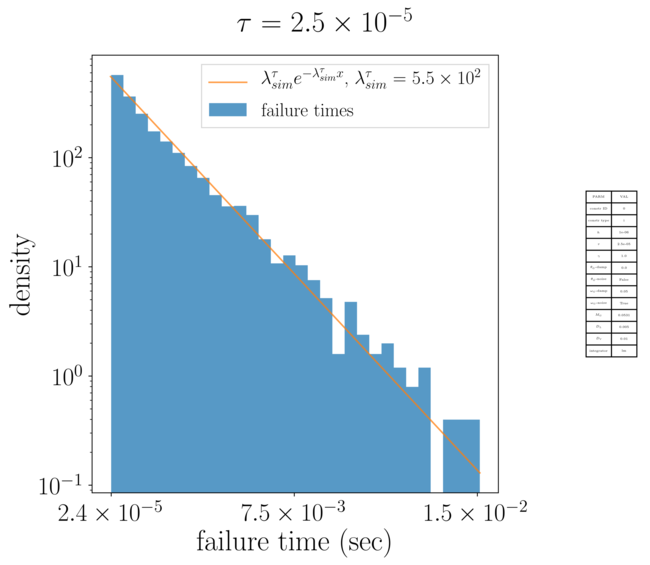}}%
	\end{subfigure}
	\begin{subfigure}[t]{\failurewidth\linewidth}%
		\centering
		\resizebox{\linewidth}{!}{\includegraphics[trim={0 0 3cm 0}, clip=true]{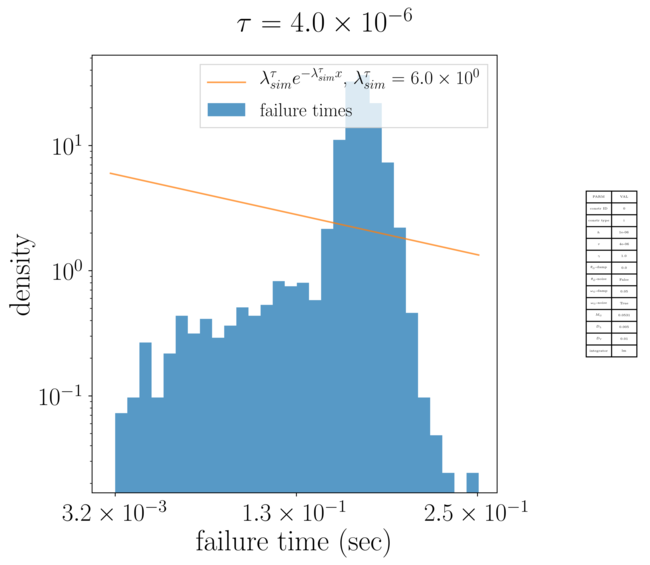}}%
	\end{subfigure}
	\addtocounter{subfigure}{-4}
	\begin{subfigure}[t]{\failurewidth\linewidth}
		\vspace{\failurecaption}%
		\renewcommand\thesubfigure{II.\alph{subfigure}}
		\caption{}
	\end{subfigure}%
	\hfill%
	\begin{subfigure}[t]{\failurewidth\linewidth}
		\vspace{\failurecaption}%
		\renewcommand\thesubfigure{II.\alph{subfigure}}
		\caption{}
	\end{subfigure}%
	\hfill%
	\begin{subfigure}[t]{\failurewidth\linewidth}
		\vspace{\failurecaption}%
		\renewcommand\thesubfigure{II.\alph{subfigure}}
		\caption{}
	\end{subfigure}%
	\hfill%
	\begin{subfigure}[t]{\failurewidth\linewidth}
		\vspace{\failurecaption}%
		\renewcommand\thesubfigure{II.\alph{subfigure}}
		\caption{}
	\end{subfigure}%
	\hfill%
	\label{fig:kmc-depth-B}
\end{minipage}
\begin{minipage}{\linewidth}
	\begin{subfigure}[t]{\failurewidth\linewidth}%
		\centering
		\resizebox{\linewidth}{!}{\includegraphics[trim={0 0 3cm 0}, clip=true]{\DepthCFirst}}%
	\end{subfigure}
	\begin{subfigure}[t]{\failurewidth\linewidth}%
		\centering
		\resizebox{\linewidth}{!}{\includegraphics[trim={0 0 3cm 0}, clip=true]{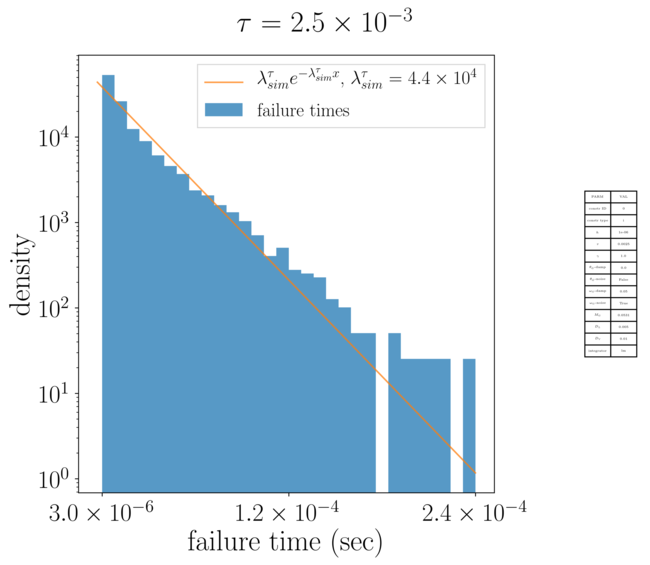}}%
	\end{subfigure}
	\begin{subfigure}[t]{\failurewidth\linewidth}%
		\centering
		\resizebox{\linewidth}{!}{\includegraphics[trim={0 0 3cm 0}, clip=true]{\DepthCThird}}%
	\end{subfigure}
	\begin{subfigure}[t]{\failurewidth\linewidth}%
		\centering
		\resizebox{\linewidth}{!}{\includegraphics[trim={0 0 3cm 0}, clip=true]{\DepthCFourth}}%
	\end{subfigure}
	\addtocounter{subfigure}{-4}
	\begin{subfigure}[t]{\failurewidth\linewidth}
		\vspace{\failurecaption}%
		\renewcommand\thesubfigure{III.\alph{subfigure}}
		\caption{}
	\end{subfigure}%
	\hfill%
	\begin{subfigure}[t]{\failurewidth\linewidth}
		\vspace{\failurecaption}%
		\renewcommand\thesubfigure{III.\alph{subfigure}}
		\caption{}
	\end{subfigure}%
	\hfill%
	\begin{subfigure}[t]{\failurewidth\linewidth}
		\vspace{\failurecaption}%
		\renewcommand\thesubfigure{III.\alph{subfigure}}
		\caption{}
	\end{subfigure}%
	\hfill%
	\begin{subfigure}[t]{\failurewidth\linewidth}
		\vspace{\failurecaption}%
		\renewcommand\thesubfigure{III.\alph{subfigure}}
		\caption{}
	\end{subfigure}%
	\hfill%
	\label{fig:kmc-depth-C}
\end{minipage}
\begin{minipage}{\linewidth}
	\begin{subfigure}[t]{\failurewidth\linewidth}%
		\centering
		\resizebox{\linewidth}{!}{\includegraphics[trim={0 0 3cm 0}, clip=true]{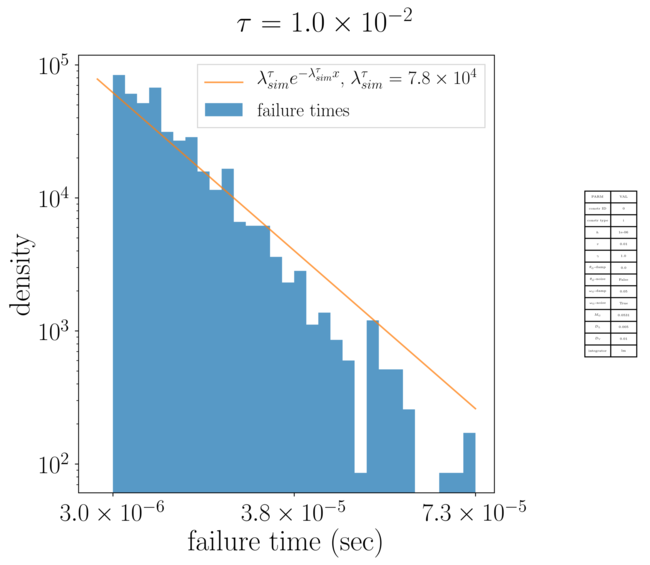}}%
	\end{subfigure}
	\begin{subfigure}[t]{\failurewidth\linewidth}%
		\centering
		\resizebox{\linewidth}{!}{\includegraphics[trim={0 0 3cm 0}, clip=true]{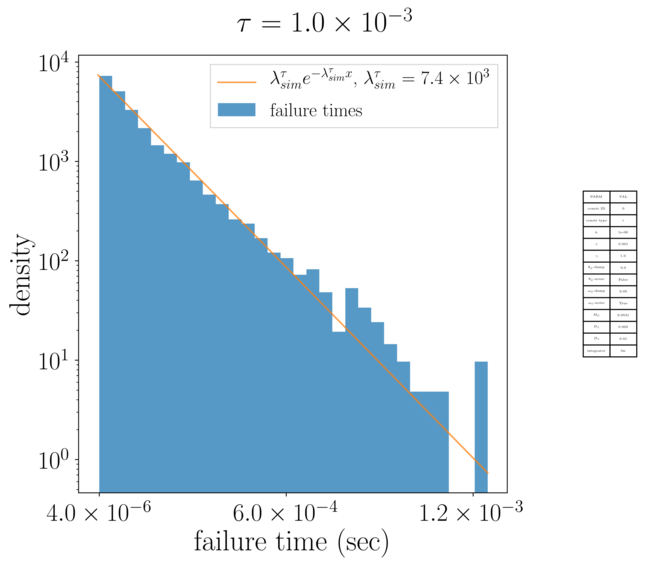}}%
	\end{subfigure}
	\begin{subfigure}[t]{\failurewidth\linewidth}%
		\centering
		\resizebox{\linewidth}{!}{\includegraphics[trim={0 0 3cm 0}, clip=true]{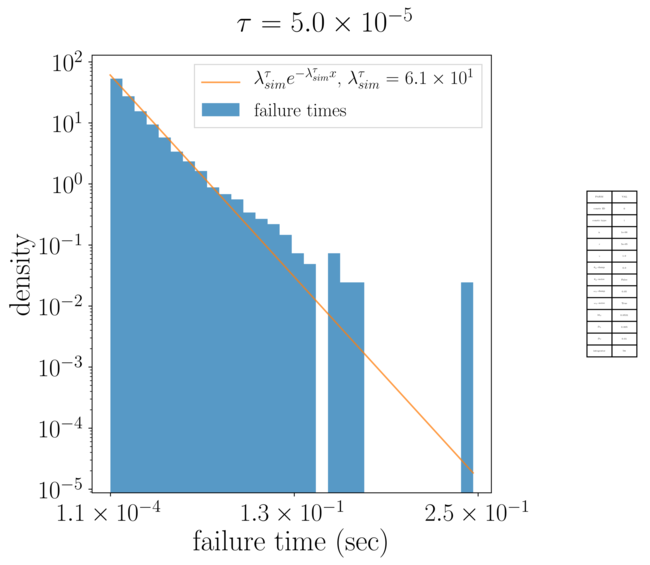}}%
	\end{subfigure}
	\begin{subfigure}[t]{\failurewidth\linewidth}%
		\centering
		\resizebox{\linewidth}{!}{\includegraphics[trim={0 0 3cm 0}, clip=true]{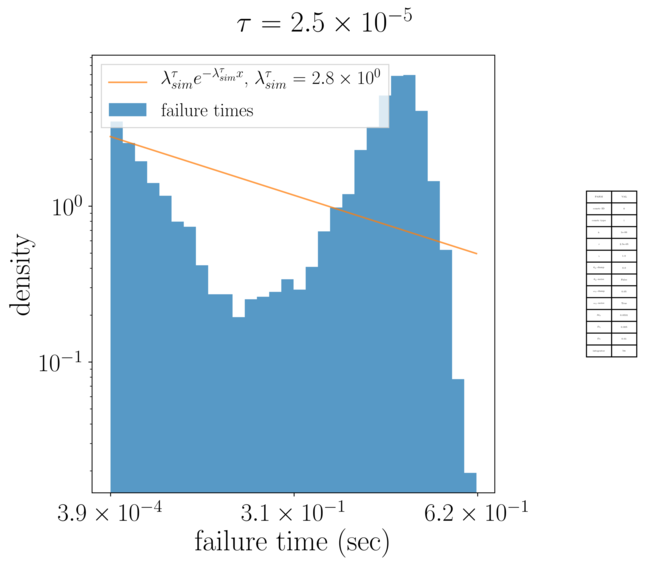}}%
	\end{subfigure}
	\addtocounter{subfigure}{-4}
	\begin{subfigure}[t]{\failurewidth\linewidth}
		\vspace{\failurecaption}%
		\renewcommand\thesubfigure{IV.\alph{subfigure}}
		\caption{}
	\end{subfigure}%
	\hfill%
	\begin{subfigure}[t]{\failurewidth\linewidth}
		\vspace{\failurecaption}%
		\renewcommand\thesubfigure{IV.\alph{subfigure}}
		\caption{}
	\end{subfigure}%
	\hfill%
	\begin{subfigure}[t]{\failurewidth\linewidth}
		\vspace{\failurecaption}%
		\renewcommand\thesubfigure{IV.\alph{subfigure}}
		\caption{}
	\end{subfigure}%
	\hfill%
	\begin{subfigure}[t]{\failurewidth\linewidth}
		\vspace{\failurecaption}%
		\renewcommand\thesubfigure{IV.\alph{subfigure}}
		\caption{}
	\end{subfigure}%
	\hfill%
	\label{fig:kmc-depth-D}
\end{minipage}
\end{figure}
\begin{figure}[H]\ContinuedFloat
\begin{minipage}{\linewidth}
	\begin{subfigure}[t]{\failurewidth\linewidth}%
		\centering
		\resizebox{\linewidth}{!}{\includegraphics[trim={0 0 3cm 0}, clip=true]{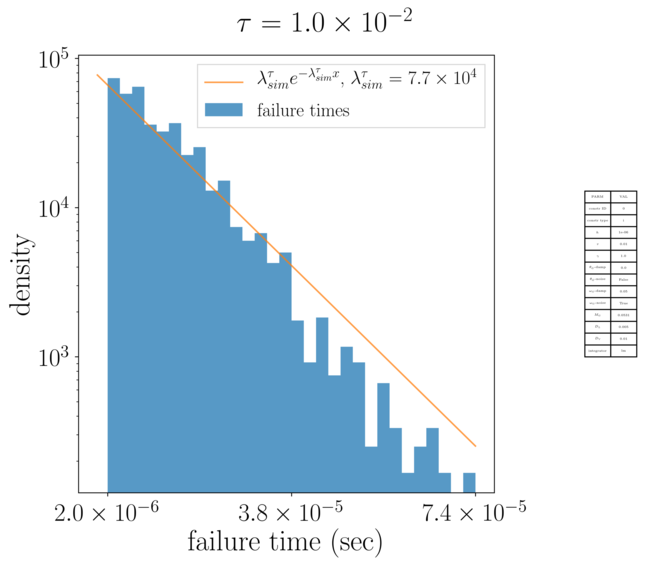}}%
	\end{subfigure}
	\begin{subfigure}[t]{\failurewidth\linewidth}%
		\centering
		\resizebox{\linewidth}{!}{\includegraphics[trim={0 0 3cm 0}, clip=true]{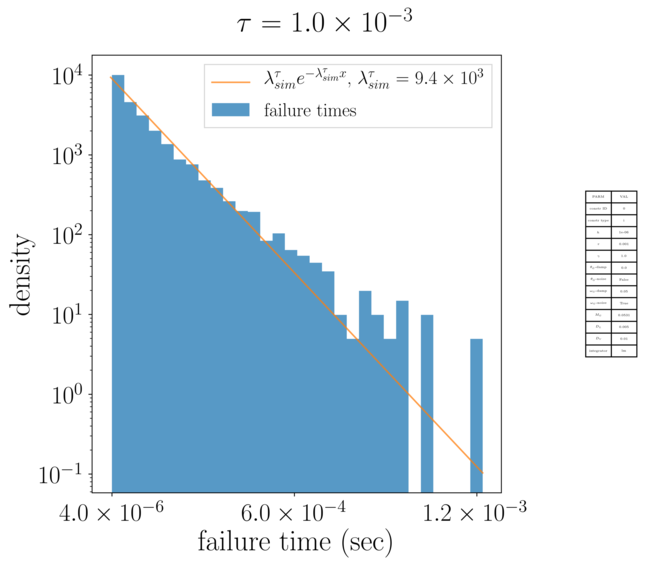}}%
	\end{subfigure}
	\begin{subfigure}[t]{\failurewidth\linewidth}%
		\centering
		\resizebox{\linewidth}{!}{\includegraphics[trim={0 0 3cm 0}, clip=true]{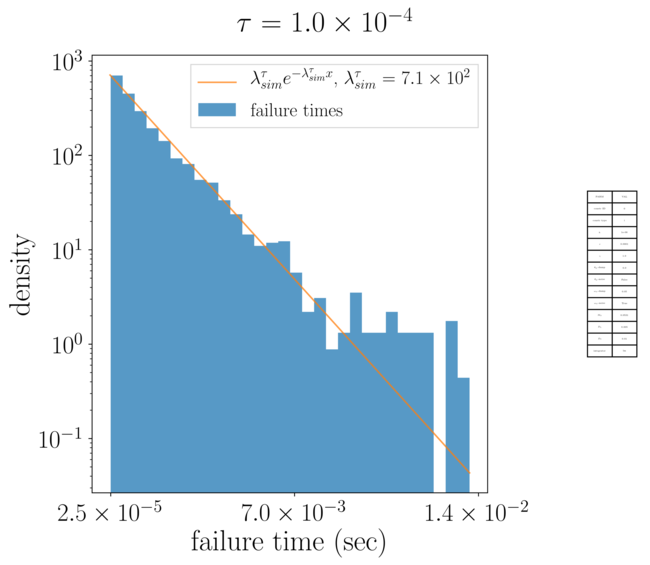}}%
	\end{subfigure}
	\begin{subfigure}[t]{\failurewidth\linewidth}%
		\centering
		\resizebox{\linewidth}{!}{\includegraphics[trim={0 0 3cm 0}, clip=true]{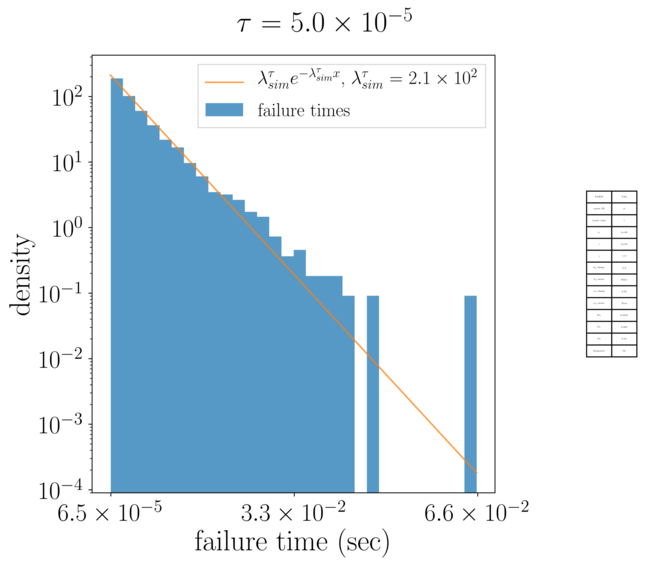}}%
	\end{subfigure}
	\addtocounter{subfigure}{-4}
	\begin{subfigure}[t]{\failurewidth\linewidth}
		\vspace{\failurecaption}%
		\renewcommand\thesubfigure{V.\alph{subfigure}}
		\caption{}
	\end{subfigure}%
	\hfill%
	\begin{subfigure}[t]{\failurewidth\linewidth}
		\vspace{\failurecaption}%
		\renewcommand\thesubfigure{V.\alph{subfigure}}
		\caption{}
	\end{subfigure}%
	\hfill%
	\begin{subfigure}[t]{\failurewidth\linewidth}
		\vspace{\failurecaption}%
		\renewcommand\thesubfigure{V.\alph{subfigure}}
		\caption{}
	\end{subfigure}%
	\hfill%
	\begin{subfigure}[t]{\failurewidth\linewidth}
		\vspace{\failurecaption}%
		\renewcommand\thesubfigure{V.\alph{subfigure}}
		\caption{}
	\end{subfigure}%
	\hfill%
	\label{fig:kmc-depth-EE}
\end{minipage}
\begin{minipage}{\linewidth}
	\begin{subfigure}[t]{\failurewidth\linewidth}%
		\centering
		\resizebox{\linewidth}{!}{\includegraphics[trim={0 0 3cm 0}, clip=true]{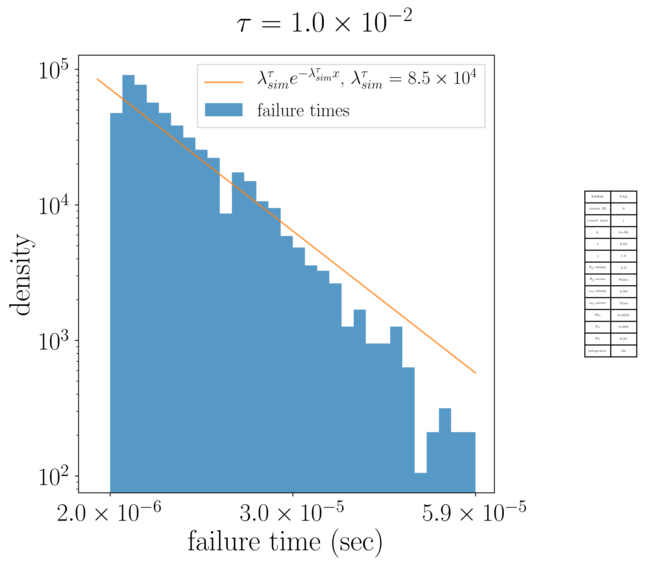}}%
	\end{subfigure}
	\begin{subfigure}[t]{\failurewidth\linewidth}%
		\centering
		\resizebox{\linewidth}{!}{\includegraphics[trim={0 0 3cm 0}, clip=true]{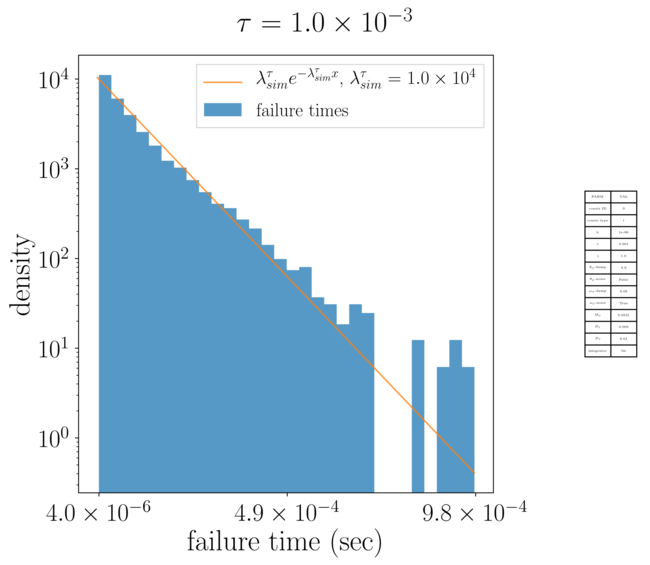}}%
	\end{subfigure}
	\begin{subfigure}[t]{\failurewidth\linewidth}%
		\centering
		\resizebox{\linewidth}{!}{\includegraphics[trim={0 0 3cm 0}, clip=true]{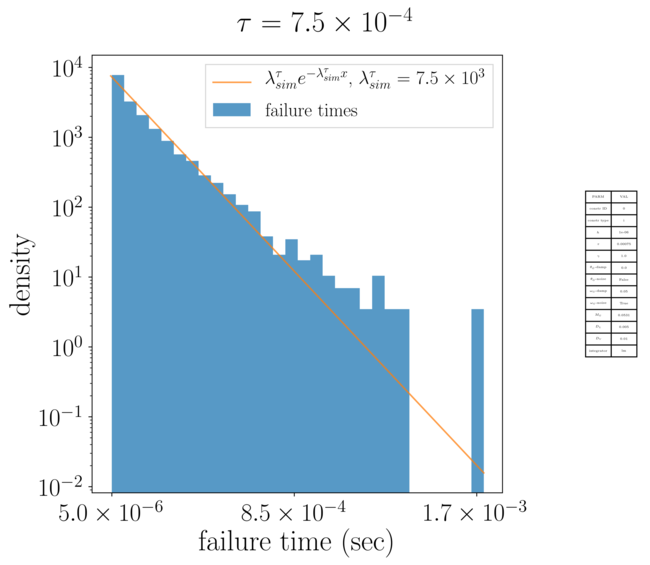}}%
	\end{subfigure}
	\begin{subfigure}[t]{\failurewidth\linewidth}%
		\centering
		\resizebox{\linewidth}{!}{\includegraphics[trim={0 0 3cm 0}, clip=true]{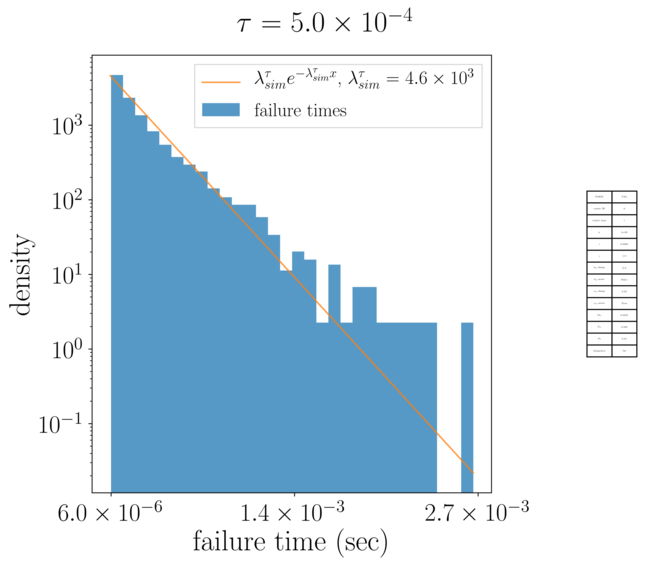}}%
	\end{subfigure}
	\addtocounter{subfigure}{-4}
	\begin{subfigure}[t]{\failurewidth\linewidth}
		\vspace{\failurecaption}%
		\renewcommand\thesubfigure{VI.\alph{subfigure}}
		\caption{}
	\end{subfigure}%
	\hfill%
	\begin{subfigure}[t]{\failurewidth\linewidth}
		\vspace{\failurecaption}%
		\renewcommand\thesubfigure{VI.\alph{subfigure}}
		\caption{}
	\end{subfigure}%
	\hfill%
	\begin{subfigure}[t]{\failurewidth\linewidth}
		\vspace{\failurecaption}%
		\renewcommand\thesubfigure{VI.\alph{subfigure}}
		\caption{}
	\end{subfigure}%
	\hfill%
	\begin{subfigure}[t]{\failurewidth\linewidth}
		\vspace{\failurecaption}%
		\renewcommand\thesubfigure{VI.\alph{subfigure}}
		\caption{}
	\end{subfigure}%
	\hfill%
	\label{fig:kmc-depth-E}
\end{minipage}
\begin{minipage}{\linewidth}
	\begin{subfigure}[t]{\failurewidth\linewidth}%
		\centering
		\resizebox{\linewidth}{!}{\includegraphics[trim={0 0 3cm 0}, clip=true]{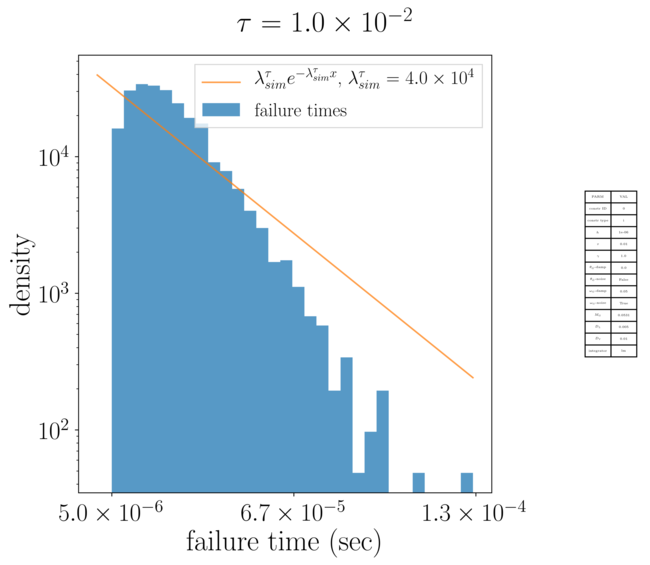}}%
	\end{subfigure}
	\begin{subfigure}[t]{\failurewidth\linewidth}%
		\centering
		\resizebox{\linewidth}{!}{\includegraphics[trim={0 0 3cm 0}, clip=true]{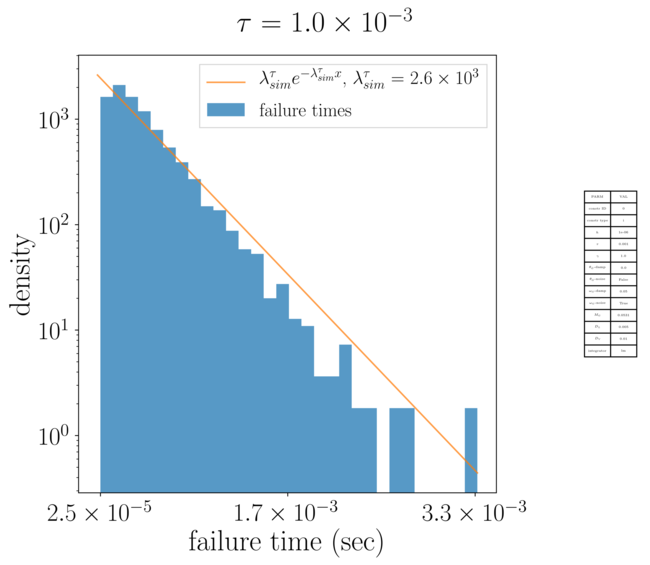}}%
	\end{subfigure}
	\begin{subfigure}[t]{\failurewidth\linewidth}%
		\centering
		\resizebox{\linewidth}{!}{\includegraphics[trim={0 0 3cm 0}, clip=true]{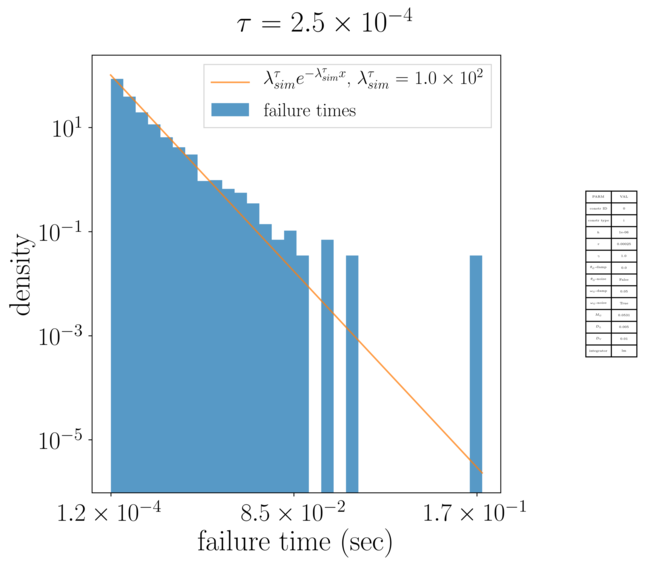}}%
	\end{subfigure}
	\begin{subfigure}[t]{\failurewidth\linewidth}%
		\centering
		\resizebox{\linewidth}{!}{\includegraphics[trim={0 0 3cm 0}, clip=true]{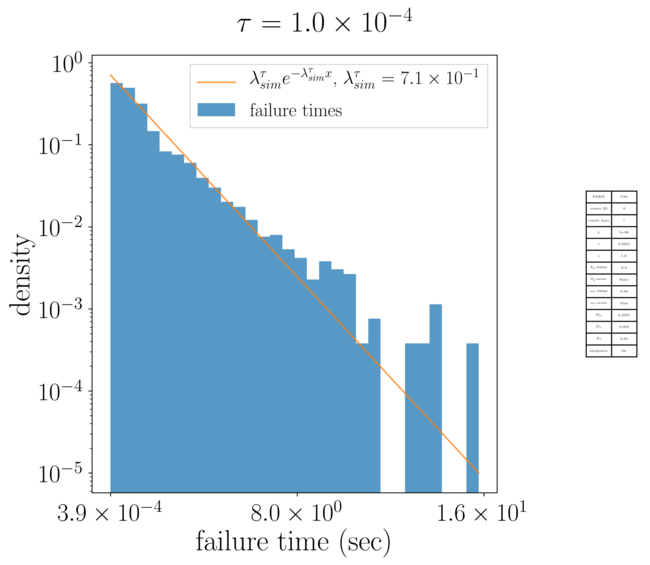}}%
	\end{subfigure}
	\addtocounter{subfigure}{-4}
	\begin{subfigure}[t]{\failurewidth\linewidth}
		\vspace{\failurecaption}%
		\renewcommand\thesubfigure{VII.\alph{subfigure}}
		\caption{}
	\end{subfigure}%
	\hfill%
	\begin{subfigure}[t]{\failurewidth\linewidth}
		\vspace{\failurecaption}%
		\renewcommand\thesubfigure{VII.\alph{subfigure}}
		\caption{}
	\end{subfigure}%
	\hfill%
	\begin{subfigure}[t]{\failurewidth\linewidth}
		\vspace{\failurecaption}%
		\renewcommand\thesubfigure{VII.\alph{subfigure}}
		\caption{}
	\end{subfigure}%
	\hfill%
	\begin{subfigure}[t]{\failurewidth\linewidth}
		\vspace{\failurecaption}%
		\renewcommand\thesubfigure{VII.\alph{subfigure}}
		\caption{}
	\end{subfigure}%
	\hfill%
	\label{fig:kmc-depth-F}
\end{minipage}
\begin{minipage}{\linewidth}
	\begin{subfigure}[t]{\failurewidth\linewidth}%
		\centering
		\resizebox{\linewidth}{!}{\includegraphics[trim={0 0 3cm 0}, clip=true]{\DepthGFirst}}%
	\end{subfigure}
	\begin{subfigure}[t]{\failurewidth\linewidth}%
		\centering
		\resizebox{\linewidth}{!}{\includegraphics[trim={0 0 3cm 0}, clip=true]{\DepthGSecond}}%
	\end{subfigure}
	\begin{subfigure}[t]{\failurewidth\linewidth}%
		\centering
		\resizebox{\linewidth}{!}{\includegraphics[trim={0 0 3cm 0}, clip=true]{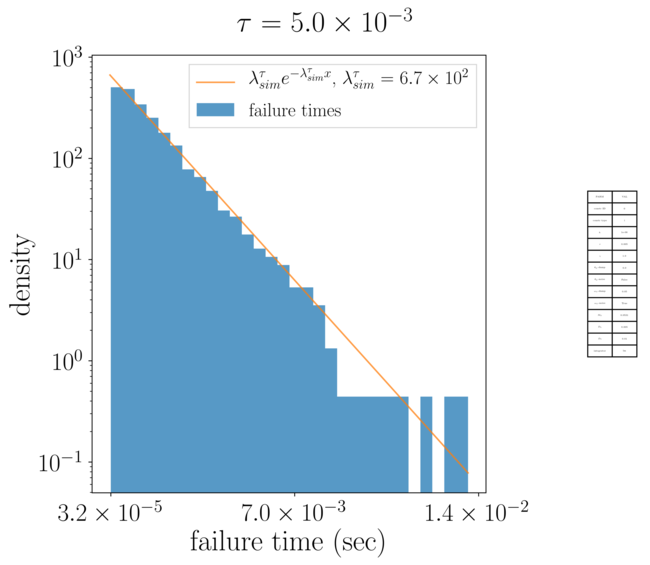}}%
	\end{subfigure}
	\begin{subfigure}[t]{\failurewidth\linewidth}%
		\centering
		\resizebox{\linewidth}{!}{\includegraphics[trim={0 0 3cm 0}, clip=true]{\DepthGFourth}}%
	\end{subfigure}
	\addtocounter{subfigure}{-4}
	\begin{subfigure}[t]{\failurewidth\linewidth}
		\vspace{\failurecaption}%
		\renewcommand\thesubfigure{VIII.\alph{subfigure}}
		\caption{}
	\end{subfigure}%
	\hfill%
	\begin{subfigure}[t]{\failurewidth\linewidth}
		\vspace{\failurecaption}%
		\renewcommand\thesubfigure{VIII.\alph{subfigure}}
		\caption{}
	\end{subfigure}%
	\hfill%
	\begin{subfigure}[t]{\failurewidth\linewidth}
		\vspace{\failurecaption}%
		\renewcommand\thesubfigure{VIII.\alph{subfigure}}
		\caption{}
	\end{subfigure}%
	\hfill%
	\begin{subfigure}[t]{\failurewidth\linewidth}
		\vspace{\failurecaption}%
		\renewcommand\thesubfigure{VIII.\alph{subfigure}}
		\caption{}
	\end{subfigure}%
	\hfill%
	\label{fig:kmc-depth-G}
\end{minipage}
\vspace{\failurecaption}
\caption{\small{Equilibration study across failure depth. (I) depth 1, (II) depth 2, (III) depth 4, (IV) depth 6, (V) depth 8, (VI) depth 10, (VII) depth 12, (VIII) depth 21.}}
\label{fig:kmc-depth-abcdefg}
\end{figure}

\newpage
\clearpage
\newgeometry{bottom=0.35in}
\subsection{Cascade experiments}
\label{sm:kmc-cascade}
In this section, we present supplementary experiments for our study of temperature and network configuration on the Zipf relationships of cascade severity.
We observe non-monotonicity of the SESPI with increasing temperature in two regards.
First, both network configurations display oscillatory patterns of SEPSI across temperature.
Second, within each Zipf plot, the failure data also demonstrate an oscillatory pattern around the power law.
\vspace{-0.25in}
\newcommand{\CascWidth}{0.32}
\newcommand{\Aa}{kmc-118bus_dc-zipf-3600-60-zipf_cascs_gens_0_0004}
\newcommand{\Ab}{kmc-118bus_dc-zipf-3600-60-zipf_cascs_gens_0_0005}
\newcommand{\Ac}{kmc-118bus_dc-zipf-3600-60-zipf_cascs_gens_0_0006}
\newcommand{\Ba}{kmc-118bus_dc-zipf-3600-60-zipf_cascs_gens_0_0007}
\newcommand{\Bb}{kmc-118bus_dc-zipf-3600-60-zipf_cascs_gens_0_0008}
\newcommand{\Bc}{kmc-118bus_dc-zipf-3600-60-zipf_cascs_gens_0_0009}
\newcommand{\Ca}{kmc-118bus_dc-zipf-3600-60-zipf_cascs_gens_0_001}
\newcommand{\Cb}{kmc-118bus_dc-zipf-3600-60-zipf_cascs_gens_0_002}
\newcommand{\Cc}{kmc-118bus_dc-zipf-3600-60-zipf_cascs_gens_0_003}
\newcommand{\Da}{kmc-118bus_dc-zipf-3600-60-zipf_cascs_gens_0_004}
\newcommand{\Db}{kmc-118bus_dc-zipf-3600-60-zipf_cascs_gens_0_005}
\newcommand{\Dc}{kmc-118bus_dc-zipf-3600-60-zipf_cascs_gens_0_006}

\newcommand{\Ea}{kmc-118bus_n1-zipf-3600-60-zipf_cascs_gens_0_0004}
\newcommand{\Eb}{kmc-118bus_n1-zipf-3600-60-zipf_cascs_gens_0_0005}
\newcommand{\Ec}{kmc-118bus_n1-zipf-3600-60-zipf_cascs_gens_0_0006}
\newcommand{\Fa}{kmc-118bus_n1-zipf-3600-60-zipf_cascs_gens_0_0007}
\newcommand{\Fb}{kmc-118bus_n1-zipf-3600-60-zipf_cascs_gens_0_0008}
\newcommand{\Fc}{kmc-118bus_n1-zipf-3600-60-zipf_cascs_gens_0_0009}
\newcommand{\Ga}{kmc-118bus_n1-zipf-3600-60-zipf_cascs_gens_0_001}
\newcommand{\Gb}{kmc-118bus_n1-zipf-3600-60-zipf_cascs_gens_0_002}
\newcommand{\Gc}{kmc-118bus_n1-zipf-3600-60-zipf_cascs_gens_0_003}
\newcommand{\Ha}{kmc-118bus_n1-zipf-3600-60-zipf_cascs_gens_0_004}
\newcommand{\Hb}{kmc-118bus_n1-zipf-3600-60-zipf_cascs_gens_0_005}
\newcommand{\Hc}{kmc-118bus_n1-zipf-3600-60-zipf_cascs_gens_0_006}

\begin{figure}[H]
	\begin{minipage}{\linewidth}
		\begin{subfigure}[t]{\CascWidth\linewidth}%
			\centering
			\resizebox{\linewidth}{!}{\includegraphics[trim={0 0 3cm 0}, clip=true]{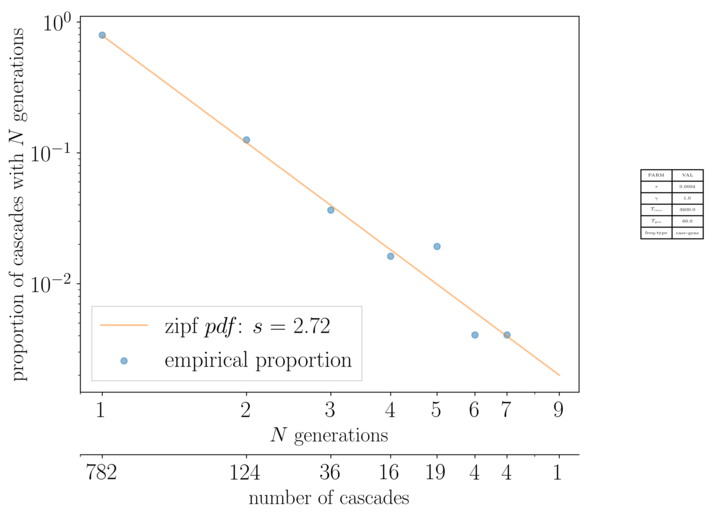}}%
		\end{subfigure}
		\begin{subfigure}[t]{\CascWidth\linewidth}%
			\centering
			\resizebox{\linewidth}{!}{\includegraphics[trim={0 0 3cm 0}, clip=true]{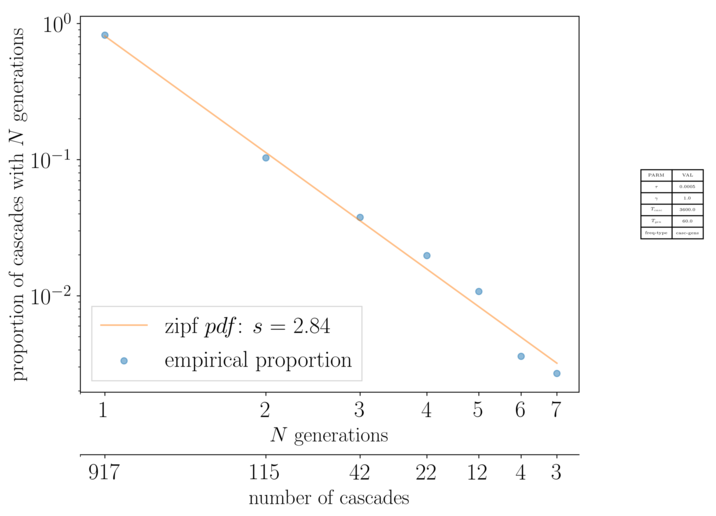}}%
		\end{subfigure}
		\begin{subfigure}[t]{\CascWidth\linewidth}%
			\centering
			\resizebox{\linewidth}{!}{\includegraphics[trim={0 0 3cm 0}, clip=true]{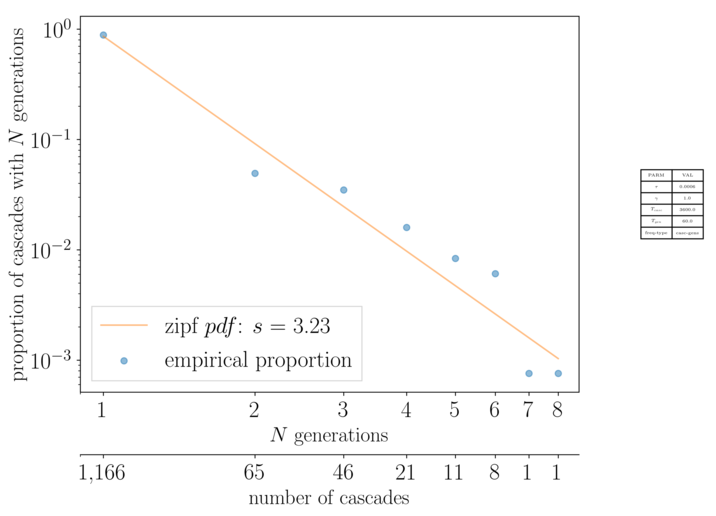}}%
		\end{subfigure}
		\begin{subfigure}[t]{\CascWidth\linewidth}
			\vspace{\failurecaption}%
			\renewcommand\thesubfigure{$\tau = 4 \times 10^{-4}$}
			\caption{}
		\end{subfigure}%
		\hfill%
		\begin{subfigure}[t]{\CascWidth\linewidth}
			\vspace{\failurecaption}%
			\renewcommand\thesubfigure{$\tau = 5 \times 10^{-4}$}
			\caption{}
		\end{subfigure}%
		\hfill%
		\begin{subfigure}[t]{\CascWidth\linewidth}
			\vspace{\failurecaption}%
			\renewcommand\thesubfigure{$\tau = 6 \times 10^{-4}$}
			\caption{}
		\end{subfigure}%
		\hfill%
		\label{fig:kmc-cascade-A}
	\end{minipage}
	\begin{minipage}{\linewidth}
	\begin{subfigure}[t]{\CascWidth\linewidth}%
		\centering
		\resizebox{\linewidth}{!}{\includegraphics[trim={0 0 3cm 0}, clip=true]{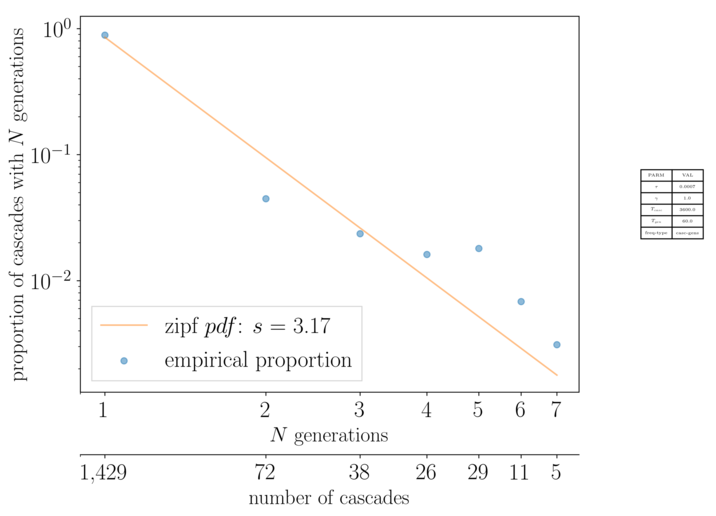}}%
	\end{subfigure}
	\begin{subfigure}[t]{\CascWidth\linewidth}%
		\centering
		\resizebox{\linewidth}{!}{\includegraphics[trim={0 0 3cm 0}, clip=true]{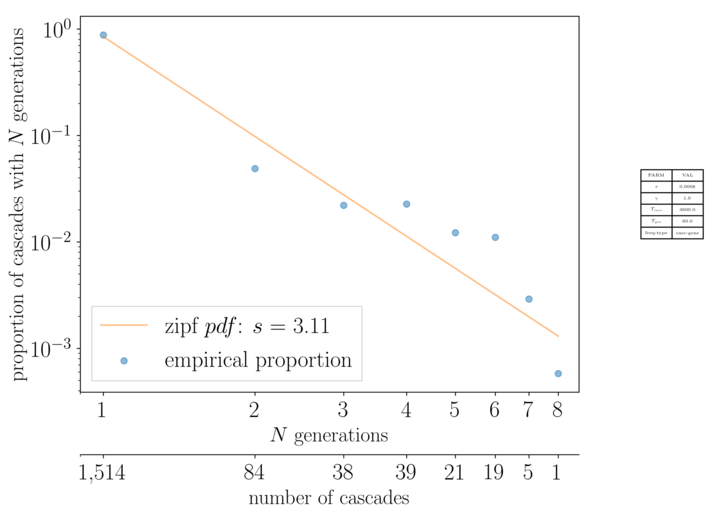}}%
	\end{subfigure}
	\begin{subfigure}[t]{\CascWidth\linewidth}%
		\centering
		\resizebox{\linewidth}{!}{\includegraphics[trim={0 0 3cm 0}, clip=true]{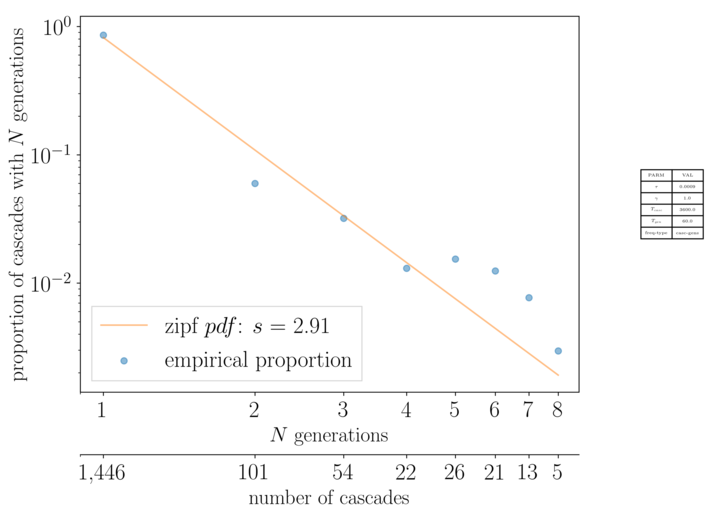}}%
	\end{subfigure}
	\begin{subfigure}[t]{\CascWidth\linewidth}
		\vspace{\failurecaption}%
		\renewcommand\thesubfigure{$\tau = 7 \times 10^{-4}$}
		\caption{}
	\end{subfigure}%
	\hfill%
	\begin{subfigure}[t]{\CascWidth\linewidth}
		\vspace{\failurecaption}%
		\renewcommand\thesubfigure{$\tau = 8 \times 10^{-4}$}
		\caption{}
	\end{subfigure}%
	\hfill%
	\begin{subfigure}[t]{\CascWidth\linewidth}
		\vspace{\failurecaption}%
		\renewcommand\thesubfigure{$\tau = 9 \times 10^{-4}$}
		\caption{}
	\end{subfigure}%
	\hfill%
	\label{fig:kmc-cascade-B}
	\end{minipage}
	\begin{minipage}{\linewidth}
	\begin{subfigure}[t]{\CascWidth\linewidth}%
		\centering
		\resizebox{\linewidth}{!}{\includegraphics[trim={0 0 3cm 0}, clip=true]{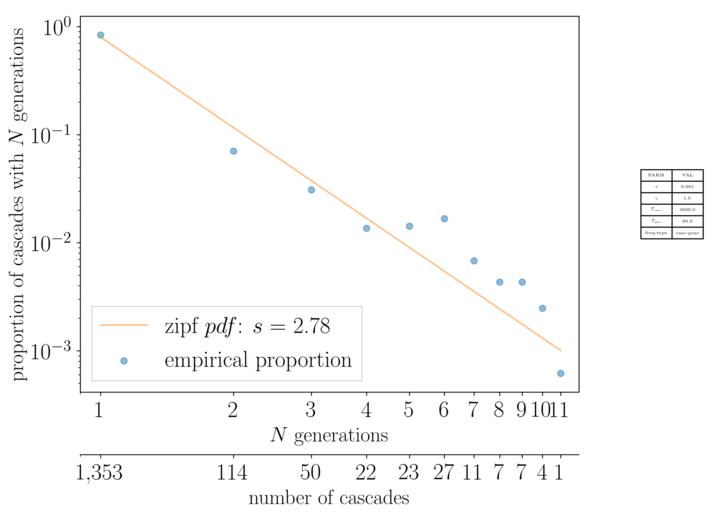}}%
	\end{subfigure}
	\begin{subfigure}[t]{\CascWidth\linewidth}%
		\centering
		\resizebox{\linewidth}{!}{\includegraphics[trim={0 0 3cm 0}, clip=true]{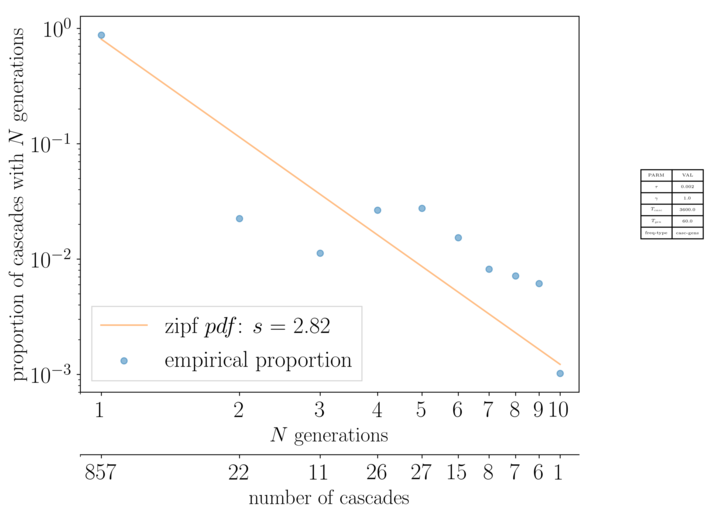}}%
	\end{subfigure}
	\begin{subfigure}[t]{\CascWidth\linewidth}%
		\centering
		\resizebox{\linewidth}{!}{\includegraphics[trim={0 0 3cm 0}, clip=true]{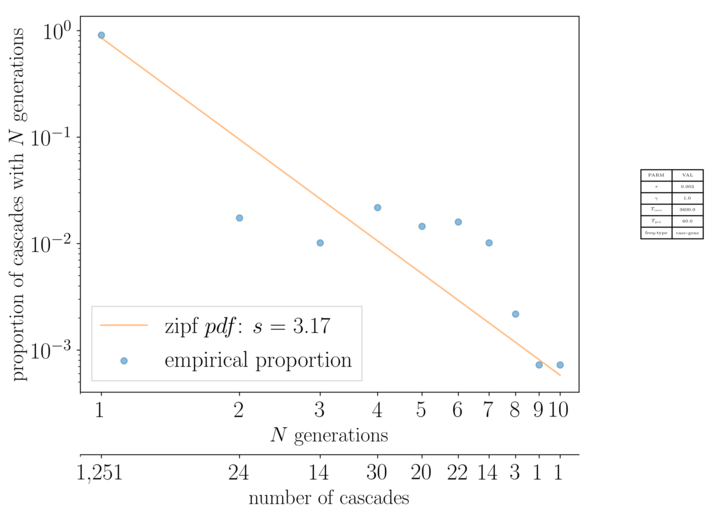}}%
	\end{subfigure}
	\begin{subfigure}[t]{\CascWidth\linewidth}
		\vspace{\failurecaption}%
		\renewcommand\thesubfigure{$\tau = 1 \times 10^{-3}$}
		\caption{}
	\end{subfigure}%
	\hfill%
	\begin{subfigure}[t]{\CascWidth\linewidth}
		\vspace{\failurecaption}%
		\renewcommand\thesubfigure{$\tau = 2 \times 10^{-3}$}
		\caption{}
	\end{subfigure}%
	\hfill%
	\begin{subfigure}[t]{\CascWidth\linewidth}
		\vspace{\failurecaption}%
		\renewcommand\thesubfigure{$\tau = 3 \times 10^{-3}$}
		\caption{}
	\end{subfigure}%
	\hfill%
	\label{fig:kmc-cascade-C}
	\end{minipage}
	\begin{minipage}{\linewidth}
	\begin{subfigure}[t]{\CascWidth\linewidth}%
		\centering
		\resizebox{\linewidth}{!}{\includegraphics[trim={0 0 3cm 0}, clip=true]{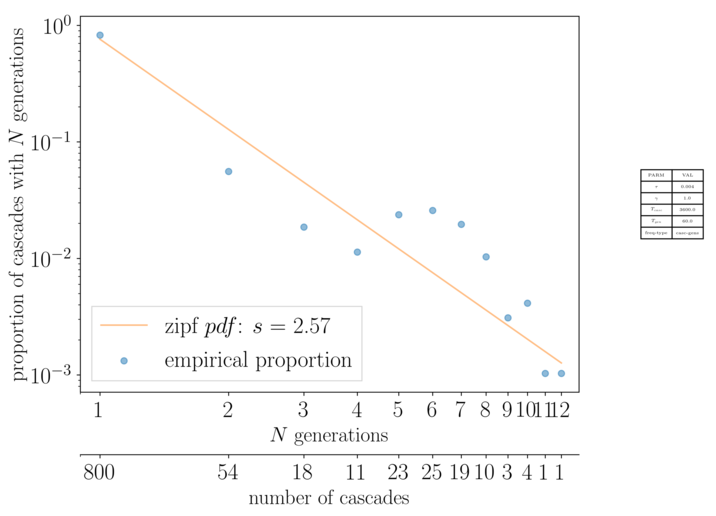}}%
	\end{subfigure}
	\begin{subfigure}[t]{\CascWidth\linewidth}%
		\centering
		\resizebox{\linewidth}{!}{\includegraphics[trim={0 0 3cm 0}, clip=true]{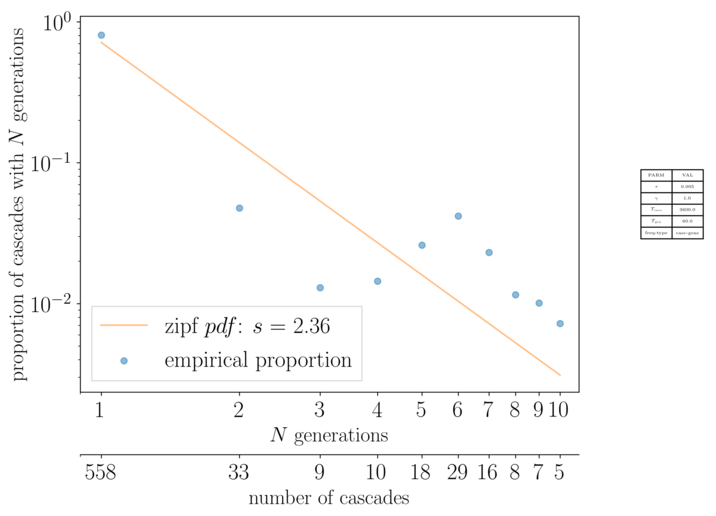}}%
	\end{subfigure}
	\begin{subfigure}[t]{\CascWidth\linewidth}%
		\centering
		\resizebox{\linewidth}{!}{\includegraphics[trim={0 0 3cm 0}, clip=true]{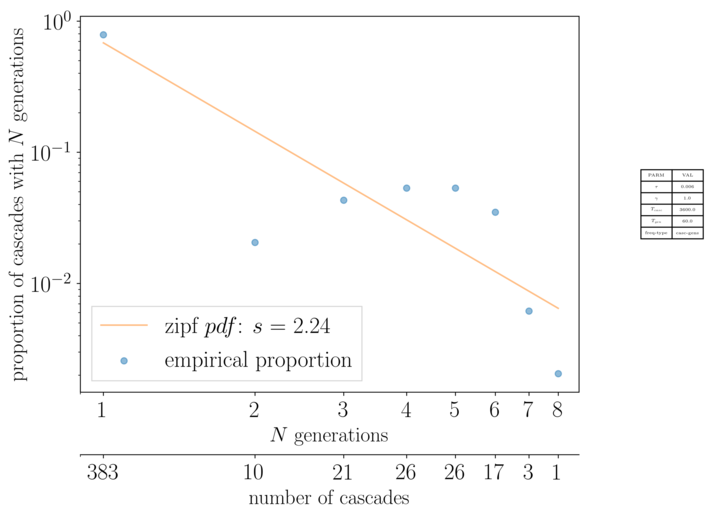}}%
	\end{subfigure}
	\begin{subfigure}[t]{\CascWidth\linewidth}
		\vspace{\failurecaption}%
		\renewcommand\thesubfigure{$\tau = 4 \times 10^{-3}$}
		\caption{}
	\end{subfigure}%
	\hfill%
	\begin{subfigure}[t]{\CascWidth\linewidth}
		\vspace{\failurecaption}%
		\renewcommand\thesubfigure{$\tau = 5 \times 10^{-3}$}
		\caption{}
	\end{subfigure}%
	\hfill%
	\begin{subfigure}[t]{\CascWidth\linewidth}
		\vspace{\failurecaption}%
		\renewcommand\thesubfigure{$\tau = 6 \times 10^{-3}$}
		\caption{}
	\end{subfigure}%
	\hfill%
	\label{fig:kmc-cascade-D}
\end{minipage}
\caption{Zipf plots of cascades with $N$ generations and DC line limits across temperature.}
\label{fig:cascade-DC-lines}
\end{figure}

\begin{figure}[H]
	\begin{minipage}{\linewidth}
		\begin{subfigure}[t]{\CascWidth\linewidth}%
			\centering
			\resizebox{\linewidth}{!}{\includegraphics[trim={0 0 3cm 0}, clip=true]{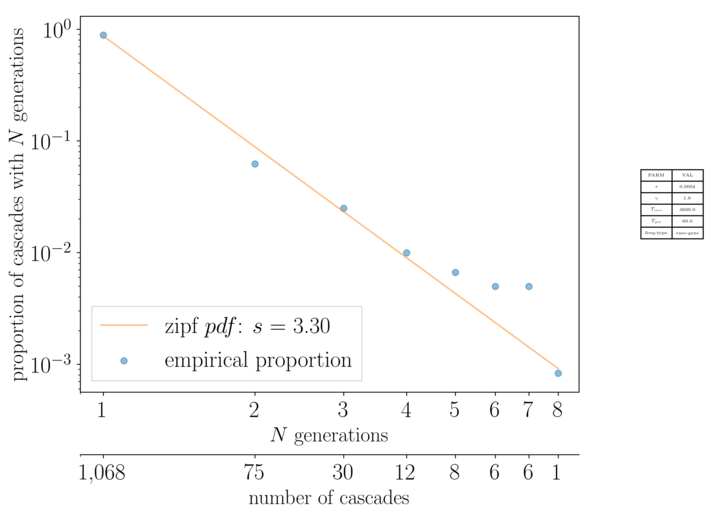}}%
		\end{subfigure}
		\begin{subfigure}[t]{\CascWidth\linewidth}%
			\centering
			\resizebox{\linewidth}{!}{\includegraphics[trim={0 0 3cm 0}, clip=true]{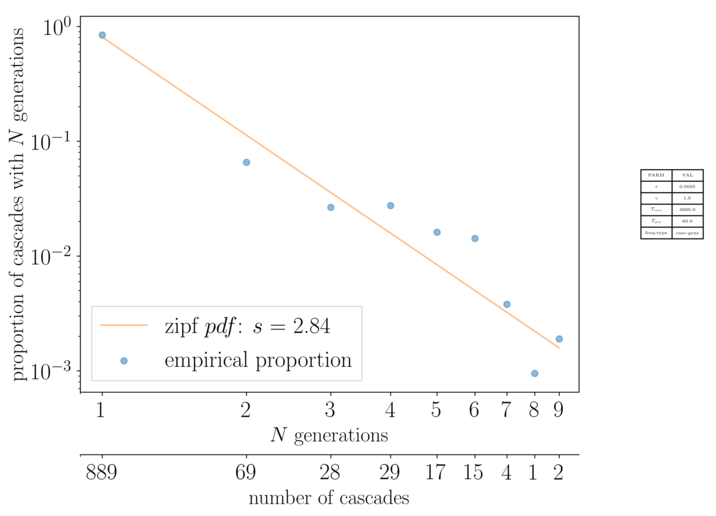}}%
		\end{subfigure}
		\begin{subfigure}[t]{\CascWidth\linewidth}%
			\centering
			\resizebox{\linewidth}{!}{\includegraphics[trim={0 0 3cm 0}, clip=true]{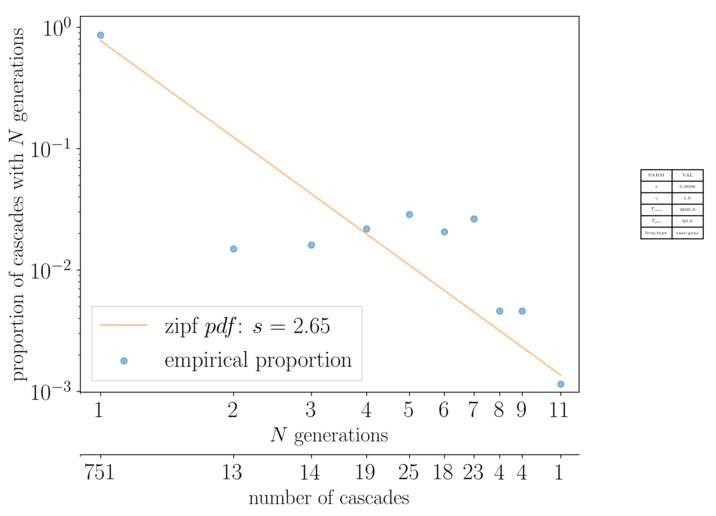}}%
		\end{subfigure}
		\begin{subfigure}[t]{\CascWidth\linewidth}
			\vspace{\failurecaption}%
			\renewcommand\thesubfigure{$\tau = 4 \times 10^{-4}$}
			\caption{}
		\end{subfigure}%
		\hfill%
		\begin{subfigure}[t]{\CascWidth\linewidth}
			\vspace{\failurecaption}%
			\renewcommand\thesubfigure{$\tau = 5 \times 10^{-4}$}
			\caption{}
		\end{subfigure}%
		\hfill%
		\begin{subfigure}[t]{\CascWidth\linewidth}
			\vspace{\failurecaption}%
			\renewcommand\thesubfigure{$\tau = 6 \times 10^{-4}$}
			\caption{}
		\end{subfigure}%
		\hfill%
		\label{fig:kmc-cascade-E}
	\end{minipage}
	\begin{minipage}{\linewidth}
		\begin{subfigure}[t]{\CascWidth\linewidth}%
			\centering
			\resizebox{\linewidth}{!}{\includegraphics[trim={0 0 3cm 0}, clip=true]{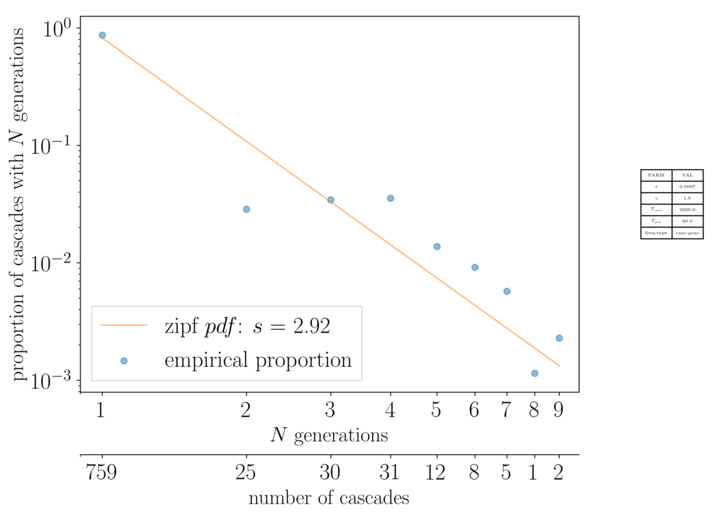}}%
		\end{subfigure}
		\begin{subfigure}[t]{\CascWidth\linewidth}%
			\centering
			\resizebox{\linewidth}{!}{\includegraphics[trim={0 0 3cm 0}, clip=true]{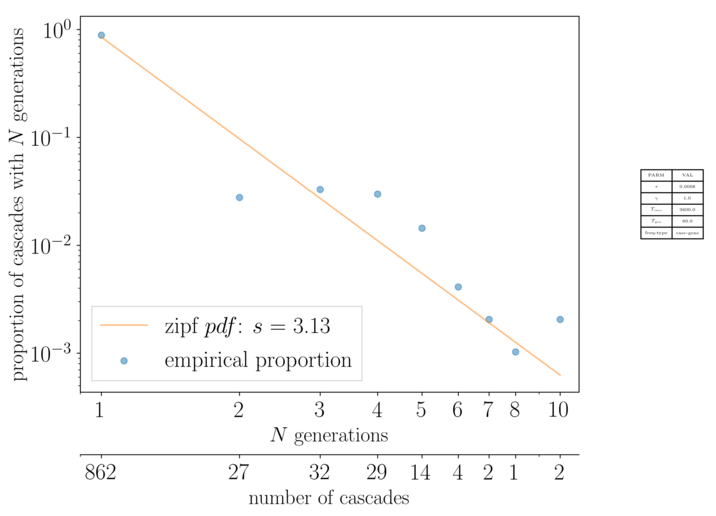}}%
		\end{subfigure}
		\begin{subfigure}[t]{\CascWidth\linewidth}%
			\centering
			\resizebox{\linewidth}{!}{\includegraphics[trim={0 0 3cm 0}, clip=true]{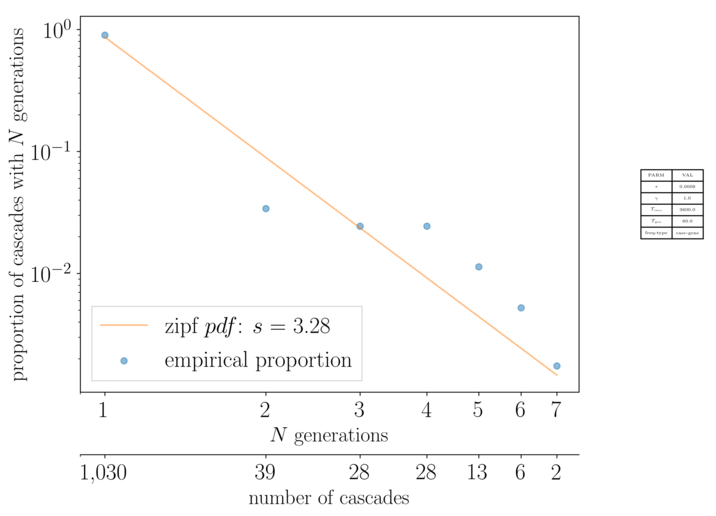}}%
		\end{subfigure}
		\begin{subfigure}[t]{\CascWidth\linewidth}
			\vspace{\failurecaption}%
			\renewcommand\thesubfigure{$\tau = 7 \times 10^{-4}$}
			\caption{}
		\end{subfigure}%
		\hfill%
		\begin{subfigure}[t]{\CascWidth\linewidth}
			\vspace{\failurecaption}%
			\renewcommand\thesubfigure{$\tau = 8 \times 10^{-4}$}
			\caption{}
		\end{subfigure}%
		\hfill%
		\begin{subfigure}[t]{\CascWidth\linewidth}
			\vspace{\failurecaption}%
			\renewcommand\thesubfigure{$\tau = 9 \times 10^{-4}$}
			\caption{}
		\end{subfigure}%
		\hfill%
		\label{fig:kmc-cascade-F}
	\end{minipage}
	\begin{minipage}{\linewidth}
		\begin{subfigure}[t]{\CascWidth\linewidth}%
			\centering
			\resizebox{\linewidth}{!}{\includegraphics[trim={0 0 3cm 0}, clip=true]{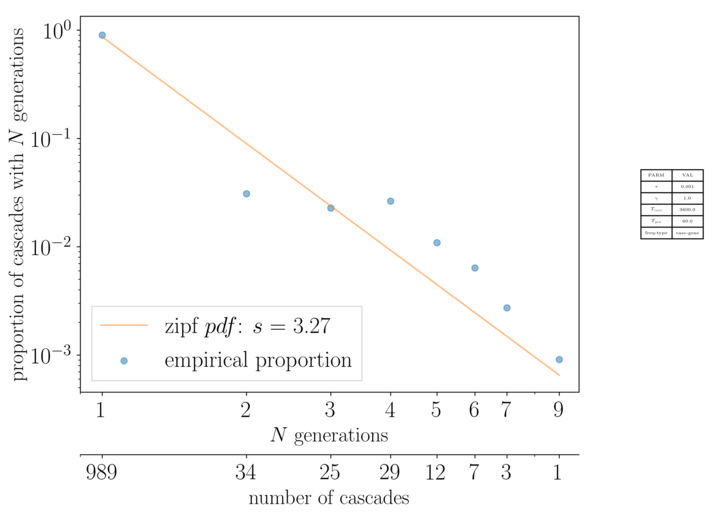}}%
		\end{subfigure}
		\begin{subfigure}[t]{\CascWidth\linewidth}%
			\centering
			\resizebox{\linewidth}{!}{\includegraphics[trim={0 0 3cm 0}, clip=true]{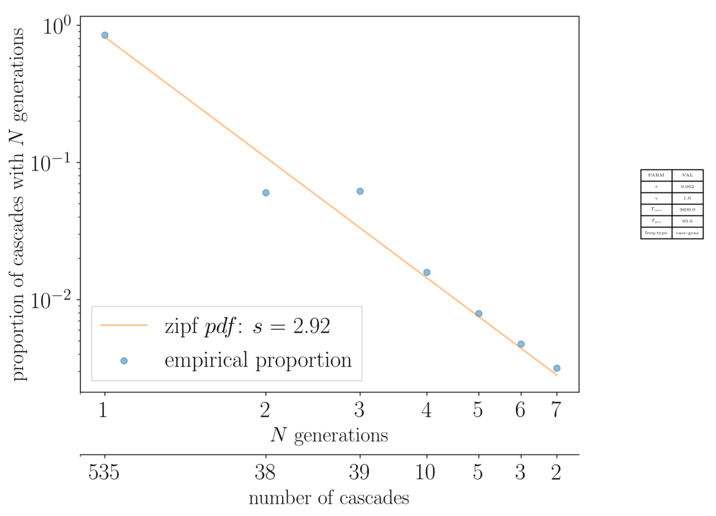}}%
		\end{subfigure}
		\begin{subfigure}[t]{\CascWidth\linewidth}%
			\centering
			\resizebox{\linewidth}{!}{\includegraphics[trim={0 0 3cm 0}, clip=true]{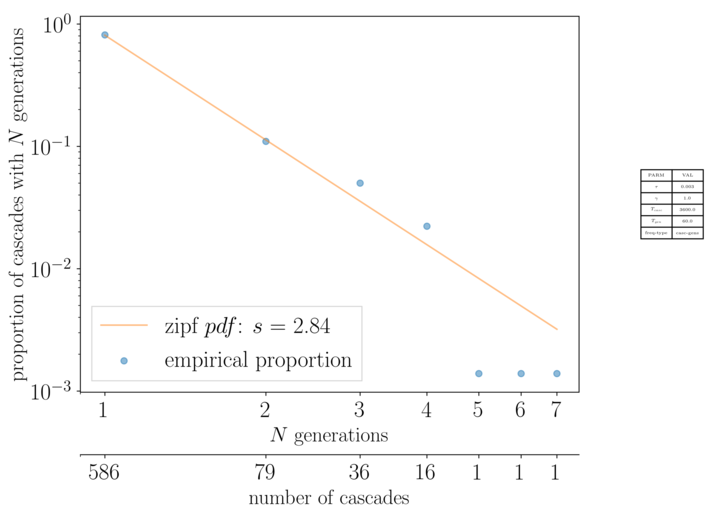}}%
		\end{subfigure}
		\begin{subfigure}[t]{\CascWidth\linewidth}
			\vspace{\failurecaption}%
			\renewcommand\thesubfigure{$\tau = 1 \times 10^{-3}$}
			\caption{}
		\end{subfigure}%
		\hfill%
		\begin{subfigure}[t]{\CascWidth\linewidth}
			\vspace{\failurecaption}%
			\renewcommand\thesubfigure{$\tau = 2 \times 10^{-3}$}
			\caption{}
		\end{subfigure}%
		\hfill%
		\begin{subfigure}[t]{\CascWidth\linewidth}
			\vspace{\failurecaption}%
			\renewcommand\thesubfigure{$\tau = 3 \times 10^{-3}$}
			\caption{}
		\end{subfigure}%
		\hfill%
		\label{fig:kmc-cascade-G}
	\end{minipage}
	\begin{minipage}{\linewidth}
		\begin{subfigure}[t]{\CascWidth\linewidth}%
			\centering
			\resizebox{\linewidth}{!}{\includegraphics[trim={0 0 3cm 0}, clip=true]{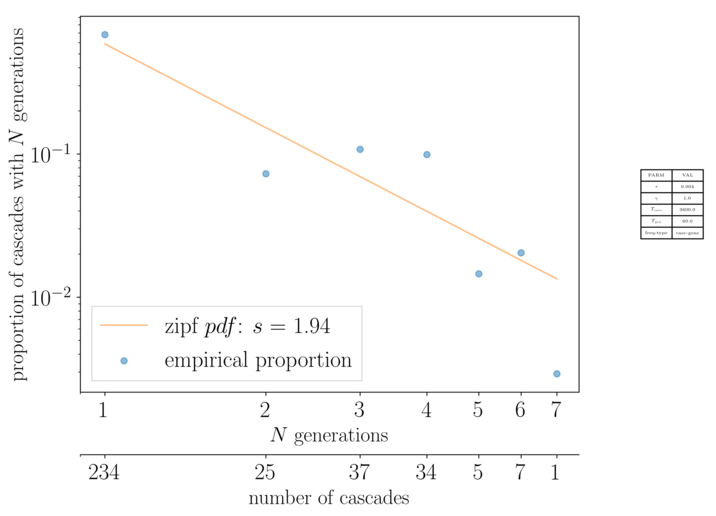}}%
		\end{subfigure}
		\begin{subfigure}[t]{\CascWidth\linewidth}%
			\centering
			\resizebox{\linewidth}{!}{\includegraphics[trim={0 0 3cm 0}, clip=true]{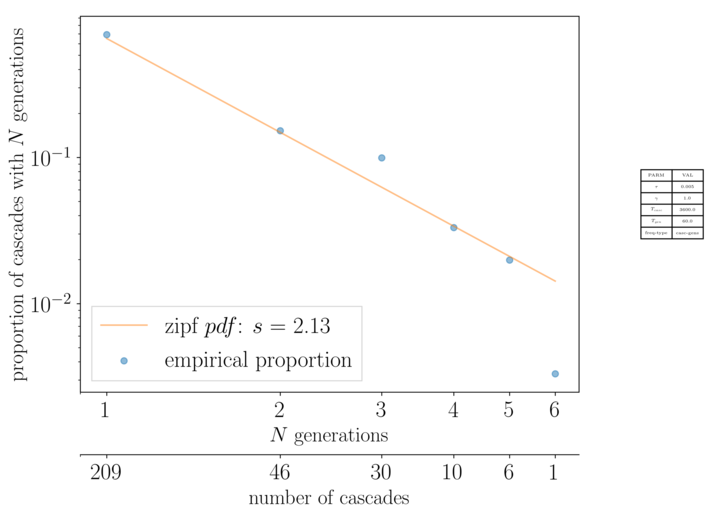}}%
		\end{subfigure}
		\begin{subfigure}[t]{\CascWidth\linewidth}%
			\centering
			\resizebox{\linewidth}{!}{\includegraphics[trim={0 0 3cm 0}, clip=true]{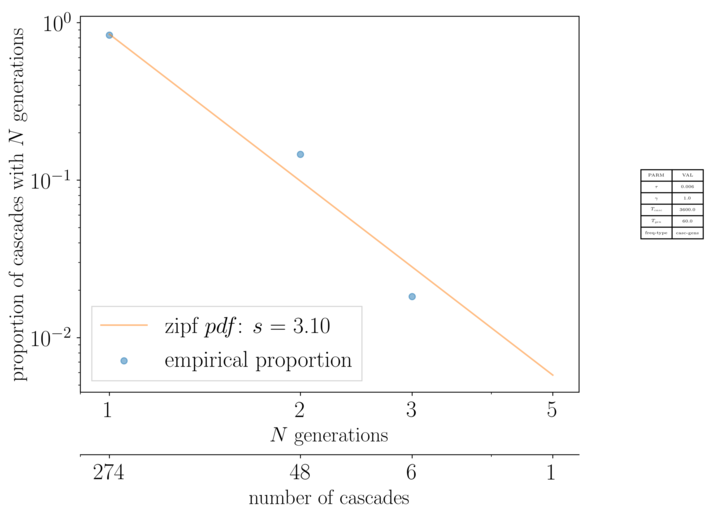}}%
		\end{subfigure}
		\begin{subfigure}[t]{\CascWidth\linewidth}
			\vspace{\failurecaption}%
			\renewcommand\thesubfigure{$\tau = 4 \times 10^{-3}$}
			\caption{}
		\end{subfigure}%
		\hfill%
		\begin{subfigure}[t]{\CascWidth\linewidth}
			\vspace{\failurecaption}%
			\renewcommand\thesubfigure{$\tau = 5 \times 10^{-3}$}
			\caption{}
		\end{subfigure}%
		\hfill%
		\begin{subfigure}[t]{\CascWidth\linewidth}
			\vspace{\failurecaption}%
			\renewcommand\thesubfigure{$\tau = 6 \times 10^{-3}$}
			\caption{}
		\end{subfigure}%
		\hfill%
		\label{fig:kmc-cascade-H}
	\end{minipage}
	\caption{Zipf plots of cascades with $N$ generations and $N$-1 line limits across temperature.}
	\label{fig:cascade-N1-lines}
\end{figure}
\restoregeometry

\bibliographystyle{siam}
\bibliography{_supplement}